\pdfoutput=1
\documentclass[a4paper]{article}

\usepackage{hyperref}
\usepackage{amsmath,amsfonts,amssymb}
\usepackage{amsthm}
\usepackage{color}
\usepackage{tikz}
\usepackage{graphicx}
\usepackage{cite}
\usepackage{array}
\usepackage{mathtools}
\usetikzlibrary {decorations.pathmorphing,shadows,fadings}

\def\ds{\displaystyle}
\def\bea{\begin{array}{c}}
\def\ea{\end{array}}
\def\be{\begin{equation}\bea\ds}
\def\ee{\ea\end{equation}}
\def\bee{\begin{equation}\begin{array}{rcl}\ds}
\def\eee{\end{array}\end{equation}}

\def\tr{{\rm Tr}\,}

\def\Lc{{\mathcal{L}}}
\def\Hc{{\mathcal{H}}}
\def\Mc{{\mathcal{M}}}
\def\Oc{{\mathcal{O}}}

\def\tr{{\rm tr}\,}
\def\Tr{{\rm Tr}\,}

\newcommand\Trule{\rule{0pt}{2.5ex}}
\newcommand\Brule{\rule[-1.ex]{0pt}{0pt}}

\title{Connectomes and Properties of Quantum Entanglement}

\author{Dmitry Melnikov}
\date{}

\begin{document}

\thispagestyle{empty}

\maketitle

\begin{center}
\textit{\small International Institute of Physics, Federal University of 
Rio Grande do Norte, \\ Campus Universit\'ario, Lagoa Nova, Natal-RN  
59078-970, Brazil}
\\ \vspace{0.6cm}
\textit{\small Institute for Theoretical and Experimental Physics, NRC Kurchatov Institute, \\ B.~Cheremushkinskaya 25, 117218 Moscow, Russia}
\\ \vspace{2cm}

\end{center}

\vspace{-1cm}

\begin{abstract}
Topological quantum field theories (TQFT) encode properties of quantum states in the topological features of abstract manifolds. One can use the topological avatars of quantum states to develop intuition about different concepts and phenomena of quantum mechanics. In this paper we focus on the class of simplest topologies provided by a specific TQFT and investigate what the corresponding states teach us about entanglement. These ``planar connectome" states are defined by graphs of simplest topology for a given adjacency matrix. In the case of bipartite systems the connectomes classify different types of entanglement matching the classification of stochastic local operations and classical communication (SLOCC). The topological realization makes explicit the nature of entanglement as a resource and makes apparent a number of its properties, including monogamy and characteristic inequalities for the entanglement entropy. It also provides tools and hints to engineer new measures of entanglement and other applications. Here the approach is used to construct purely topological versions of the dense coding and quantum teleportation protocols, giving diagrammatic interpretation of the role of entanglement in quantum computation and communication. Finally, the topological concepts of entanglement and quantum teleportation are employed in a simple model of information retrieval from a causally disconnected region, similar to the interior of an evaporating black hole.

\end{abstract} 

\section{Introduction}

Topological Quantum Field Theories (TQFT) are instances of ordinary, often finite-dimensional, quantum mechanical systems~\cite{Witten:1988ze,Atiyah:1989vu}. A special feature of these instances is the possibility to cast quantum states and quantum operations as topological spaces, which realize the defining properties of Hilbert spaces. In other words, in TQFT quantum states and quantum operations (operators) can be visualized by drawing diagrams of topological spaces and manipulating them.

Diagrammatic representation of linear operations is a very natural way to visualize matrix multiplication, widely used in different areas of physics,  mathematics and computer science, including tensor networks and computer algorithms. However, in most cases such representations are symbolic and do not literally reflect the physical processes in the manipulated systems. In TQFT, on the contrary, the diagrams can be understood literally, as quantum evolution of the system's elements, such as particles, qubits, or general media composing the systems.

In this paper we will work with a topological version of tensor networks based on the axiomatic definition of TQFT~\cite{Atiyah:1989vu}. Similar constructions are well-known under the name of topological quantum computation~\cite{Kitaev:1997wr,Kitaev:2000nmw,Freedman:2000rc,Nayak:2008zza}, but the focus of this paper will be in different aspects of the construction. In particular, we will discuss what quantum entanglement is from the point of view of topology and show how topology highlights some of its fundamental properties. In this sense the work continues some previous discussion in the literature (e.g~\cite{Salton:2016qpp,Balasubramanian:2016sro,Balasubramanian:2018por,Melnikov:2017bjb,Melnikov:2018zfn,Kauffman:2019top} and earlier works by Kauffman et al. cited in~\cite{Kauffman:2013bh}) inspired by the question of the connection between topological and quantum entanglement raised explicitly in~\cite{Aravind:1997}.

We will consider a specific example of a TQFT, that is Chern-Simons theory with gauge group $SU(2)$ in spaces, which are locally $\mathbb{R}^3$, with boundaries carved as disjoint 2-spheres. We will allow the 2-spheres to have punctures -- point-like defects that have to be extended in the three-dimensional bulk as one-dimensional lines. These are the Wilson lines from the point of view of the Chern-Simons theory. Three-dimensional manifolds with boundaries and Wilson lines represent states and more general tensors of the quantum theory, with boundary $S^2$ corresponding to elementary subsystems, that is Hilbert spaces of individual ``particles''~\cite{Witten:1988hf}. The key observation in this construction is that spheres connected by a sufficient number of Wilson lines correspond to entangled subsystems~\cite{Melnikov:2018zfn}. So, entanglement corresponds to wiring of the bulk space with Wilson lines. We refer to such wirings as ``connectomes'', borrowing the term from neuroscience.

Obviously the same space can be wired in different ways, so the first question is what kind of entanglement different wirings describe. Here we focus on the choice of wirings with the simplest topology. The essential information that such wirings should contain is what is connected to what. Such objects can be characterized by classes labeled by the adjacency matrices of graphs, whose nodes are associated with the subsystems and edges -- with the Wilson lines. The classes contain infinite number of elements and we would like to consider only the simplest representatives, which do not have non-trivial knotting and tangling of the Wilson lines, the planar (trivial) connectomes. 

In this work we will mostly focus on these simplest connectome quantum states and use them to illustrate different features and applications of quantum entanglement. We will find that such states have some distinctive features. The entanglement entropy for such states share the properties with the holographic states\footnote{That is states that are expected to have a dual gravity description.}: the entropy of a single subsystem is given by a discrete version of the minimal area law, while for many subsystems, the entropy satisfies a number of inequalities beyond subadditivity, which are also satisfied by the holographic states. For bipartite entanglement the planar connectome states are equivalent to the classes of states in the classification provided by the action of stochastic local operations and classical communication (SLOCC)~\cite{Dur:2000zz,Melnikov:2022qyt}. For multipartite entanglement they are similar to either full multipartite GHZ states or to the embeddings of lower rank GHZ states. 

The main advantage of the topological approach to description of entanglement is that it makes the properties of entanglement very intuitive: the states are entangled if the topological spaces are properly connected; shared Wilson lines is the entanglement resource shared between the parties; impossibility of sharing this resource with several parties simultaneously is the monogamy of entanglement. One of the goals of this work is to show that the topological interpretation can motivate new tools for study and applications of entanglement. In this work we use the topology argument to construct a new measure of multipartite entanglement. We also find that basic quantum algorithms, such as dense coding and quantum teleportation have purely topological interpretation, which makes visual the role of different aspects of entanglement in quantum computation and communication. 

As another application of the topological method we reflect on the recent progress in the understanding of black holes and propose a toy model that contains the salient features of an evaporating black hole in the context of the information paradox~\cite{Hawking:1976ra}. In this model we build upon our experience with the planar connectome states, which are supposedly similar to the holographic ones. The way in which the Hawking radiation gets entangled with the interior and later destroys the entanglement, allowing the information to escape, is analogous to the topological realization of quantum teleportation~\cite{Melnikov:2022vij}.

The paper is organized as follows. In section~\ref{sec:TQFT} we give a short introduction in TQFT and describe a specific realization of quantum mechanics (qubits and entanglement) in Chern-Simons theory with $S^2$ boundaries. In section~\ref{sec:classification} we introduce the connectome states and review the classification of bipartite entanglement via connectomes. In section~\ref{sec:multipartite} we briefly discuss connectomes in the multipartite situation. Section~\ref{sec:applications} contains the main discussion of entanglement properties and applications. In section~\ref{sec:measures} we discuss the topological expression for the entanglement entropy and introduce a new measure, which detects multipartite entanglement. In section~\ref{sec:properties} we discuss inequalities for the entanglement entropy focusing on the planar connectome states. In section~\ref{sec:algorithms} we propose topological cartoons of the dense coding and quantum teleportation protocols. Finally, section~\ref{sec:blackhole} describes a model of unitary evaporation in a topological interpretation of the black hole information paradox.

Some of the results discussed in this paper were previously reported in~\cite{Melnikov:2022qyt} and~\cite{Melnikov:2022vij}. This paper makes a more detailed discussion of those results and offers a number of new ones. For example, the discussion of a new entanglement measure, the inequalities for the entanglement entropy and the dense coding protocol are the main new results. Moreover, this paper is written for a more general audience and is expected to be self-contained.


\section{Topological quantum field theory}
\label{sec:TQFT}

\subsection{Definitions}
\label{sec:defs}

Following the axiomatic definition of TQFT~\cite{Atiyah:1989vu} we will consider an $n$ dimensional orientable hypersurface $\Sigma$ as a Hilbert space $\Hc_{\Sigma}$. Then any $n+1$-dimensional topological space $\Mc$, such that $\Sigma$ is a boundary of $\Mc$, that is $\Sigma = \partial\Mc$, corresponds to a vector in $\Hc_\Sigma$, as illustrated by the following diagram,
\be
\label{TQFTstate}
\begin{array}{c}
    \includegraphics[scale=0.15]{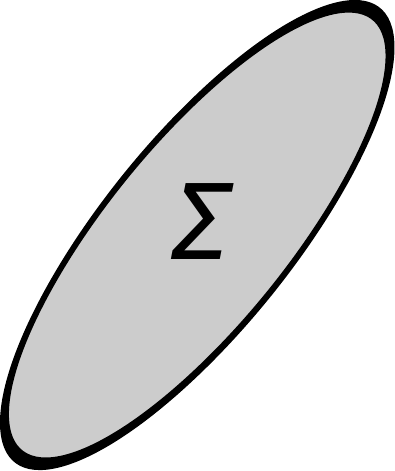}
    \end{array} \ \longrightarrow \ \Hc_\Sigma\,, \qquad \qquad
\begin{array}{c}
    \includegraphics[scale=0.12]{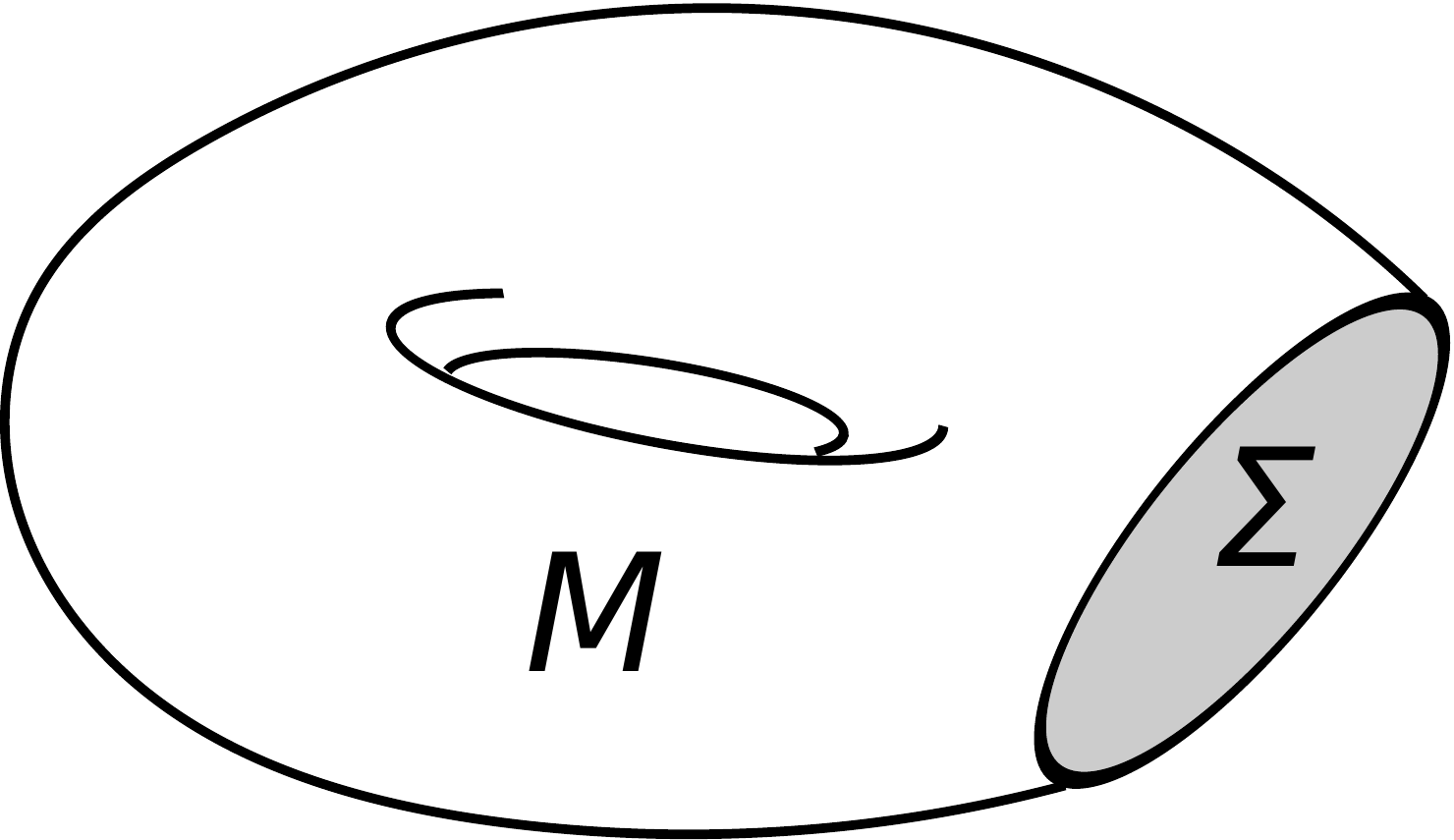}
    \end{array}\ \longrightarrow \ |\Psi\rangle\ \in \ \Hc_\Sigma\,.
\ee
Note that spaces homeomorphic to each other, that is continuously deformable into each other, preserving the topology, define equivalent states.

We can also consider an $n+1$-dimensional space $\Oc$ with boundaries $\partial\Oc = \Sigma\cup \overline{\Sigma}$ as an evolution of the Hilbert space $\Hc_\Sigma$, that is an operator acting on $\Hc_\Sigma$. Here ${\Sigma}$ and $\overline{\Sigma}$ differ by choice of orientation, so one may think of $\overline{\Sigma}$ as representing the dual vector space $\Hc_{\overline{\Sigma}}=\Hc_\Sigma^\ast$. More generally, space $\Oc$ with boundary $\partial\Oc=\Sigma_1\cup \Sigma_2$ can be viewed as a linear operator that acts on the respective Hilbert spaces. Such space is also called a cobordism of $\Sigma_1$ and $\Sigma_2$.

Application of an operator on a state is realized by gluing boundary $\Sigma$ of the state with boundary $\overline{\Sigma}$ of the operator, as the following diagram shows. 
\be
\begin{array}{c}
    \includegraphics[scale=0.25]{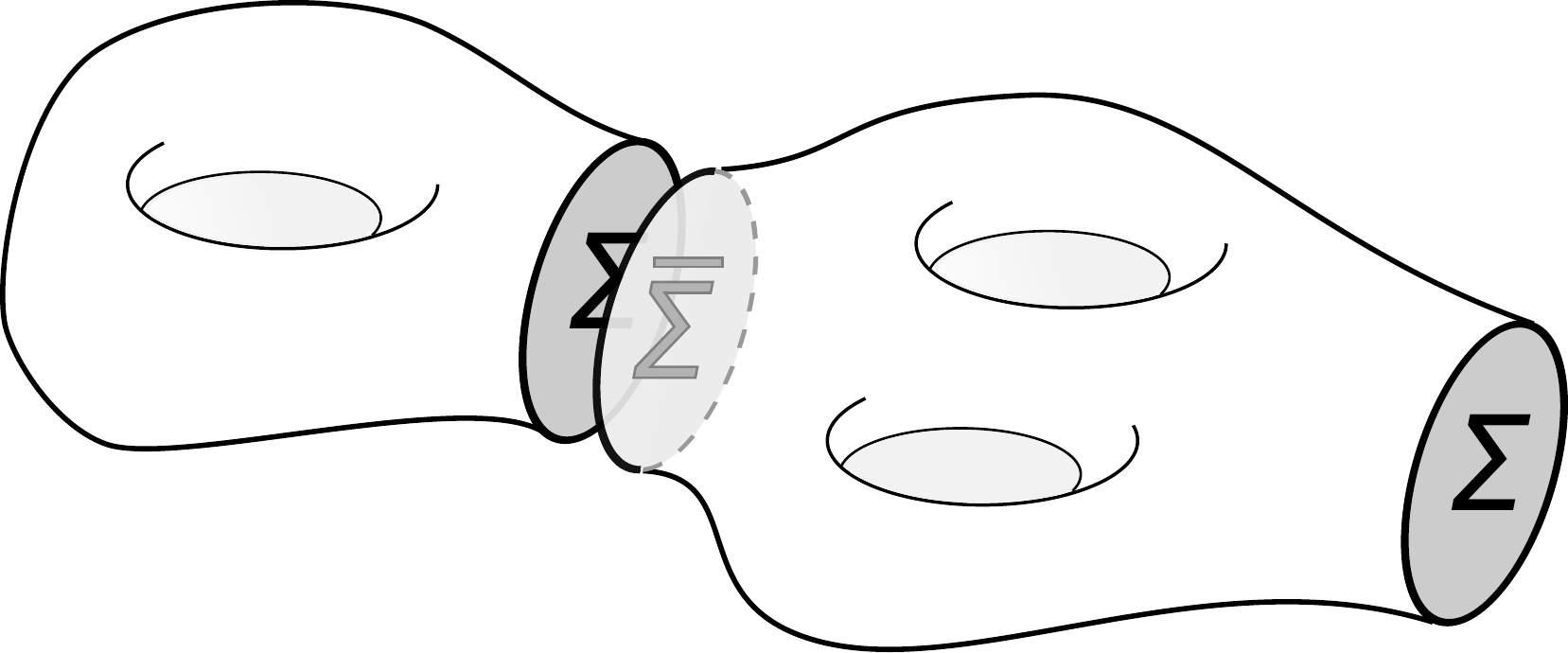}
    \end{array} \quad \longrightarrow \quad  O\cdot|\Psi\rangle\,.
\ee
The result of this operation is obviously another state in $\Hc_\Sigma$. Similarly, composition of operators is a concatenation of the latter.

For completeness of the axiomatic definition we need to spell out a few other properties. It should be obvious that gluing together two homeomorphic spaces-states should correspond to the square of the norm of the state. More generally, gluing two inequivalent spaces should produce a diagram of the internal product in the Hilbert space,
\be
\begin{array}{c}
 \includegraphics[scale=0.2]{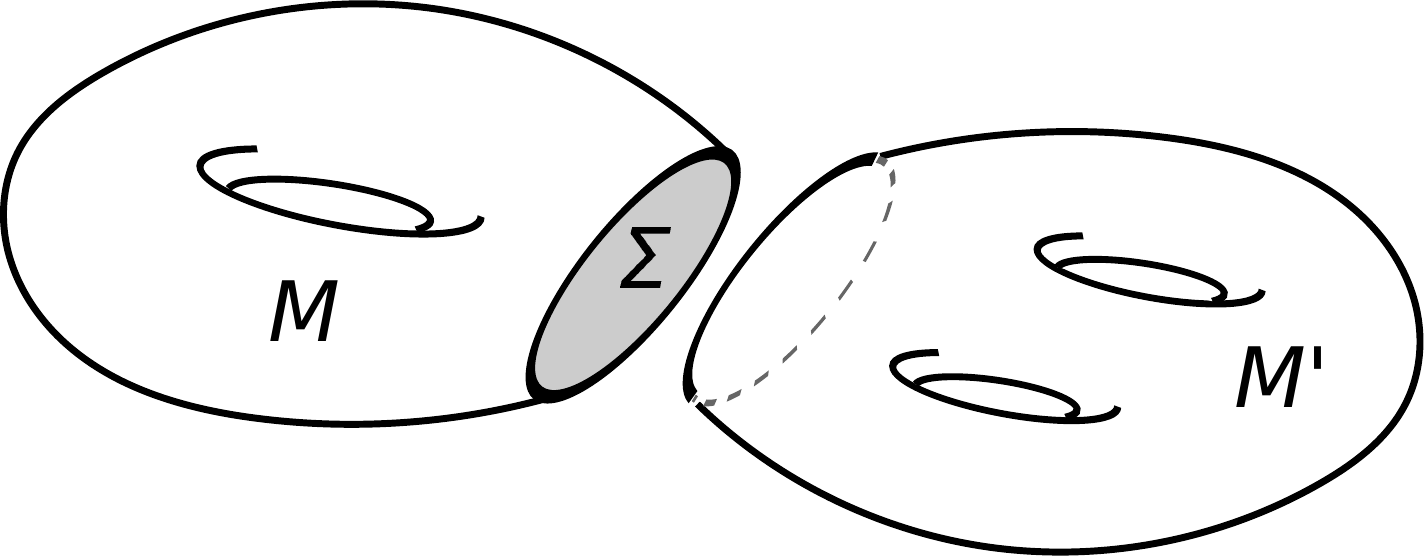}
 \end{array}
 \ \longrightarrow \quad   
 \begin{array}{c}
 \includegraphics[scale=0.2]{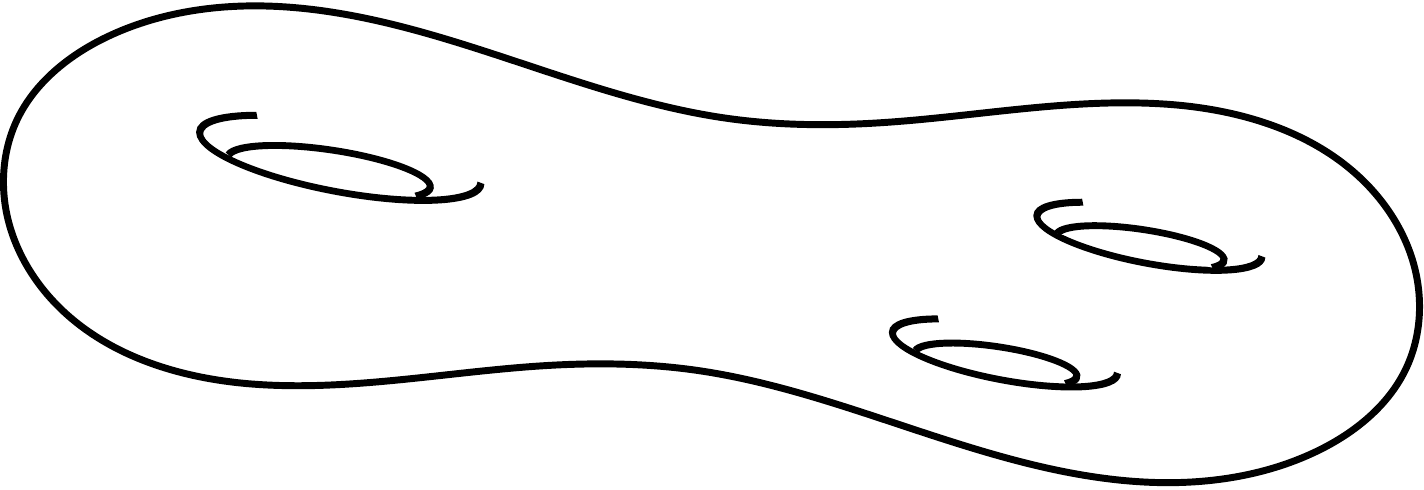}
 \end{array}
  \ \longrightarrow \ \langle\Psi|\Psi'\rangle
\ee
The result of the gluing is a space without boundary. Spaces without boundary thus correspond to zero-dimensional Hilbert spaces or $\mathbb{C}$-numbers. We also need to define the diagrammatic analog of the identity operator. The latter should represent a trivial evolution of the Hilbert space, which is achieved by gluing a featureless cylinder $\Sigma\otimes I$ ($I$ being an interval) or any space homeomorphic to it.

Finally the last property that plays the major role in this paper is the consequence of separability of topological spaces. Namely, if $\Sigma$ consists of disconnected components $\Sigma=\Sigma_1\cup \Sigma_2$ ($\Sigma_1\cap\Sigma_2=\emptyset$) then it corresponds to a direct product of Hilbert spaces $\Hc_\Sigma=\Hc_{\Sigma_1}\otimes \Hc_{\Sigma_2}$.

The map of topological spaces to linear spaces described above is realized by path integrals of metric-independent quantum field theories, with the primary example provided by the three-dimension Chern-Simons theory~\cite{Witten:1988hf}. This example motivated the general definition of TQFT~\cite{Witten:1988ze}. In particular, states in quantum Chern-Simons theory are path integrals of the on manifolds $\Mc$ with prescribed boundary conditions for the fields on boundary $\Sigma$.

\subsection{Topological qubits}
\label{sec:qubits}

Now let us describe a specific realization of TQFT axioms and a particular class of Hilbert spaces. In other words, we describe the qubits and qudits that we will be working with in this paper.

We will consider spaces $\Sigma$ that are disjoint unions of two-spheres $S^2$ and assume the underlying CFT to be Chern-Simons theory with gauge group $SU(2)$ and coupling constant (level) $k$. The main information that we will need here about this theory is the dimension of the Hilbert spaces on the spheres and a few rules of assigning matrix operators to topological diagrams, or equivalently computing scalar products of states. These will be introduced in due course.

It turns out that in the Chern-Simons theory, the Hilbert space of a featureless two-sphere is one-dimensional~\cite{Witten:1988hf}. In order to have a non-trivial Hilbert space one has to consider spheres with punctures. From the point of view of the Chern-Simons theory punctures are special points of the Chern-Simons field and can be viewed as external particles coupled to the field. In the $SU(2)$ theory these particles are characterized by a spin $j$ (to be precise, by ``integrable'' representations $R$ of Kac-Moody algebra $su(2)_k$). In order to consistently put $n$ particles with spins $j_1$, $j_2$, \ldots $j_n$ on a sphere these particles should form a spin singlet (tensor product of irreps $R_i$ should contain a trivial irrep). The dimension of the Hilbert space of $S^2$ is defined by the number of possible ways of forming a singlet from $n$ spins (number of trivial irreps in the tensor product of $R_i$) with the following caveat. In $su(2)_k$ Kac-Moody algebra there is only a finite number of integrable representations labeled by spin $0\leq j\leq k/2$. As a result, the dimension of the Hilbert space may be less than the number of singlets formed by the tensor product of the puncture spins, unless $k/2$ is larger than any of the spins that can appear in the tensor products. In the rest of the paper we will assume $k$ to be large, so that the dimension is precisely the number of singlets in the tensor product.

Obviously a sphere cannot have a single puncture, unless it corresponds to a particle of spin zero. The latter case is equivalent to absence of any punctures. For two punctures, in order to have a Hilbert space, the associated particles must have equal spins. There is only one way to form a singlet of two spins, so the dimension of the Hilbert space is at most one, as for the zero-puncture case. We will be mostly interested in the situation of the punctures-particles with spin $1/2$. Three particles of spin half cannot form a singlet. The minimal non-trivial example of a Hilbert space comes with four punctures of spin $1/2$. There are two ways of forming a singlet from four spin-half particles, by pairwise forming singlets or triplets, so $S^2$ with four spin-half punctures correspond to a Hilbert space of dimension two -- the qubit.

\begin{figure}
    \centering
    \includegraphics[width=0.2\linewidth]{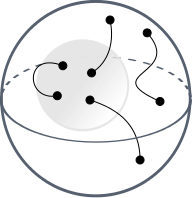}
    \qquad\qquad
    \includegraphics[width=0.15\linewidth]{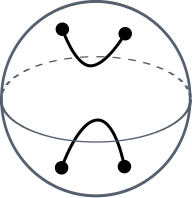}
    \quad 
    \includegraphics[width=0.15\linewidth]{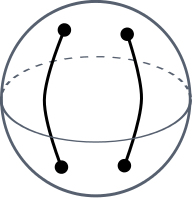}
    \caption{(Left) An example of a three-dimensional space bounded by a pair of two-spheres $S^2$. Each sphere has four punctures. In the three-dimensional space the punctures are connected by one dimensional defects (Wilson lines). (Right) A pair of basis states for a TQFT on $S^2$.}
    \label{fig:s2punctures}
\end{figure}

In a three-dimensional bulk space $\Mc$ punctures are extended to one-dimensional defects. In the Chern-Simons language these defects are trajectories of the boundary particles, called Wilson lines. Wilson lines cannot end in the bulk: they either end on the same two-sphere or on two different spheres included in $\Sigma$. An example of a two-qubit state with different options for the Wilson lines is shown in figure~\ref{fig:s2punctures} (left). Besides, the bulk of $\Mc$ can contain closed Wilson lines -- Wilson loops.

The next question is choice of a basis in the Hilbert space. We start by choosing two inequivalent extensions of the punctures in the bulk, as shown by figure~\ref{fig:s2punctures} (right). In the following sections, we will not draw the spheres explicitly, but will rather group the punctures assuming that each group belongs to the surface of a two-sphere. We will not draw the three-dimensional spaces either, assuming they have a topology of $S^3$, so the pair of states shown in figure~\ref{fig:s2punctures} (right) will be simply denoted as
\be
\label{4pointbasis0}
|e_1\rangle \ = \ \begin{array}{c}
\begin{tikzpicture}[thick]
\fill[black] (0,0.0) circle (0.05cm);
\fill[black] (0,0.3) circle (0.05cm);
\fill[black] (0,0.6) circle (0.05cm);
\fill[black] (0,0.9) circle (0.05cm);
\draw (0,0) -- (0.4,0) arc (-90:90:0.15cm) -- (0,0.3);
\draw (0,0.6) -- (0.4,0.6) arc (-90:90:0.15cm) -- (0,0.9);
\end{tikzpicture} 
\end{array}\,, \qquad |e_2\rangle \ = \ \begin{array}{c}
\begin{tikzpicture}[thick]
\fill[black] (0,0.0) circle (0.05cm);
\fill[black] (0,0.3) circle (0.05cm);
\fill[black] (0,0.6) circle (0.05cm);
\fill[black] (0,0.9) circle (0.05cm);
\draw (0,0) -- (0.1,0) arc (-90:90:0.45cm) -- (0,0.9);
\draw (0,0.6) -- (0.1,0.6) arc (90:-90:0.15cm) -- (0,0.3);
\end{tikzpicture} 
\end{array}\,.
\ee
These two diagrams are not homeomorphic (one has to assume the position of the punctures fixed), so the two diagrams correspond to two linearly independent states. The states are not orthogonal though. To construct an orthogonal basis from the above pair of states one needs to know how to compute scalar products in this Hilbert space.

From the definitions in section~\ref{sec:defs} one has the following three overlaps to compute,
\be
\langle e_1|e_1\rangle \ = \ \begin{array}{c}
\begin{tikzpicture}[thick]
\fill[black] (0,0.0) circle (0.05cm);
\fill[black] (0,0.3) circle (0.05cm);
\fill[black] (0,0.6) circle (0.05cm);
\fill[black] (0,0.9) circle (0.05cm);
\draw (0,0) -- (0.4,0) arc (-90:90:0.15cm) -- (0,0.3);
\draw (0,0.6) -- (0.4,0.6) arc (-90:90:0.15cm) -- (0,0.9);
\draw (0,0) -- (-0.4,0) arc (-90:-270:0.15cm) -- (0,0.3);
\draw (0,0.6) -- (-0.4,0.6) arc (-90:-270:0.15cm) -- (0,0.9);
\end{tikzpicture} 
\end{array}\,, \qquad \langle e_1|e_2\rangle \ = \ \begin{array}{c}
\begin{tikzpicture}[thick]
\fill[black] (0,0.0) circle (0.05cm);
\fill[black] (0,0.3) circle (0.05cm);
\fill[black] (0,0.6) circle (0.05cm);
\fill[black] (0,0.9) circle (0.05cm);
\draw (0,0) -- (0.1,0) arc (-90:90:0.45cm) -- (0,0.9);
\draw (0,0.6) -- (0.1,0.6) arc (90:-90:0.15cm) -- (0,0.3);
\draw (0,0) -- (-0.4,0) arc (-90:-270:0.15cm) -- (0,0.3);
\draw (0,0.6) -- (-0.4,0.6) arc (-90:-270:0.15cm) -- (0,0.9);
\end{tikzpicture} 
\end{array}
\,, \qquad \langle e_2|e_2\rangle \ = \ \begin{array}{c}
\begin{tikzpicture}[thick]
\fill[black] (0,0.0) circle (0.05cm);
\fill[black] (0,0.3) circle (0.05cm);
\fill[black] (0,0.6) circle (0.05cm);
\fill[black] (0,0.9) circle (0.05cm);
\draw (0,0) -- (0.1,0) arc (-90:90:0.45cm) -- (0,0.9);
\draw (0,0.6) -- (0.1,0.6) arc (90:-90:0.15cm) -- (0,0.3);
\draw (0,0) -- (-0.1,0) arc (-90:-270:0.45cm) -- (0,0.9);
\draw (0,0.6) -- (-0.1,0.6) arc (90:270:0.15cm) -- (0,0.3);
\end{tikzpicture} 
\end{array}
\ee
In this exercise one glues two three-balls along a two-sphere. The result is a three-sphere $S^3$ with the above Wilson loops inside. The celebrated result of Witten~\cite{Witten:1988hf} is that such partition functions of the $SU(2)$ Chern-Simons theory are computed by Jones polynomials of the corresponding knots (links) of variable 
\be
\label{qdef}
q\ = \ \exp\left(\frac{2\pi i}{k+2}\right).
\ee

In order to compute overlaps of quantum states that are given by knots and links in $S^3$ it will be sufficient to use the skein relations of Conway, which we summarize here using the conventions of Kauffman~\cite{Kauffman:1987sta}. Let us denote by $J(\Lc)$ the Jones polynomial of link $\Lc$. First, we normalize the polynomial in such a way that $J(\emptyset)=1$, that is empty $S^3$ corresponds to the trivial polynomial. Second, the Jones polynomial of a diagram with a topologically trivial loop (the one that can be contracted to a point) is equal to the Jones polynomial of the same diagram, but with the loop removed, times a numerical factor,
\be
\label{Jonesrules}
J(\begin{tikzpicture}[thick]
\draw[black] (0,0.0) circle (0.15cm);
\end{tikzpicture} 
\cup \Lc) \ = \ d\cdot J(\Lc)\,, \qquad d = - A^2 - A^{-2}\,, \qquad A\ = \ q^{1/4}\,.  
\ee
Here $A$ is the variable of Kauffman's ``bracket polynomial'', which we will mostly use. These two rules are sufficient to compute the above overlap, but if the diagram contains a non-trivial knot then to compute it one also needs the skein relation
\be
\label{skein}
\begin{tikzpicture}[baseline=0]
\draw[thick] (0.5,-0.1) -- (0.3,-0.1) -- (0.15,0);
\draw[thick] (-0.15,0.2) -- (-0.3,0.3) -- (-0.5,0.3);
\draw[thick] (-0.5,-0.1) -- (-0.3,-0.1) -- (0.3,0.3) -- (0.5,0.3);
\end{tikzpicture}  \quad = \ A\quad \begin{tikzpicture}[baseline=0]
\draw[thick] (-0.5,-0.1) -- (-0.3,-0.1) arc (-90:90:0.2) -- (-0.5,0.3);
\draw[thick] (0.5,-0.1) -- (0.3,-0.1) arc (-90:-270:0.2) -- (0.5,0.3);
\end{tikzpicture} \quad +\ A^{-1}\quad \begin{tikzpicture}[baseline=0]
\draw[thick] (0.5,-0.1) -- (-0.5,-0.1);
\draw[thick] (0.5,0.3) -- (-0.5,0.3);
\end{tikzpicture}
\ee
This relation tells that any diagram with a crossing can be replaced by a linear combination of diagrams with no crossings. The Jones polynomial acts linearly on the linear combination of diagrams. 

With the above rules, one can now construct an orthonormal basis for the qubit,
\be
\label{4pointbasis}
|0\rangle \ = \ \frac{1}{d}|e_1\rangle\,, \qquad |1\rangle \ = \ \frac{1}{\sqrt{d^2-1}}\left(|e_2\rangle - \frac{1}{d} |e_1\rangle\right)\,. 
\ee
Note that for $k=1$ parameter $d^2-1$ vanishes, which means that $|e_1\rangle$ and $|e_2\rangle$ are linearly dependent, which is a consequence of the fact that the dimension of the Hilbert space has a non-trivial dependence of $k$.

It is useful to note that loop factorization and skein relations apply not only to calculations in $S^3$, but also to any three-dimensional manifold with boundaries. This will be useful in the discussion of many particle states and operators. The rules for three-manifolds with different global structure follow automatically from the above, since any manifold can be constructed by gluing three-spheres with $S^2$ boundaries.

To construct a basis in the case of an arbitrary even number of punctures one needs to consider all diagrams that connect the points without intersecting lines. For $2n$ points, any such diagram can be mapped to an element of the Temperley-Lieb algebra $TL_n$. The dimension of the Temperley-Lieb algebra is given by the Catalan numbers, so the dimension of the Hilbert space of a two-sphere with $2n$ punctures in $SU(2)$ Chern-Simons theory with sufficiently large level $k$ is given by
\be
\dim \Hc_{2n} \ = \ C_n \ = \ \frac{(2n)!}{(n+1)!n!}\,, \qquad k>n-1\,.
\ee

\subsection{Quantum entanglement}

In this paper we discuss properties of quantum entanglement as seen by the topological description of quantum mechanics. The map from topological to linear spaces offers a very natural interpretation for entanglement: separability of spaces implies separability of wavefunctions. We can illustrate this by the following heuristic diagrams, 
\be
\label{entanglement}
\begin{array}{c}
     \includegraphics[scale=0.2]{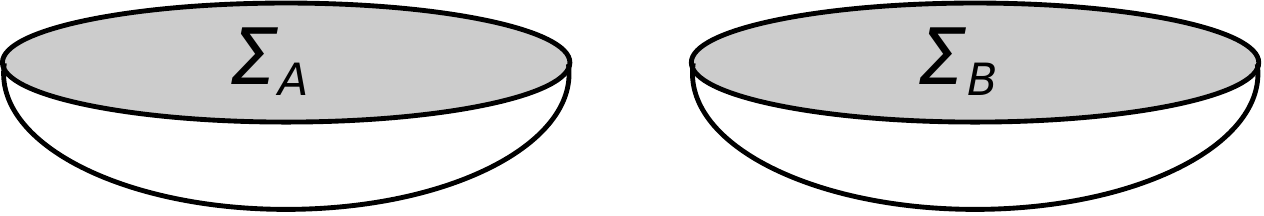}
\end{array} \quad \longrightarrow \quad \text{separable,} \qquad\qquad
\begin{array}{c}
\includegraphics[scale=0.2]{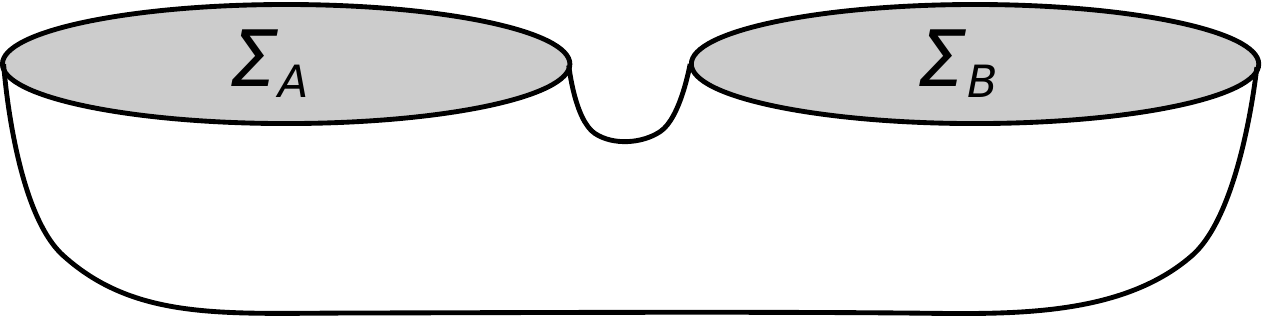}
\end{array} \quad \longrightarrow \quad \text{entangled.} 
\ee
Here $\Sigma_A$ and $\Sigma_B$ may represent a pair of two spheres. In the latter case the first diagram can be a pair of three-balls, and the second -- a space between two concentric $S^2$, but in principle the diagrams are supposed to illustrate the most general situation.\footnote{Another interesting example is the case of $\Sigma=T^2$, a two-dimensional torus. Entanglement for states with linked $T^2$ boundaries (knot complement states) was originally discussed in~\cite{Balasubramanian:2016sro}.} 

One can formally prove the correspondence described by diagrams~(\ref{entanglement}), for example, by computing the von Neumann entropies~\cite{Melnikov:2018zfn}, but there are some subtleties. The second diagram in~(\ref{entanglement}) is in general an entangled state, but there are situations, in which it is actually separable. One such situation happens when either $\Sigma_A$ or $\Sigma_B$ correspond to Hilbert spaces of dimension one. Therefore, for entanglement one needs non-trivial $\Sigma$. Moreover, if we understand the diagram as an evolution $\Sigma[t]$ with $\Sigma[0]=\Sigma_A$ and $\Sigma[1]=\Sigma_B$, at no intermediate $0\leq t\leq 1$ can the dimension of $\Hc_\Sigma$ be non-trivial, otherwise the diagram corresponds to a separable state.

For two-spheres this means that we will always need Wilson lines to support entanglement. For Wilson lines in the representation $j=1/2$, we will need at least four lines crossing any section of the three-dimensional space $\Mc$ that breaks it into a disconnected pair of three-manifolds. Therefore, we have the following refinement of the simple classification~(\ref{entanglement}),
\be
\label{entanglement2}
\begin{array}{c}
\includegraphics[scale=0.2]{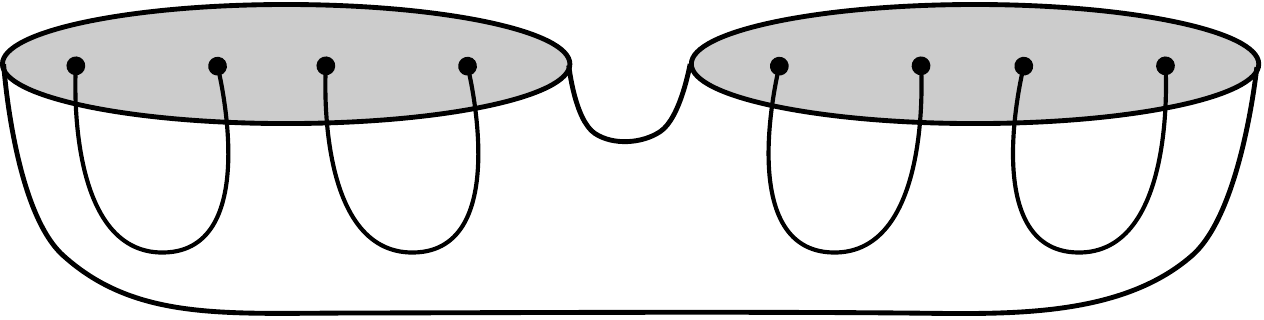}
\end{array} \quad \longrightarrow \quad \text{separable}\,, \qquad\qquad 
\begin{array}{c}
     \includegraphics[scale=0.2]{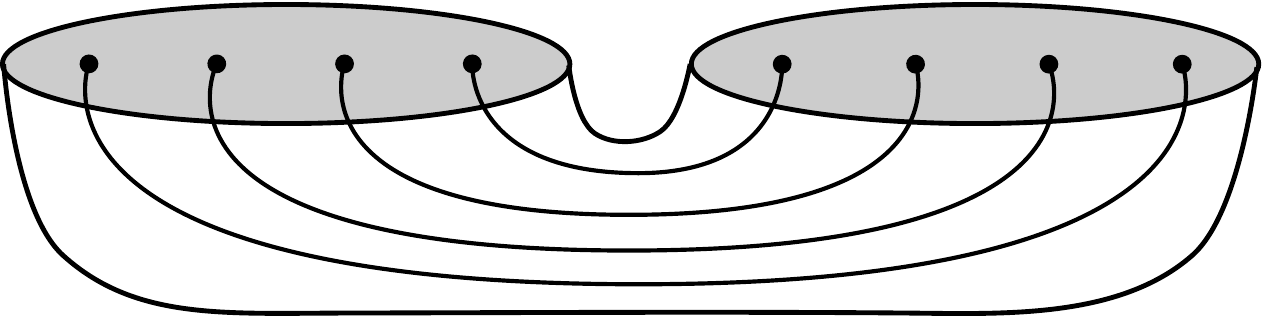}
\end{array} \quad \longrightarrow \quad \text{entangled.}
 \ee

Since the Wilson lines are fundamental for creating entanglement between 2-spheres, in this paper we will focus on the Wilson-line wiring of three-dimensional spaces. As already mentioned before we will not draw spheres, where the lines end, nor we will draw the three-dimensional spaces, essentially reducing the study to $S^3$ topologies. Classification~(\ref{entanglement2}) will be cast as
\be
\begin{array}{c}
\begin{tikzpicture}[thick]
\fill[black] (0,0.0) circle (0.05cm);
\fill[black] (0,0.3) circle (0.05cm);
\fill[black] (0,0.6) circle (0.05cm);
\fill[black] (0,0.9) circle (0.05cm);
\draw (0,0) -- (0.4,0) arc (-90:90:0.15cm) -- (0,0.3);
\draw (0,0.6) -- (0.4,0.6) arc (-90:90:0.15cm) -- (0,0.9);
\fill[black] (1.2,0.0) circle (0.05cm);
\fill[black] (1.2,0.3) circle (0.05cm);
\fill[black] (1.2,0.6) circle (0.05cm);
\fill[black] (1.2,0.9) circle (0.05cm);
\draw (1.2,0) -- (0.8,0) arc (-90:-270:0.15cm) -- (1.2,0.3);
\draw (1.2,0.6) -- (0.8,0.6) arc (-90:-270:0.15cm) -- (1.2,0.9);
\end{tikzpicture} 
\end{array}
\quad \longrightarrow \quad \text{separable}\,, \qquad \qquad  
\begin{array}{c}
\begin{tikzpicture}[thick]
\fill[black] (0,0.0) circle (0.05cm);
\fill[black] (0,0.3) circle (0.05cm);
\fill[black] (0,0.6) circle (0.05cm);
\fill[black] (0,0.9) circle (0.05cm);
\draw (0,0) -- (1.2,0);
\draw (0,0.6) -- (1.2,0.6);
\fill[black] (1.2,0.0) circle (0.05cm);
\fill[black] (1.2,0.3) circle (0.05cm);
\fill[black] (1.2,0.6) circle (0.05cm);
\fill[black] (1.2,0.9) circle (0.05cm);
\draw (1.2,0.3) -- (0,0.3);
\draw (1.2,0.9) -- (0,0.9);
\end{tikzpicture} 
\end{array} \quad \longrightarrow \quad \text{entangled.}
\ee
In fact, non-trivial global topologies can be replaced by linear combinations of $S^3$ topologies with additional Wilson loops via the ``surgery operation''~\cite{Witten:1988hf}. This means, as we will demonstrate below, that such topologies correspond to weaker entanglement, as compared to simply connected 3D topologies.

We have already seen that subtleties of quantum Chern-Simons theories can make images of some non-homeomorphic topological spaces linearly dependent. Therefore one has to keep in mind that the TQFT map is not always faithful, or more generally, only a finite set of quantum states is available for the topological description for some integer values of the Chern-Simons coupling constant $k$. A specific version of this problem is known as non-universality of quantum computation with Ising anyons~\cite{Nayak:2008zza}. The problem can be avoided if $k$ is taken sufficiently large, or more generally if one makes an analytic continuation to generic values of $k$ or $q$. We will assume either of this loopholes in the remaining discussion.

\section{Connectome classification of bipartite entanglement}
\label{sec:classification}

\subsection{Connectomes}
\label{sec:connectome}

Quantum entanglement is studied by the Quantum Resource Theory, which views it as a resource for quantum computation. States entangled in different ways are suitable for different quantum tasks, so classification of different types of entanglement is an important problem. Topological picture discussed in this paper provide an intuitive classification of entanglement in terms of topology: wiring of Wilson lines, and more general connectivity of topological spaces. This classification, although discrete, is still too detailed, and one is typically interested in a coarser one, with a finite number of classes. 

One very well known classification in Quantum Resource Theory is by stochastic local operations and classical communication (SLOCC)~\cite{Dur:2000zz}. This classification identifies quantum states that differ by the action of local invertible operators (not necessarily the unitary ones), and produces a finite number of classes for bipartite entanglement. Specifically, if a quantum system is split into subsystems $A$ and $B$, such that $\Hc_A\simeq \mathbb{C}^m$ and $\Hc_B\simeq \mathbb{C}^n$, with $m\geq n$, then there are $n$ possible classes that cannot be mixed by action of invertible operators on $\Hc_A$ and $\Hc_B$. This result can be obtained by applying Gram-Schmidt decomposition of the reduced density matrix, which has at most $n$ independent terms (rank), with the rank invariant under the action of local invertible operators. For a pair of qubits this gives just two classes: separable states and entangled states.

Let us introduce a simple class of topological diagrams that characterize different types of entanglement and also produce a finite number of classes, equivalent to the SLOCC ones for bipartite entanglement. We will assume that each sphere represents a party. So, for bipartite entanglement we will consider different wiring of a pair of two-spheres. 

We expect that different ways of wiring of spheres with Wilson lines should produce different ways of entangling quantum states. We can label different choices of wiring by graphs, which only distinguish what is connected to what, but ignore the specific 3D topology of connections. That is the information encoded by the graph is $A_{ij}$ (the adjacency matrix) -- the number of connections between sphere $i$ and sphere $j$, including self-connections. Graphs split the infinite number of wiring options in a finite number of classes. Let us focus on the simplest class representative that are given by planar graphs, which can be drawn on a plane without line intersections. For a pair of spheres planar graphs include representatives of all the classes. We can also identify graphs that are connected by local permutations. Such permutations should not affect entanglement.

For two qubits, that is a pair of two-spheres with four punctures, our wiring prescription leads to three inequivalent options,
\be
\label{2quadruples}
\begin{array}{c}
\begin{tikzpicture}[thick]
\fill[black] (0,0.0) circle (0.05cm);
\fill[black] (0,0.3) circle (0.05cm);
\fill[black] (0,0.6) circle (0.05cm);
\fill[black] (0,0.9) circle (0.05cm);
\draw (0,0) -- (0.4,0) arc (-90:90:0.15cm) -- (0,0.3);
\draw (0,0.6) -- (0.4,0.6) arc (-90:90:0.15cm) -- (0,0.9);
\fill[black] (1.2,0.0) circle (0.05cm);
\fill[black] (1.2,0.3) circle (0.05cm);
\fill[black] (1.2,0.6) circle (0.05cm);
\fill[black] (1.2,0.9) circle (0.05cm);
\draw (1.2,0) -- (0.8,0) arc (-90:-270:0.15cm) -- (1.2,0.3);
\draw (1.2,0.6) -- (0.8,0.6) arc (-90:-270:0.15cm) -- (1.2,0.9);
\end{tikzpicture} 
\end{array}
\,, \qquad 
\begin{array}{c}
\begin{tikzpicture}[thick]
\fill[black] (0,0.0) circle (0.05cm);
\fill[black] (0,0.3) circle (0.05cm);
\fill[black] (0,0.6) circle (0.05cm);
\fill[black] (0,0.9) circle (0.05cm);
\draw (0,0) -- (0.4,0) -- (1.2,0.0);
\draw (0,0.6) -- (0.4,0.6) arc (-90:90:0.15cm) -- (0,0.9);
\fill[black] (1.2,0.0) circle (0.05cm);
\fill[black] (1.2,0.3) circle (0.05cm);
\fill[black] (1.2,0.6) circle (0.05cm);
\fill[black] (1.2,0.9) circle (0.05cm);
\draw (1.2,0.3) -- (0,0.3);
\draw (1.2,0.6) -- (0.8,0.6) arc (-90:-270:0.15cm) -- (1.2,0.9);
\end{tikzpicture} 
\end{array}
\,,\qquad \text{and}\qquad  
\begin{array}{c}
\begin{tikzpicture}[thick]
\fill[black] (0,0.0) circle (0.05cm);
\fill[black] (0,0.3) circle (0.05cm);
\fill[black] (0,0.6) circle (0.05cm);
\fill[black] (0,0.9) circle (0.05cm);
\draw (0,0) -- (1.2,0);
\draw (0,0.6) -- (1.2,0.6);
\fill[black] (1.2,0.0) circle (0.05cm);
\fill[black] (1.2,0.3) circle (0.05cm);
\fill[black] (1.2,0.6) circle (0.05cm);
\fill[black] (1.2,0.9) circle (0.05cm);
\draw (1.2,0.3) -- (0,0.3);
\draw (1.2,0.9) -- (0,0.9);
\end{tikzpicture} 
\end{array}\,,
\ee
considered up to local permutation of punctures. Absence of either local or non-local braiding makes the diagrams equivalent to elements of Temperley-Lieb algebra $TL_4$. We will generally refer to representatives of the adjacency matrix classes as \emph{connectomes}, specifying whether they are planar or not whenever necessary.

One can note that the second diagram in~(\ref{2quadruples}) possesses a section crossed by only two lines, so it in fact represents a separable state. Indeed, using basis~(\ref{4pointbasis}) one can show that the first and the second diagram correspond to the same normalized wavefunction. Therefore there are two inequivalent connectome diagrams describing entanglement of two qubits: separable state and entangled state. Moreover, the diagram of the entangled state describes the case of maximal entanglement. As expected, the collection of parallel lines is equivalent to the identity operator, which in the present case describes both the wavefunction and the reduced density matrix. To confirm this, expand in basis~(\ref{4pointbasis}),
\be
\label{Bellstate}
\begin{array}{c}
     \includegraphics[scale=0.2]{lines.pdf}
\end{array} \quad \equiv \quad \begin{array}{c}
\begin{tikzpicture}[thick]
\fill[black] (0,0.0) circle (0.05cm);
\fill[black] (0,0.3) circle (0.05cm);
\fill[black] (0,0.6) circle (0.05cm);
\fill[black] (0,0.9) circle (0.05cm);
\draw (0,0) -- (1.2,0);
\draw (0,0.6) -- (1.2,0.6);
\fill[black] (1.2,0.0) circle (0.05cm);
\fill[black] (1.2,0.3) circle (0.05cm);
\fill[black] (1.2,0.6) circle (0.05cm);
\fill[black] (1.2,0.9) circle (0.05cm);
\draw (1.2,0.3) -- (0,0.3);
\draw (1.2,0.9) -- (0,0.9);
\end{tikzpicture} 
\end{array} \quad = \ |0\rangle\langle 0| + |1\rangle\langle 1|\,.
\ee
One can also note that the equivalence between the reduced density matrix and the wavefunction is true for all the connectome states.

Let us now extend the connectome classification to pairs of qudits. 


\subsection{Bipartite entanglement}
\label{sec:entanglement}

To systematically construct qudit Hilbert spaces, that is $\Hc=\mathbb{C}^n$ with $n>2$, we will make use of the fact that tensor product of representations of spin $j_1$ and $j_2$ expands in the sum of $j_1+j_2-|j_1-j_2|+1$ representations with spin varying between $|j_1-j_2|$ and $j_1+j_2$. Since we would like to work with $S^2$ boundaries with $j=1/2$ punctures, representations of any spin $j_1$ and $j_2$ can be obtained by fusing several $j=1/2$ representations. In other words, for spin $j_1$ ($j_2$), we will consider groups of $2j_1$ ($2j_2$) regular punctures and project out all the irrelevant irreps $j<j_1$ ($j<j_2$) that appear in the tensor product of $2j_1$ ($2j_2$) fundamental irreps. In our case the projection can be organized through the Jones-Wenzl symmetrizers (projectors)~\cite{Jones:1985dw,Wenzl:1985seq}, which we will now introduce. 

The Jones-Wenzl projectors are elements of the Temperley-Lieb algebra, which can be defined recursively as follows. For $TL_{n+2}$ the projector on the space of spin $n/2+1$ can be obtained from the projector on the space of spin $(n+1)/2$ of $TL_{n+1}$ through relation~\cite{Kauffman:2013bh}
\be
\label{JWprojector}
\begin{tikzpicture}[baseline=2]
\draw[thick] (0,0) rectangle (1,0.5);
\draw[ultra thick] (0.2,1) node[anchor=east] {$n$} -- (0.2,0.5);
\draw[ultra thick] (0.2,0) -- (0.2,-0.5);
\draw[thick] (0.5,-0.5) -- (0.5,0);
\draw[thick] (0.5,1) -- (0.5,0.5);
\draw[thick] (0.8,-0.5) -- (0.8,0);
\draw[thick] (0.8,1) -- (0.8,0.5);
\end{tikzpicture}\quad \ = \ \quad 
\begin{tikzpicture}[baseline=2]
\draw[thick] (0,0) rectangle (1,0.5);
\draw[ultra thick] (0.3,-0.5) -- (0.3,0);
\draw[ultra thick] (0.3,1) node[anchor=east] {$n$} -- (0.3,0.5);
\draw[thick] (0.7,-0.5) -- (0.7,0);
\draw[thick] (0.7,1) -- (0.7,0.5);
\draw[thick] (1.2,-0.5) -- (1.2,1);
\end{tikzpicture}\quad \ - \ \frac{\Delta_n}{\Delta_{n+1}} \quad \begin{tikzpicture}[baseline=2]
\draw[thick] (0,0.5) rectangle (1,0.8);
\draw[ultra thick] (0.3,-0.) -- (0.3,0.5);
\draw[ultra thick] (0.3,-0.5) -- (0.3,-0.3);
\draw[ultra thick] (0.3,1) node[anchor=east] {$n$} -- (0.3,0.8);
\draw[thick] (0.7,-0.)  arc (180:0:0.2) -- (1.1,-0.5);
\draw[thick] (0.7,1) -- (0.7,0.8);
\draw[thick] (0.7,-0.5) -- (0.7,-0.3);
\draw[thick] (1.1,1) -- (1.1,0.5) arc (0:-180:0.2);
\draw[thick] (0,-0.3) rectangle (1,0.);
\end{tikzpicture}\quad \,.
\ee
Here the thick line with label $n$ substitutes $n$ ordinary lines. Parameters $\Delta_n$ are also defined recursively,
\be
\Delta_{-1}\ = \ 0\,,\qquad \Delta_{0} \ = \ 1\,, \qquad \Delta_{n+1} \ = \ d\Delta_{n} - \Delta_{n-1}\,.
\ee

For example, for $TL_2$ the following is the projector on the subspace of $j=1$ in the tensor product of two spins $j=1/2$, 
\be
\label{P3}
\begin{tikzpicture}[baseline=2]
\draw[thick] (0,0) rectangle (1,0.5);
\draw[thick] (0.3,-0.5) -- (0.3,0);
\draw[thick] (0.3,1) -- (0.3,0.5);
\draw[thick] (0.7,-0.5) -- (0.7,0);
\draw[thick] (0.7,1) -- (0.7,0.5);
\end{tikzpicture}
\quad \ = \ \quad 
\begin{tikzpicture}[baseline=2]
\draw[thick] (0.3,-0.5) -- (0.3,1);
\draw[thick] (0.7,-0.5) -- (0.7,1);
\end{tikzpicture} \ - \ \frac{1}{d} \quad \begin{tikzpicture}[baseline=2]
\draw[thick] (0.7,-0.)  arc (180:0:0.2) -- (1.1,-0.5);
\draw[thick] (0.7,1) -- (0.7,0.5);
\draw[thick] (0.7,-0.5) -- (0.7,0);
\draw[thick] (1.1,1) -- (1.1,0.5) arc (0:-180:0.2);
\end{tikzpicture}\quad\,.
\ee
Note that the last term in this expression is an orthogonal projector on the $j=0$ subspace, and the two projectors are equivalent to states $|0\rangle$ and $|1\rangle$ in basis~(\ref{4pointbasis}). We can use this projector to construct the basis in the Hilbert space of a qutrit.

The qutrit can be obtained in the tensor product of two $j=1$ spins. Recall that in the large $k$ regime the dimension of the Hilbert space in the TQFT construction is counted by the number of singlet representations appearing in the expansion of the tensor product of the spins of the punctures. If we take four spin $j=1$ punctures, there will be three ways to form a singlet and the dimension of the Hilbert space will be three. 

One can convince oneself that the following is a basis in the sought Hilbert space,
\be
|e_1\rangle \ = \ \begin{array}{c}
\begin{tikzpicture}[thick]
\fill[black] (0,0.0) circle (0.05cm);
\fill[black] (0,0.3) circle (0.05cm);
\fill[black] (0,0.6) circle (0.05cm);
\fill[black] (0,0.9) circle (0.05cm);
\draw (0,0) -- (0.3,0) arc (-90:90:0.45cm) -- (0,0.9);
\draw (0,0.6) -- (0.3,0.6) arc (90:-90:0.15cm) -- (0,0.3);
\fill[white] (0.1,-0.1) rectangle (0.3,0.4);
\draw (0.1,-0.1) rectangle (0.3,0.4);
\fill[white] (0.1,0.5) rectangle (0.3,1);
\draw (0.1,0.5) rectangle (0.3,1);
\newcommand\y{1.2};
\fill[black] (0,0.0+\y) circle (0.05cm);
\fill[black] (0,0.3+\y) circle (0.05cm);
\fill[black] (0,0.6+\y) circle (0.05cm);
\fill[black] (0,0.9+\y) circle (0.05cm);
\draw (0,0+\y) -- (0.3,0+\y) arc (-90:90:0.45cm) -- (0,0.9+\y);
\draw (0,0.6+\y) -- (0.3,0.6+\y) arc (90:-90:0.15cm) -- (0,0.3+\y);
\fill[white] (0.1,-0.1+\y) rectangle (0.3,0.4+\y);
\draw (0.1,-0.1+\y) rectangle (0.3,0.4+\y);
\fill[white] (0.1,0.5+\y) rectangle (0.3,1+\y);
\draw (0.1,0.5+\y) rectangle (0.3,1+\y);
\end{tikzpicture} 
\end{array}\,, \qquad
|e_2\rangle \ = \ \begin{array}{c}
\begin{tikzpicture}[thick]
\newcommand\y{1.2};
\fill[black] (0,0.0) circle (0.05cm);
\fill[black] (0,0.3) circle (0.05cm);
\fill[black] (0,0.6) circle (0.05cm);
\fill[black] (0,0.9) circle (0.05cm);
\draw (0,0) -- (0.3,0) arc (-90:90:1.05cm) -- (0,0.9+\y);
\draw (0,0.6) -- (0.3,0.6) arc (90:-90:0.15cm) -- (0,0.3);
\fill[black] (0,0.0+\y) circle (0.05cm);
\fill[black] (0,0.3+\y) circle (0.05cm);
\fill[black] (0,0.6+\y) circle (0.05cm);
\fill[black] (0,0.9+\y) circle (0.05cm);
\draw (0,0+\y) -- (0.9,0+\y) arc (90:-90:0.15cm) -- (0,0.9);
\draw (0,0.6+\y) -- (0.3,0.6+\y) arc (90:-90:0.15cm) -- (0,0.3+\y);
\fill[white] (0.1,-0.1) rectangle (0.3,0.4);
\draw (0.1,-0.1) rectangle (0.3,0.4);
\fill[white] (0.1,0.5) rectangle (0.3,1);
\draw (0.1,0.5) rectangle (0.3,1);
\fill[white] (0.1,-0.1+\y) rectangle (0.3,0.4+\y);
\draw (0.1,-0.1+\y) rectangle (0.3,0.4+\y);
\fill[white] (0.1,0.5+\y) rectangle (0.3,1+\y);
\draw (0.1,0.5+\y) rectangle (0.3,1+\y);
\end{tikzpicture} 
\end{array}\,, \qquad
|e_3\rangle \ = \ \begin{array}{c}
\begin{tikzpicture}[thick]
\newcommand\y{1.2};
\fill[black] (0,0.0) circle (0.05cm);
\fill[black] (0,0.3) circle (0.05cm);
\fill[black] (0,0.6) circle (0.05cm);
\fill[black] (0,0.9) circle (0.05cm);
\draw (0,0) -- (0.3,0) arc (-90:90:1.05cm) -- (0,0.9+\y);
\draw (0,0.6) -- (0.3,0.6) arc (-90:90:0.45cm) -- (0,0.3+\y);
\fill[black] (0,0.0+\y) circle (0.05cm);
\fill[black] (0,0.3+\y) circle (0.05cm);
\fill[black] (0,0.6+\y) circle (0.05cm);
\fill[black] (0,0.9+\y) circle (0.05cm);
\draw (0,0+\y) -- (0.3,0+\y) arc (90:-90:0.15cm) -- (0,0.9);
\draw (0,0.6+\y) -- (0.3,0.6+\y) arc (90:-90:0.75cm) -- (0,0.3);
\fill[white] (0.1,-0.1) rectangle (0.3,0.4);
\draw (0.1,-0.1) rectangle (0.3,0.4);
\fill[white] (0.1,0.5) rectangle (0.3,1);
\draw (0.1,0.5) rectangle (0.3,1);
\fill[white] (0.1,-0.1+\y) rectangle (0.3,0.4+\y);
\draw (0.1,-0.1+\y) rectangle (0.3,0.4+\y);
\fill[white] (0.1,0.5+\y) rectangle (0.3,1+\y);
\draw (0.1,0.5+\y) rectangle (0.3,1+\y);
\end{tikzpicture} 
\end{array}\,.
\label{qutritbasis}
\ee
Here each projector groups a pair of $j=1/2$ particles in a $j=1$ particle and we search for all possible ways of connecting the outputs of the projects without intersections. Apart from the standard property of a projector to square to itself, the Jones-Wenzl symmetrizers have the property that a closure of any pair of adjacent input, or output lines, annihilates the projector,
\be
\label{platclosure}
\begin{tikzpicture}[baseline=2]
\draw[thick] (0,0) rectangle (1.5,0.5);
\draw[thick] (0.2,0.5) -- (0.2,0.7) arc (180:0:0.15) -- (0.5,0.5);
\draw[thick] (0.8,0.5) -- (0.8,0.85);
\draw[thick] (0.5,0) -- (0.5,-0.35);
\draw[thick] (0.2,0) -- (0.2,-0.35);
\draw[thick] (0.8,0) -- (0.8,-0.35);
\draw (1.2,0.7) node {\bf $\cdots$};
\draw (1.2,-0.2) node {\bf $\cdots$};
\end{tikzpicture}
\ = \  \begin{tikzpicture}[baseline=2]
\draw[thick] (0,0) rectangle (1.5,0.5);
\draw[thick] (0.5,0.5) -- (0.5,0.7) arc (180:0:0.15) -- (0.8,0.5);
\draw[thick] (0.2,0.5) -- (0.2,0.85);
\draw[thick] (0.5,0) -- (0.5,-0.35);
\draw[thick] (0.2,0) -- (0.2,-0.35);
\draw[thick] (0.8,0) -- (0.8,-0.35);
\draw (1.2,0.7) node {\bf $\cdots$};
\draw (1.2,-0.2) node {\bf $\cdots$};
\end{tikzpicture} \ = \ \begin{tikzpicture}[baseline=2]
\draw[thick] (0,0) rectangle (1.5,0.5);
\draw[thick] (0.2,0.5) -- (0.2,0.85);
\draw[thick] (0.8,0.5) -- (0.8,0.85);
\draw[thick] (0.5,0.5) -- (0.5,0.85);
\draw[thick] (0.2,0) -- (0.2,-0.2) arc (-180:0:0.15) -- (0.5,0);
\draw[thick] (0.8,0) -- (0.8,-0.35);
\draw (1.2,0.7) node {\bf $\cdots$};
\draw (1.2,-0.2) node {\bf $\cdots$};
\end{tikzpicture} \ = \ 0\,.
\ee
This obviously follows from the fact that the symmetrizers are orthogonal to the subspaces of a lower spin. Assuming this property and ignoring permutations of the projectors, the diagrams in~(\ref{qutritbasis}) are the only valid ways of connecting the outputs.

Basis~~(\ref{qutritbasis}) now looks a natural generalization of the qubit case, where instead of connecting punctures, we connect bunches of punctures. The cables that we use to connect the bunches have fibers, which give more options for wiring. As in the case of the qubit, we can orthogonalize the basis. For example,
\be
\label{8pointbasis}
|0\rangle \ = \ \frac{1}{\Delta_2}\,|e_1\rangle\,, \qquad |1\rangle \ = \ \frac{d}{(\Delta_2-1)\sqrt{\Delta_2}}|\hat{e}_2\rangle\,, \qquad |2\rangle \ = \ \frac{1}{\sqrt{\Delta_2^2-\Delta_2-1}}\left(|e_3\rangle - |0\rangle-\sqrt{\Delta_2}|1\rangle\right).
\ee
Here, for convenience, we redefined the state $|e_2\rangle$,
\be
|\hat{e}_2\rangle \ = \ |e_2\rangle -\frac{1}{d}|e_1\rangle \ = \ \begin{array}{c}
\begin{tikzpicture}[thick]
\newcommand\y{1.2};
\fill[black] (0,0.0) circle (0.05cm);
\fill[black] (0,0.3) circle (0.05cm);
\fill[black] (0,0.6) circle (0.05cm);
\fill[black] (0,0.9) circle (0.05cm);
\draw (0,0) -- (0.3,0) arc (-90:90:1.05cm) -- (0,0.9+\y);
\draw (0,0.6) -- (0.3,0.6) arc (90:-90:0.15cm) -- (0,0.3);
\fill[black] (0,0.0+\y) circle (0.05cm);
\fill[black] (0,0.3+\y) circle (0.05cm);
\fill[black] (0,0.6+\y) circle (0.05cm);
\fill[black] (0,0.9+\y) circle (0.05cm);
\draw (0,0+\y) -- (0.9,0+\y) arc (90:-90:0.15cm) -- (0,0.9);
\draw (0,0.6+\y) -- (0.3,0.6+\y) arc (90:-90:0.15cm) -- (0,0.3+\y);
\fill[white] (0.1,-0.1) rectangle (0.3,0.4);
\draw (0.1,-0.1) rectangle (0.3,0.4);
\fill[white] (0.1,0.5) rectangle (0.3,1);
\draw (0.1,0.5) rectangle (0.3,1);
\fill[white] (0.1,-0.1+\y) rectangle (0.3,0.4+\y);
\draw (0.1,-0.1+\y) rectangle (0.3,0.4+\y);
\fill[white] (0.1,0.5+\y) rectangle (0.3,1+\y);
\draw (0.1,0.5+\y) rectangle (0.3,1+\y);
\fill[white] (0.9,1.0) rectangle (1.5,1.1);
\draw (0.9,1.0) rectangle (1.5,1.1);
\end{tikzpicture} 
\end{array}\,.
\ee

At the next step we discuss possible entanglement types for a pair of qutrits described by Hilbert spaces spanned by~(\ref{qutritbasis}). Taking into account~(\ref{platclosure}) and considering the diagrams up to permutations of the projectors and permutations of outputs within each projector one should conclude that the following three diagrams make a complete set of inequivalent wirings of two qutrits,
\be
\label{8pointdiags}
\begin{array}{c}
\begin{tikzpicture}[thick]
\fill[pink] (-0.2,1.1) rectangle (0.1,1.6);
\fill[pink] (-0.2,-0.1) rectangle (0.1,0.4);
\fill[pink] (-0.2,0.5) rectangle (0.1,1.0);
\fill[pink] (-0.2,1.7) rectangle (0.1,2.2);
\fill[cyan] (1.1,1.1) rectangle (1.3,1.6);
\fill[cyan] (1.1,-0.1) rectangle (1.3,0.4);
\fill[cyan] (1.1,0.5) rectangle (1.3,1.0);
\fill[cyan] (1.1,1.7) rectangle (1.3,2.2);
\fill[black] (0,0.0) circle (0.05cm);
\fill[black] (0,0.3) circle (0.05cm);
\fill[black] (0,0.6) circle (0.05cm);
\fill[black] (0,0.9) circle (0.05cm);
\draw (0,0) -- (0.4,0) -- (1.2,0.0);
\draw (0,1.2) -- (0.1,1.2) arc (-90:90:0.45cm) -- (0,2.1);
\fill[black] (1.2,0.0) circle (0.05cm);
\fill[black] (1.2,0.3) circle (0.05cm);
\fill[black] (1.2,0.6) circle (0.05cm);
\fill[black] (1.2,0.9) circle (0.05cm);
\draw (1.2,0.3) -- (0,0.3);
\draw (1.2,1.2) -- (1.1,1.2) arc (-90:-270:0.45cm) -- (1.2,2.1);
\fill[black] (1.2,1.2) circle (0.05cm);
\fill[black] (1.2,1.5) circle (0.05cm);
\fill[black] (0,1.2) circle (0.05cm);
\fill[black] (0,1.5) circle (0.05cm);
\draw (0,0.6) -- (1.2,0.6);
\draw (1.2,0.9) -- (0,0.9);
\fill[black] (1.2,1.8) circle (0.05cm);
\fill[black] (1.2,2.1) circle (0.05cm);
\fill[black] (0,1.8) circle (0.05cm);
\fill[black] (0,2.1) circle (0.05cm);
\draw (0,1.8) -- (0.1,1.8) arc (90:-90:0.15) -- (0,1.5);
\draw (1.2,1.8) -- (1.1,1.8) arc (90:270:0.15) -- (1.2,1.5);
\end{tikzpicture} 
\end{array}
\,,\qquad \begin{array}{c}
\begin{tikzpicture}[thick]
\fill[pink] (-0.2,1.1) rectangle (0.1,1.6);
\fill[pink] (-0.2,-0.1) rectangle (0.1,0.4);
\fill[pink] (-0.2,0.5) rectangle (0.1,1.0);
\fill[pink] (-0.2,1.7) rectangle (0.1,2.2);
\fill[cyan] (1.1,1.1) rectangle (1.3,1.6);
\fill[cyan] (1.1,-0.1) rectangle (1.3,0.4);
\fill[cyan] (1.1,0.5) rectangle (1.3,1.0);
\fill[cyan] (1.1,1.7) rectangle (1.3,2.2);
\fill[black] (0,0.0) circle (0.05cm);
\fill[black] (0,0.3) circle (0.05cm);
\fill[black] (0,0.6) circle (0.05cm);
\fill[black] (0,0.9) circle (0.05cm);
\draw (0,0) -- (0.4,0) -- (1.2,0.0);
\draw (0,2.1) -- (1.2,2.1);
\fill[black] (1.2,0.0) circle (0.05cm);
\fill[black] (1.2,0.3) circle (0.05cm);
\fill[black] (1.2,0.6) circle (0.05cm);
\fill[black] (1.2,0.9) circle (0.05cm);
\draw (1.2,0.3) -- (0,0.3);
\draw (1.2,0.6) -- (0,0.6);
\fill[black] (1.2,1.2) circle (0.05cm);
\fill[black] (1.2,1.5) circle (0.05cm);
\fill[black] (0,1.2) circle (0.05cm);
\fill[black] (0,1.5) circle (0.05cm);
\draw (0,0.9) -- (1.2,0.9);
\draw (0,1.2) -- (1.2,1.2);
\fill[black] (1.2,1.8) circle (0.05cm);
\fill[black] (1.2,2.1) circle (0.05cm);
\fill[black] (0,1.8) circle (0.05cm);
\fill[black] (0,2.1) circle (0.05cm);
\draw (0,1.8) -- (0.1,1.8) arc (90:-90:0.15) -- (0,1.5);
\draw (1.2,1.8) -- (1.1,1.8) arc (90:270:0.15) -- (1.2,1.5);
\end{tikzpicture} 
\end{array}
\,, \qquad \text{and}\qquad  
\begin{array}{c}
\begin{tikzpicture}[thick]
\fill[pink] (-0.2,1.1) rectangle (0.1,1.6);
\fill[pink] (-0.2,-0.1) rectangle (0.1,0.4);
\fill[pink] (-0.2,0.5) rectangle (0.1,1.0);
\fill[pink] (-0.2,1.7) rectangle (0.1,2.2);
\fill[cyan] (1.1,1.1) rectangle (1.3,1.6);
\fill[cyan] (1.1,-0.1) rectangle (1.3,0.4);
\fill[cyan] (1.1,0.5) rectangle (1.3,1.0);
\fill[cyan] (1.1,1.7) rectangle (1.3,2.2);
\fill[black] (0,0.0) circle (0.05cm);
\fill[black] (0,0.3) circle (0.05cm);
\fill[black] (0,0.6) circle (0.05cm);
\fill[black] (0,0.9) circle (0.05cm);
\draw (0,0) -- (1.2,0);
\draw (0,0.6) -- (1.2,0.6);
\fill[black] (1.2,0.0) circle (0.05cm);
\fill[black] (1.2,0.3) circle (0.05cm);
\fill[black] (1.2,0.6) circle (0.05cm);
\fill[black] (1.2,0.9) circle (0.05cm);
\draw (1.2,0.3) -- (0,0.3);
\draw (1.2,0.9) -- (0,0.9);
\fill[black] (1.2,1.2) circle (0.05cm);
\fill[black] (1.2,1.5) circle (0.05cm);
\fill[black] (0,1.2) circle (0.05cm);
\fill[black] (0,1.5) circle (0.05cm);
\draw (0,1.2) -- (1.2,1.2);
\draw (1.2,1.5) -- (0,1.5);
\fill[black] (1.2,1.8) circle (0.05cm);
\fill[black] (1.2,2.1) circle (0.05cm);
\fill[black] (0,1.8) circle (0.05cm);
\fill[black] (0,2.1) circle (0.05cm);
\draw (0,1.8) -- (1.2,1.8);
\draw (1.2,2.1) -- (0,2.1);
\end{tikzpicture} 
\end{array}\,.
\ee
Note that in our choice the lower half of each diagram is the same. As in the case of the qubit, breaking of a half of the connections between the two sides already leads to a separable state. Breaking more connections would just give linearly dependent states. This allows to choose the lower half of all the diagrams to have the maximal possible number of connections and only consider variations in the upper half. Using basis~(\ref{8pointbasis}) one can check that the corresponding matrices have ranks one, two and three respectively.

In summary, a state of two qutrits admits three types of connectomes corresponding to reduced density matrices of ranks one, two and three. Ignoring the non-informative halves of the diagrams, these connectomes are
\be
\begin{array}{c}
\begin{tikzpicture}[thick]
\fill[pink] (-0.2,-0.1) rectangle (0.1,0.4);
\fill[pink] (-0.2,0.5) rectangle (0.1,1.0);
\fill[cyan] (1.1,-0.1) rectangle (1.3,0.4);
\fill[cyan] (1.1,0.5) rectangle (1.3,1.0);
\fill[black] (0,0.0) circle (0.05cm);
\fill[black] (0,0.3) circle (0.05cm);
\fill[black] (0,0.6) circle (0.05cm);
\fill[black] (0,0.9) circle (0.05cm);
\draw (0,0) -- (0.1,0) arc (-90:90:0.45) -- (0.,0.9);
\draw (0,0.6) -- (0.1,0.6) arc (90:-90:0.15cm) -- (0,0.3);
\fill[black] (1.2,0.0) circle (0.05cm);
\fill[black] (1.2,0.3) circle (0.05cm);
\fill[black] (1.2,0.6) circle (0.05cm);
\fill[black] (1.2,0.9) circle (0.05cm);
\draw (1.2,0.9) -- (1.1,0.9) arc (90:270:0.45) -- (1.2,0);
\draw (1.2,0.6) -- (1.1,0.6) arc (90:270:0.15cm) -- (1.2,0.3);
\end{tikzpicture} 
\end{array}
\,,\qquad  
\begin{array}{c}
\begin{tikzpicture}[thick]
\fill[pink] (-0.2,-0.1) rectangle (0.1,0.4);
\fill[pink] (-0.2,0.5) rectangle (0.1,1.0);
\fill[cyan] (1.1,-0.1) rectangle (1.3,0.4);
\fill[cyan] (1.1,0.5) rectangle (1.3,1.0);
\fill[black] (0,0.0) circle (0.05cm);
\fill[black] (0,0.3) circle (0.05cm);
\fill[black] (0,0.6) circle (0.05cm);
\fill[black] (0,0.9) circle (0.05cm);
\draw (0,0) -- (0.4,0) -- (1.2,0.0);
\draw (0,0.6) -- (0.1,0.6) arc (90:-90:0.15cm) -- (0,0.3);
\fill[black] (1.2,0.0) circle (0.05cm);
\fill[black] (1.2,0.3) circle (0.05cm);
\fill[black] (1.2,0.6) circle (0.05cm);
\fill[black] (1.2,0.9) circle (0.05cm);
\draw (1.2,0.9) -- (0,0.9);
\draw (1.2,0.6) -- (1.1,0.6) arc (90:270:0.15cm) -- (1.2,0.3);
\end{tikzpicture} 
\end{array}
\,,\qquad  
\begin{array}{c}
\begin{tikzpicture}[thick]
\fill[pink] (-0.2,-0.1) rectangle (0.1,0.4);
\fill[pink] (-0.2,0.5) rectangle (0.1,1.0);
\fill[cyan] (1.1,-0.1) rectangle (1.3,0.4);
\fill[cyan] (1.1,0.5) rectangle (1.3,1.0);
\fill[black] (0,0.0) circle (0.05cm);
\fill[black] (0,0.3) circle (0.05cm);
\fill[black] (0,0.6) circle (0.05cm);
\fill[black] (0,0.9) circle (0.05cm);
\draw (0,0) -- (1.2,0);
\draw (0,0.6) -- (1.2,0.6);
\fill[black] (1.2,0.0) circle (0.05cm);
\fill[black] (1.2,0.3) circle (0.05cm);
\fill[black] (1.2,0.6) circle (0.05cm);
\fill[black] (1.2,0.9) circle (0.05cm);
\draw (1.2,0.3) -- (0,0.3);
\draw (1.2,0.9) -- (0,0.9);
\end{tikzpicture} 
\end{array}\,.
\ee
The three cases are in one-to-one correspondence with the SLOCC classes.

It is straightforward to generalize this construction to a pair of qudits with the same dimension $n$. The Hilbert space of a qudit is constructed as a subspace of the Hilbert space of $S^2$ with $4(n-1)$ spin $j=1/2$ punctures. The subspace is obtained by applying Jones-Wenzl $(n-1)$-projectors on each of the four groups of $n-1$ punctures. Next one has to draw all possible connectomes of $2(n-1)$ points, modulo permutations of points within the projectors and permutations of the projectors. For each of the two considered projectors of the left qudit the lines can only connect to either the other projector of the left qubit, or to a projector of the right qudit, so the number of diagrams is equal to the maximal number of lines that can connect two projectors plus one, that  is $n$. So there are $n$ inequivalent classes of diagrams, or $n$ classes of entanglement. For $n=2,3,4$, one has the following connectomes,
\begin{eqnarray}
n=2\,: & & \begin{array}{c}
\begin{tikzpicture}[thick]
\fill[black] (0,0.6) circle (0.05cm);
\fill[black] (0,0.9) circle (0.05cm);
\draw (0,0.6) -- (0.4,0.6) arc (-90:90:0.15cm) -- (0,0.9);
\fill[black] (1.2,0.6) circle (0.05cm);
\fill[black] (1.2,0.9) circle (0.05cm);
\draw (1.2,0.6) -- (0.8,0.6) arc (-90:-270:0.15cm) -- (1.2,0.9);
\end{tikzpicture} 
\end{array}
\,,\qquad 
\begin{array}{c}
\begin{tikzpicture}[thick]
\fill[black] (0,0.6) circle (0.05cm);
\fill[black] (0,0.9) circle (0.05cm);
\draw (0,0.6) -- (1.2,0.6);
\fill[black] (1.2,0.6) circle (0.05cm);
\fill[black] (1.2,0.9) circle (0.05cm);
\draw (1.2,0.9) -- (0,0.9);
\end{tikzpicture} 
\end{array}\,, \\ [2mm]
n=3\,: & & \begin{array}{c}
\begin{tikzpicture}[thick]
\fill[black] (0,0.0) circle (0.05cm);
\fill[black] (0,0.3) circle (0.05cm);
\fill[black] (0,0.6) circle (0.05cm);
\fill[black] (0,0.9) circle (0.05cm);
\draw (0,0) -- (0.1,0) arc (-90:90:0.45) -- (0.,0.9);
\draw (0,0.6) -- (0.1,0.6) arc (90:-90:0.15cm) -- (0,0.3);
\fill[black] (1.2,0.0) circle (0.05cm);
\fill[black] (1.2,0.3) circle (0.05cm);
\fill[black] (1.2,0.6) circle (0.05cm);
\fill[black] (1.2,0.9) circle (0.05cm);
\draw (1.2,0.9) -- (1.1,0.9) arc (90:270:0.45) -- (1.2,0);
\draw (1.2,0.6) -- (1.1,0.6) arc (90:270:0.15cm) -- (1.2,0.3);
\end{tikzpicture} 
\end{array}
\,,\qquad  
\begin{array}{c}
\begin{tikzpicture}[thick]
\fill[black] (0,0.0) circle (0.05cm);
\fill[black] (0,0.3) circle (0.05cm);
\fill[black] (0,0.6) circle (0.05cm);
\fill[black] (0,0.9) circle (0.05cm);
\draw (0,0) -- (0.4,0) -- (1.2,0.0);
\draw (0,0.6) -- (0.1,0.6) arc (90:-90:0.15cm) -- (0,0.3);
\fill[black] (1.2,0.0) circle (0.05cm);
\fill[black] (1.2,0.3) circle (0.05cm);
\fill[black] (1.2,0.6) circle (0.05cm);
\fill[black] (1.2,0.9) circle (0.05cm);
\draw (1.2,0.9) -- (0,0.9);
\draw (1.2,0.6) -- (1.1,0.6) arc (90:270:0.15cm) -- (1.2,0.3);
\end{tikzpicture} 
\end{array}
\,,\qquad  
\begin{array}{c}
\begin{tikzpicture}[thick]
\fill[black] (0,0.0) circle (0.05cm);
\fill[black] (0,0.3) circle (0.05cm);
\fill[black] (0,0.6) circle (0.05cm);
\fill[black] (0,0.9) circle (0.05cm);
\draw (0,0) -- (1.2,0);
\draw (0,0.6) -- (1.2,0.6);
\fill[black] (1.2,0.0) circle (0.05cm);
\fill[black] (1.2,0.3) circle (0.05cm);
\fill[black] (1.2,0.6) circle (0.05cm);
\fill[black] (1.2,0.9) circle (0.05cm);
\draw (1.2,0.3) -- (0,0.3);
\draw (1.2,0.9) -- (0,0.9);
\end{tikzpicture} 
\end{array}\,,\\ [2mm]
n=4\,: & & \begin{array}{c}
\begin{tikzpicture}[thick]
\fill[black] (0,0.0) circle (0.05cm);
\fill[black] (0,0.3) circle (0.05cm);
\fill[black] (0,0.6) circle (0.05cm);
\fill[black] (0,0.9) circle (0.05cm);
\draw (0,0.0) -- (0.1,0.0) arc (-90:0:0.45cm) -- (0.55,1.05) arc (0:90:0.45) -- (0,1.5);
\draw (0,1.2) -- (0.1,1.2) arc (90:0:0.3cm) -- (0.4,0.6) arc (0:-90:0.3) -- (0,0.3);
\fill[black] (0,1.2) circle (0.05cm);
\fill[black] (0,1.5) circle (0.05cm);
\draw (0,0.6) -- (0.1,0.6) arc (-90:90:0.15cm) -- (0,0.9);
\fill[black] (1.2,0.0) circle (0.05cm);
\fill[black] (1.2,0.3) circle (0.05cm);
\fill[black] (1.2,0.6) circle (0.05cm);
\fill[black] (1.2,0.9) circle (0.05cm);
\fill[black] (1.2,1.2) circle (0.05cm);
\fill[black] (1.2,1.5) circle (0.05cm);
\draw (1.2,0.0) -- (1.1,0.0) arc (270:180:0.45cm) -- (0.65,1.05) arc (180:90:0.45) -- (1.2,1.5);
\draw (1.2,1.2) -- (1.1,1.2) arc (90:180:0.3cm) -- (0.8,0.6) arc (180:270:0.3) -- (1.2,0.3);
\draw (1.2,0.6) -- (1.1,0.6) arc (270:90:0.15cm) -- (1.2,0.9);
\end{tikzpicture} 
\end{array}
\,,\qquad \begin{array}{c}
\begin{tikzpicture}[thick]
\fill[black] (0,0.0) circle (0.05cm);
\fill[black] (0,0.3) circle (0.05cm);
\fill[black] (0,0.6) circle (0.05cm);
\fill[black] (0,0.9) circle (0.05cm);
\draw (0,0) -- (0.1,0) arc (-90:90:0.45) -- (0.,0.9);
\draw (0,0.6) -- (0.1,0.6) arc (90:-90:0.15cm) -- (0,0.3);
\fill[black] (1.2,0.0) circle (0.05cm);
\fill[black] (1.2,0.3) circle (0.05cm);
\fill[black] (1.2,0.6) circle (0.05cm);
\fill[black] (1.2,0.9) circle (0.05cm);
\draw (1.2,0.9) -- (1.1,0.9) arc (90:270:0.45) -- (1.2,0);
\draw (1.2,0.6) -- (1.1,0.6) arc (90:270:0.15cm) -- (1.2,0.3);
\fill[black] (0,-0.3) circle (0.05cm);
\fill[black] (0,1.2) circle (0.05cm);
\fill[black] (1.2,-0.3) circle (0.05cm);
\fill[black] (1.2,1.2) circle (0.05cm);
\draw (1.2,-0.3) -- (0,-0.3);
\draw (1.2,1.2) -- (0,1.2);
\end{tikzpicture} 
\end{array}
\,,\qquad  
\begin{array}{c}
\begin{tikzpicture}[thick]
\fill[black] (0,0.0) circle (0.05cm);
\fill[black] (0,0.3) circle (0.05cm);
\fill[black] (0,0.6) circle (0.05cm);
\fill[black] (0,0.9) circle (0.05cm);
\draw (0,0) -- (0.4,0) -- (1.2,0.0);
\draw (0,0.6) -- (0.1,0.6) arc (90:-90:0.15cm) -- (0,0.3);
\fill[black] (1.2,0.0) circle (0.05cm);
\fill[black] (1.2,0.3) circle (0.05cm);
\fill[black] (1.2,0.6) circle (0.05cm);
\fill[black] (1.2,0.9) circle (0.05cm);
\draw (1.2,0.9) -- (0,0.9);
\draw (1.2,0.6) -- (1.1,0.6) arc (90:270:0.15cm) -- (1.2,0.3);
\fill[black] (0,-0.3) circle (0.05cm);
\fill[black] (0,1.2) circle (0.05cm);
\fill[black] (1.2,-0.3) circle (0.05cm);
\fill[black] (1.2,1.2) circle (0.05cm);
\draw (1.2,-0.3) -- (0,-0.3);
\draw (1.2,1.2) -- (0,1.2);
\end{tikzpicture} 
\end{array}
\,,\qquad  
\begin{array}{c}
\begin{tikzpicture}[thick]
\fill[black] (0,0.0) circle (0.05cm);
\fill[black] (0,0.3) circle (0.05cm);
\fill[black] (0,0.6) circle (0.05cm);
\fill[black] (0,0.9) circle (0.05cm);
\draw (0,0) -- (1.2,0);
\draw (0,0.6) -- (1.2,0.6);
\fill[black] (1.2,0.0) circle (0.05cm);
\fill[black] (1.2,0.3) circle (0.05cm);
\fill[black] (1.2,0.6) circle (0.05cm);
\fill[black] (1.2,0.9) circle (0.05cm);
\draw (1.2,0.3) -- (0,0.3);
\draw (1.2,0.9) -- (0,0.9);
\fill[black] (0,-0.3) circle (0.05cm);
\fill[black] (0,1.2) circle (0.05cm);
\fill[black] (1.2,-0.3) circle (0.05cm);
\fill[black] (1.2,1.2) circle (0.05cm);
\draw (1.2,-0.3) -- (0,-0.3);
\draw (1.2,1.2) -- (0,1.2);
\end{tikzpicture} 
\end{array}\,.
\end{eqnarray}

Finally, we have to consider the situation of two inequivalent qudits, with dimensions $m$ and $n$. Each of these qudits is represented by four projectors on spin $(m-1)/2$ and $(n-1)/2$ representations respectively. As for the case of identical qudits it is enough to consider connections between two pairs of projectors, assuming that the other pair has the maximal connection rank.  

Let us assume that $m>n$. Then one can have at most $4(n-1)$ lines connecting the two parties ($2(n-1)$ in the reduced diagrams). Considering only a half of the system, each $(m-1)$ projector must receive the same number of lines from the $(n-1)$ projectors. Equally, each $(m-1)$ projector can receive lines from only one $(n-1)$ projector. Then there are obviously $n$ inequivalent diagrams, which correspond to $n$ SLOCC classes. We can illustrate this with an example of entanglement classes of a qutrit and a qudit with $m=5$,
\be
\begin{array}{c}
      n=3 \\
      m=5
\end{array} \, :\qquad
\begin{array}{c}
\begin{tikzpicture}[thick]
\fill[pink] (-0.2,1.1) rectangle (0.1,1.6);
\fill[pink] (-0.2,-0.7) rectangle (0.1,-0.2);
\fill[cyan] (1.1,0.5) rectangle (1.3,1.6);
\fill[cyan] (1.1,-0.7) rectangle (1.3,0.4);
\fill[black] (0,-0.6) circle (0.05cm);
\fill[black] (0,-0.3) circle (0.05cm);
\fill[black] (0,1.5) circle (0.05cm);
\fill[black] (1.2,-0.6) circle (0.05cm);
\fill[black] (1.2,-0.3) circle (0.05cm);
\fill[black] (1.2,0.0) circle (0.05cm);
\fill[black] (1.2,0.3) circle (0.05cm);
\fill[black] (1.2,0.6) circle (0.05cm);
\fill[black] (1.2,0.9) circle (0.05cm);
\draw (0,-0.3) -- (0.1,-0.3) arc (-90:0:0.15cm) -- (0.25,1.05) arc (0:90:0.15) -- (0,1.2);
\draw (0,-0.6) -- (0.1,-0.6) arc (-90:0:0.3cm) -- (0.4,1.2) arc (0:90:0.3) -- (0,1.5);
\fill[black] (0,1.2) circle (0.05cm);
\fill[black] (0,1.5) circle (0.05cm);
\fill[black] (1.2,0.0) circle (0.05cm);
\fill[black] (1.2,0.3) circle (0.05cm);
\fill[black] (1.2,0.6) circle (0.05cm);
\fill[black] (1.2,0.9) circle (0.05cm);
\fill[black] (1.2,1.2) circle (0.05cm);
\fill[black] (1.2,1.5) circle (0.05cm);
\draw (1.2,-0.6) -- (1.1,-0.6) arc (270:180:0.6cm) -- (0.5,0.9) arc (180:90:0.6) -- (1.2,1.5);
\draw (1.2,-0.3) -- (1.1,-0.3) arc (270:180:0.45cm) -- (0.65,0.75) arc (180:90:0.45) -- (1.2,1.2);
\draw (1.2,0.9) -- (1.1,0.9) arc (90:180:0.3cm) -- (0.8,0.3) arc (180:270:0.3) -- (1.2,0.0);
\draw (1.2,0.3) -- (1.1,0.3) arc (270:90:0.15cm) -- (1.2,0.6);
\end{tikzpicture} 
\end{array} \,, \qquad 
\begin{array}{c}
\begin{tikzpicture}[thick]
\fill[pink] (-0.2,1.1) rectangle (0.1,1.6);
\fill[pink] (-0.2,-0.7) rectangle (0.1,-0.2);
\fill[cyan] (1.1,0.5) rectangle (1.3,1.6);
\fill[cyan] (1.1,-0.7) rectangle (1.3,0.4);
\fill[black] (0,-0.6) circle (0.05cm);
\fill[black] (0,-0.3) circle (0.05cm);
\fill[black] (0,1.5) circle (0.05cm);
\fill[black] (1.2,-0.6) circle (0.05cm);
\fill[black] (1.2,-0.3) circle (0.05cm);
\fill[black] (1.2,0.0) circle (0.05cm);
\fill[black] (1.2,0.3) circle (0.05cm);
\fill[black] (1.2,0.6) circle (0.05cm);
\fill[black] (1.2,0.9) circle (0.05cm);
\draw (0,-0.3) -- (0.1,-0.3) arc (-90:0:0.3cm) -- (0.4,0.9) arc (0:90:0.3) -- (0,1.2);
\fill[black] (0,1.2) circle (0.05cm);
\fill[black] (0,1.5) circle (0.05cm);
\fill[black] (1.2,0.0) circle (0.05cm);
\fill[black] (1.2,0.3) circle (0.05cm);
\fill[black] (1.2,0.6) circle (0.05cm);
\fill[black] (1.2,0.9) circle (0.05cm);
\fill[black] (1.2,1.2) circle (0.05cm);
\fill[black] (1.2,1.5) circle (0.05cm);
\draw (1.2,-0.3) -- (1.1,-0.3) arc (270:180:0.45cm) -- (0.65,0.75) arc (180:90:0.45) -- (1.2,1.2);
\draw (1.2,0.9) -- (1.1,0.9) arc (90:180:0.3cm) -- (0.8,0.3) arc (180:270:0.3) -- (1.2,0.0);
\draw (1.2,0.3) -- (1.1,0.3) arc (270:90:0.15cm) -- (1.2,0.6);
\draw (1.2,-0.6) -- (0,-0.6);
\draw (1.2,1.5) -- (0,1.5);
\end{tikzpicture} 
\end{array} \,, \qquad 
\begin{array}{c}
\begin{tikzpicture}[thick]
\fill[pink] (-0.2,1.1) rectangle (0.1,1.6);
\fill[pink] (-0.2,-0.7) rectangle (0.1,-0.2);
\fill[cyan] (1.1,0.5) rectangle (1.3,1.6);
\fill[cyan] (1.1,-0.7) rectangle (1.3,0.4);
\fill[black] (0,-0.6) circle (0.05cm);
\fill[black] (0,1.5) circle (0.05cm);
\fill[black] (1.2,-0.6) circle (0.05cm);
\fill[black] (1.2,0.0) circle (0.05cm);
\fill[black] (1.2,0.3) circle (0.05cm);
\fill[black] (1.2,0.6) circle (0.05cm);
\fill[black] (1.2,0.9) circle (0.05cm);
\draw (1.2,0.9) -- (1.1,0.9) arc (90:270:0.45) -- (1.2,0);
\draw (1.2,0.6) -- (1.1,0.6) arc (90:270:0.15cm) -- (1.2,0.3);
\fill[black] (0,-0.3) circle (0.05cm);
\fill[black] (0,1.2) circle (0.05cm);
\fill[black] (1.2,-0.3) circle (0.05cm);
\fill[black] (1.2,1.2) circle (0.05cm);
\fill[black] (1.2,1.5) circle (0.05cm);
\draw (1.2,-0.3) -- (0,-0.3);
\draw (1.2,1.2) -- (0,1.2);
\draw (1.2,-0.6) -- (0,-0.6);
\draw (1.2,1.5) -- (0,1.5);
\end{tikzpicture} 
\end{array}\,.
\ee
The last diagram with all the qutrit lines connected to the qudit gives an example of a connection of maximal rank. Altogether, this analysis is a diagrammatic illustration of the Gram-Schmidt theorem. Diagram with $4(n-1)$ connections between the left and right subsystems corresponds to a matrix of rank $n$.

Let us note that the above construction of qudit spaces is based on embedding them into higher dimensional Hilbert spaces, which corresponds to spheres with a large number of punctures. While the number of punctures and the dimension of the qudit spaces grow linearly, the dimension of the large Hilbert spaces grows exponentially. Grouping the lines in the Jones-Wenzl projectors specifies the embedding of the qudit space. The existence of an ambient space makes some diagrams equivalent when projected onto the qudit subspace, in particular, some diagrams are null when acting in that subspace, like the ones that connect outputs of the same projector. The diagrams considered above also take into account such an identification.

One natural question however, is why not identify qudit spaces directly with the spaces of spheres with punctures, without using the projectors. One answer to this question is that spaces of fundamental $j=1/2$ punctures do not grow linearly with their number, so it is not straightforward to generalize the construction to arbitrary qudits. Working with punctures of different representations requires implementation of fusion operations, which is solved by the Jones-Wenzl projectors. 

A more fundamental issue is that planar connectome diagrams for Hilbert spaces of punctures do not capture all the entanglement classes. Let us consider a pair of spheres with six punctures. The dimension of the corresponding Hilbert space is $C_3=5$ (it is straightforward to construct the diagrammatic basis for this space, see~\cite{Melnikov:2022qyt} for example). One can choose the following set of independent planar connectome diagrams for this space, 
\be
\label{6pointdiags}
\begin{array}{c}
\begin{tikzpicture}[thick]
\fill[black] (0,0.0) circle (0.05cm);
\fill[black] (0,0.3) circle (0.05cm);
\fill[black] (0,0.6) circle (0.05cm);
\fill[black] (0,0.9) circle (0.05cm);
\draw (0,0) -- (0.4,0) -- (1.2,0.0);
\draw (0,0.6) -- (0.4,0.6) arc (-90:90:0.15cm) -- (0,0.9);
\fill[black] (1.2,0.0) circle (0.05cm);
\fill[black] (1.2,0.3) circle (0.05cm);
\fill[black] (1.2,0.6) circle (0.05cm);
\fill[black] (1.2,0.9) circle (0.05cm);
\draw (1.2,0.3) -- (0,0.3);
\draw (1.2,0.6) -- (0.8,0.6) arc (-90:-270:0.15cm) -- (1.2,0.9);
\fill[black] (1.2,1.2) circle (0.05cm);
\fill[black] (1.2,1.5) circle (0.05cm);
\fill[black] (0,1.2) circle (0.05cm);
\fill[black] (0,1.5) circle (0.05cm);
\draw (1.2,1.2) -- (0.8,1.2) arc (-90:-270:0.15cm) -- (1.2,1.5);
\draw (0,1.2) -- (0.4,1.2) arc (-90:90:0.15cm) -- (0,1.5);
\end{tikzpicture} 
\end{array}
\,, \qquad 
\begin{array}{c}
\begin{tikzpicture}[thick]
\fill[black] (0,0.0) circle (0.05cm);
\fill[black] (0,0.3) circle (0.05cm);
\fill[black] (0,0.6) circle (0.05cm);
\fill[black] (0,0.9) circle (0.05cm);
\draw (0,0) -- (0.4,0) -- (1.2,0.0);
\draw (0,1.2) -- (0.4,1.2) arc (-90:90:0.15cm) -- (0,1.5);
\fill[black] (1.2,0.0) circle (0.05cm);
\fill[black] (1.2,0.3) circle (0.05cm);
\fill[black] (1.2,0.6) circle (0.05cm);
\fill[black] (1.2,0.9) circle (0.05cm);
\draw (1.2,0.3) -- (0,0.3);
\draw (1.2,1.2) -- (0.8,1.2) arc (-90:-270:0.15cm) -- (1.2,1.5);
\fill[black] (1.2,1.2) circle (0.05cm);
\fill[black] (1.2,1.5) circle (0.05cm);
\fill[black] (0,1.2) circle (0.05cm);
\fill[black] (0,1.5) circle (0.05cm);
\draw (0,0.6) -- (1.2,0.6);
\draw (1.2,0.9) -- (0,0.9);
\end{tikzpicture} 
\end{array}
\,,\qquad \text{and}\qquad  
\begin{array}{c}
\begin{tikzpicture}[thick]
\fill[black] (0,0.0) circle (0.05cm);
\fill[black] (0,0.3) circle (0.05cm);
\fill[black] (0,0.6) circle (0.05cm);
\fill[black] (0,0.9) circle (0.05cm);
\draw (0,0) -- (1.2,0);
\draw (0,0.6) -- (1.2,0.6);
\fill[black] (1.2,0.0) circle (0.05cm);
\fill[black] (1.2,0.3) circle (0.05cm);
\fill[black] (1.2,0.6) circle (0.05cm);
\fill[black] (1.2,0.9) circle (0.05cm);
\draw (1.2,0.3) -- (0,0.3);
\draw (1.2,0.9) -- (0,0.9);
\fill[black] (1.2,1.2) circle (0.05cm);
\fill[black] (1.2,1.5) circle (0.05cm);
\fill[black] (0,1.2) circle (0.05cm);
\fill[black] (0,1.5) circle (0.05cm);
\draw (0,1.2) -- (1.2,1.2);
\draw (1.2,1.5) -- (0,1.5);
\end{tikzpicture} 
\end{array}\,.
\ee
So, there are only three planar connectome diagrams for the expected five SLOCC entanglement classes. 

Meanwhile, embedding qudits in higher dimensional spaces allows to make explicit the correlations responsible for entanglement. Fusion, or fibered structure of higher spin representations in larger Hilbert spaces provide more options to entangle two parties. In terms of the matrix representation of wavefunctions and density matrices, embedding in higher dimensions allows factorizing the correlations, while in the low-dimensional representation the information about the correlations is packaged in the corresponding matrix in a not so obvious way.

\subsection{Non-local tangling and entanglement}
\label{sec:maxent}

In our topological description of bipartite entanglement we avoided using complex space connectivity, caused in particular by crossings of Wilson lines. It turned out sufficient in order to talk about entanglement at the level of the SLOCC classification. Reduction of the connectome classes to planar connectomes produced an equivalent classification. In this section we briefly discuss the properties of more complex connectomes, which contain crossing and tangling of the lines. 

We would like to distinguish local and non-local tangling caused by the crossings. Local tangling operations result in a local change of basis and by themselves are not important for entanglement. Non-local tangling is a result of a non-local permutation (exchange) of punctures. The following diagrams give an example of a local and non-local tangling,
\be
\begin{array}{c}
\begin{tikzpicture}[thick]
\draw[rounded corners=4] (0,0.0) -- (0.5,0.0) -- (0.7,0.3) -- (1.2,0.3);
\draw[rounded corners=4,draw=white,double=black] (0,0.3) -- (0.5,0.3) -- (0.7,0.0) -- (1.2,0.0);
\fill[black] (1.2,0.0) circle (0.05cm);
\fill[black] (1.2,0.3) circle (0.05cm);
\fill[black] (0,0.0) circle (0.05cm);
\fill[black] (0,0.3) circle (0.05cm);
\end{tikzpicture}
\end{array}\quad \longrightarrow \quad \text{local}\,,
\qquad
\begin{array}{c}
\begin{tikzpicture}[thick]
\fill[black] (0,0.0) circle (0.05cm);
\fill[black] (0,0.3) circle (0.05cm);
\draw (0,0) -- (0.5,0) arc (-90:30:0.15cm);
\draw (0,0.3) -- (0.5,0.3) arc (90:60:0.15cm);
\fill[black] (1.2,0.0) circle (0.05cm);
\fill[black] (1.2,0.3) circle (0.05cm);
\draw (1.2,0) -- (0.7,0) arc (-90:-120:0.15cm);
\draw (1.2,0.3) -- (0.7,0.3) arc (90:210:0.15cm);
\end{tikzpicture}
\end{array} \quad \longrightarrow \quad \text{non-local}\,.
\label{localnlocal}
\ee
The local tangling can be undone by a permutation of a pair of points on either side of the diagram. For the non-local one, one can apply skein relation~(\ref{skein}) to express the diagram as a linear combination of a fully connected and a fully disconnected diagrams (see the example below). In a more general setup this would imply a linear combination of states with a stronger and a weaker entanglement. This means in particular, that such a diagram cannot represent maximal entanglement and the correlations will be even weaker for a more complex tangling. 

Without further elaborating on this point we consider an example~\cite{Melnikov:2022qyt} of an entangled two-qubit state with two pairs of punctures connected through a non-local tangle~(\ref{localnlocal}). Using skein relations we reduce this state to a linear combination of connectome diagrams,
\be
\label{chainedstate0}
\begin{array}{c}
\begin{tikzpicture}[thick]
\fill[black] (0,0.0) circle (0.05cm);
\fill[black] (0,0.3) circle (0.05cm);
\fill[black] (0,0.6) circle (0.05cm);
\fill[black] (0,0.9) circle (0.05cm);
\draw (0,0) -- (0.5,0) arc (-90:30:0.15cm);
\draw (0,0.3) -- (0.5,0.3) arc (90:60:0.15cm);
\draw (0,0.6) -- (0.5,0.6) arc (-90:30:0.15cm);
\draw (0,0.9) -- (0.5,0.9) arc (90:60:0.15cm);
\fill[black] (1.2,0.0) circle (0.05cm);
\fill[black] (1.2,0.3) circle (0.05cm);
\fill[black] (1.2,0.6) circle (0.05cm);
\fill[black] (1.2,0.9) circle (0.05cm);
\draw (1.2,0) -- (0.7,0) arc (-90:-120:0.15cm);
\draw (1.2,0.3) -- (0.7,0.3) arc (90:210:0.15cm);
\draw (1.2,0.6) -- (0.7,0.6) arc (-90:-120:0.15cm);
\draw (1.2,0.9) -- (0.7,0.9) arc (90:210:0.15cm);
\end{tikzpicture} 
\end{array}
=  {A^{4}} 
\begin{array}{c}
\begin{tikzpicture}[thick]
\fill[black] (0,0.0) circle (0.05cm);
\fill[black] (0,0.3) circle (0.05cm);
\fill[black] (0,0.6) circle (0.05cm);
\fill[black] (0,0.9) circle (0.05cm);
\draw (0,0) -- (0.4,0) arc (-90:90:0.15cm) -- (0,0.3);
\draw (0,0.6) -- (0.4,0.6) arc (-90:90:0.15cm) -- (0,0.9);
\fill[black] (1.2,0.0) circle (0.05cm);
\fill[black] (1.2,0.3) circle (0.05cm);
\fill[black] (1.2,0.6) circle (0.05cm);
\fill[black] (1.2,0.9) circle (0.05cm);
\draw (1.2,0) -- (0.8,0) arc (-90:-270:0.15cm) -- (1.2,0.3);
\draw (1.2,0.6) -- (0.8,0.6) arc (-90:-270:0.15cm) -- (1.2,0.9);
\end{tikzpicture} 
\end{array}
 + (A^2-A^{-2}) \left(
\begin{array}{c}
\begin{tikzpicture}[thick]
\fill[black] (0,0.0) circle (0.05cm);
\fill[black] (0,0.3) circle (0.05cm);
\fill[black] (0,0.6) circle (0.05cm);
\fill[black] (0,0.9) circle (0.05cm);
\draw (0,0) -- (0.4,0) -- (1.2,0.0);
\draw (0,0.6) -- (0.4,0.6) arc (-90:90:0.15cm) -- (0,0.9);
\fill[black] (1.2,0.0) circle (0.05cm);
\fill[black] (1.2,0.3) circle (0.05cm);
\fill[black] (1.2,0.6) circle (0.05cm);
\fill[black] (1.2,0.9) circle (0.05cm);
\draw (1.2,0.3) -- (0,0.3);
\draw (1.2,0.6) -- (0.8,0.6) arc (-90:-270:0.15cm) -- (1.2,0.9);
\end{tikzpicture} 
\end{array}
 +  
\begin{array}{c}
\begin{tikzpicture}[thick]
\fill[black] (0,0.0) circle (0.05cm);
\fill[black] (0,0.3) circle (0.05cm);
\fill[black] (0,0.6) circle (0.05cm);
\fill[black] (0,0.9) circle (0.05cm);
\draw (0,0) -- (0.4,0) arc (-90:90:0.15cm) -- (0,0.3);
\draw (0,0.6) -- (1.2,0.6);
\fill[black] (1.2,0.0) circle (0.05cm);
\fill[black] (1.2,0.3) circle (0.05cm);
\fill[black] (1.2,0.6) circle (0.05cm);
\fill[black] (1.2,0.9) circle (0.05cm);
\draw (1.2,0) -- (0.8,0) arc (-90:-270:0.15cm) -- (1.2,0.3);
\draw (1.2,0.9) -- (0,0.9);
\end{tikzpicture} 
\end{array} \right)
 +  (1-A^{-4})^2  
\begin{array}{c}
\begin{tikzpicture}[thick]
\fill[black] (0,0.0) circle (0.05cm);
\fill[black] (0,0.3) circle (0.05cm);
\fill[black] (0,0.6) circle (0.05cm);
\fill[black] (0,0.9) circle (0.05cm);
\draw (0,0) -- (1.2,0);
\draw (0,0.6) -- (1.2,0.6);
\fill[black] (1.2,0.0) circle (0.05cm);
\fill[black] (1.2,0.3) circle (0.05cm);
\fill[black] (1.2,0.6) circle (0.05cm);
\fill[black] (1.2,0.9) circle (0.05cm);
\draw (1.2,0.3) -- (0,0.3);
\draw (1.2,0.9) -- (0,0.9);
\end{tikzpicture} 
\end{array}.
\ee
Using basis~(\ref{4pointbasis}) this state can be cast in an algebraic form,
\be
\begin{array}{c}
\begin{tikzpicture}[thick]
\fill[black] (0,0.0) circle (0.05cm);
\fill[black] (0,0.3) circle (0.05cm);
\fill[black] (0,0.6) circle (0.05cm);
\fill[black] (0,0.9) circle (0.05cm);
\draw (0,0) -- (0.5,0) arc (-90:30:0.15cm);
\draw (0,0.3) -- (0.5,0.3) arc (90:60:0.15cm);
\draw (0,0.6) -- (0.5,0.6) arc (-90:30:0.15cm);
\draw (0,0.9) -- (0.5,0.9) arc (90:60:0.15cm);
\fill[black] (1.2,0.0) circle (0.05cm);
\fill[black] (1.2,0.3) circle (0.05cm);
\fill[black] (1.2,0.6) circle (0.05cm);
\fill[black] (1.2,0.9) circle (0.05cm);
\draw (1.2,0) -- (0.7,0) arc (-90:-120:0.15cm);
\draw (1.2,0.3) -- (0.7,0.3) arc (90:210:0.15cm);
\draw (1.2,0.6) -- (0.7,0.6) arc (-90:-120:0.15cm);
\draw (1.2,0.9) -- (0.7,0.9) arc (90:210:0.15cm);
\end{tikzpicture} 
\end{array} \ = \ (A^4+A^{-4})^2|00\rangle + (1-A^{-4})^2|11\rangle \,.
\ee
In terms of the SLOCC classification, the diagram is an example of an operation that is invertible, but not unitary. It can be applied locally to the maximally entangled state to produce a more generic state of the Bell class. Although it is an invertible operation there is no obvious diagrammatic presentation for a local inverse, which is reminiscent of the fact that the maximally entangled state can be obtained from a generic Bell class state only with finite probability of success, while the inverse transformation can be performed with certainty~\cite{Vidal:1999vh}.  

\begin{figure}
    \centering
    \includegraphics[width=0.45\linewidth]{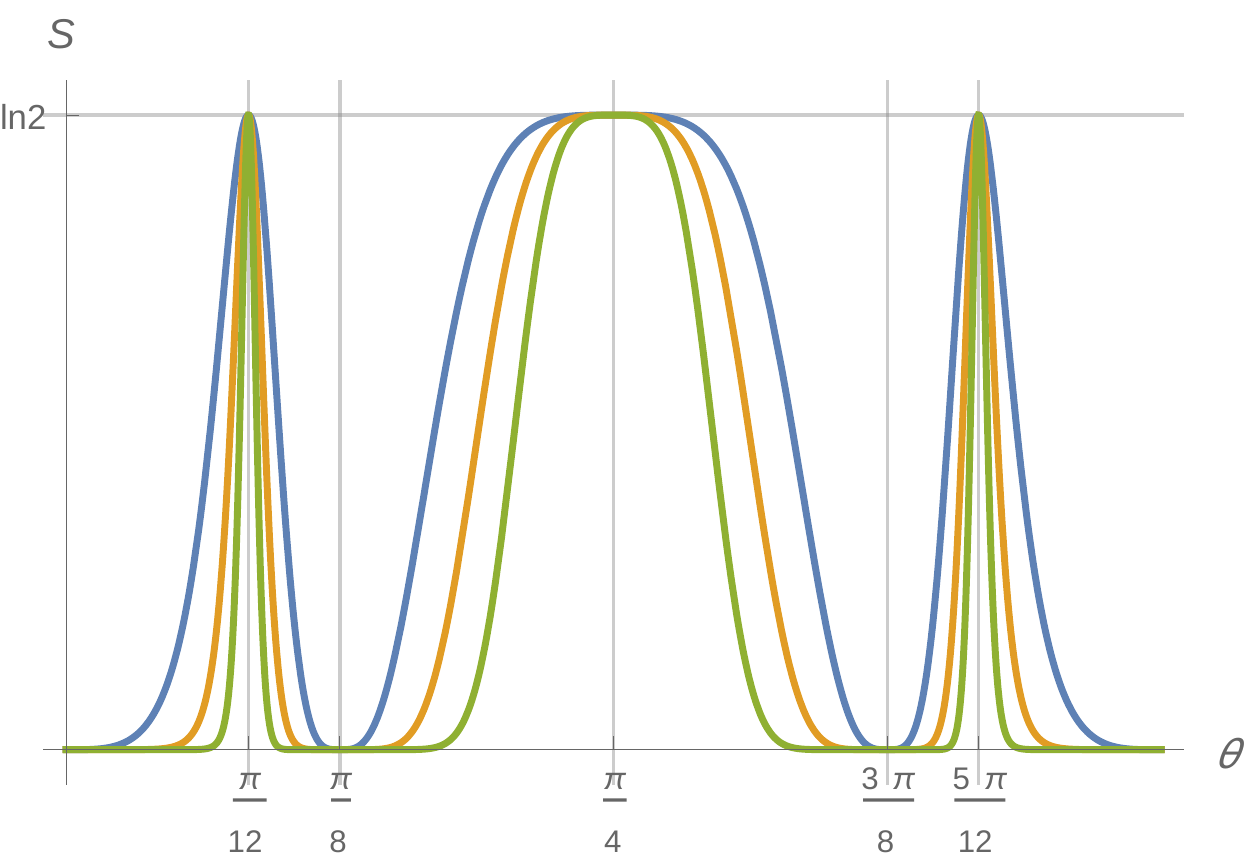}
    \caption{Entanglement entropy of the family of chain states~(\ref{chainstate}) as a function of $\theta=-i\log A$ for $\ell=0,1,2$. Apart from a few special values of $\theta$ the entropy of these states is lower than the maximal value for a pair of qubits. With increasing complexity of the tangle the entanglement entropy decreases, with the single chain lock ($\ell=0$, blue) showing the highest entropy in the family and others showing a monotonous decrease with the number of chain segments.}
    \label{fig:chainent}
\end{figure}

We can investigate the entanglement entropy of this state and a family of similar states with increasing complexity of the tangle. Let us consider the following family
\be
\label{chainstate}
\begin{array}{c}
\begin{tikzpicture}[thick]
\fill[black] (0,0.0) circle (0.05cm);
\fill[black] (0,0.3) circle (0.05cm);
\fill[black] (0,0.6) circle (0.05cm);
\fill[black] (0,0.9) circle (0.05cm);
\draw (0,0) -- (0.5,0) arc (-90:30:0.15cm);
\draw (0,0.3) -- (0.5,0.3) arc (90:60:0.15cm);
\draw (0,0.6) -- (0.5,0.6) arc (-90:30:0.15cm);
\draw (0,0.9) -- (0.5,0.9) arc (90:60:0.15cm);
\fill[black] (1.2,0.0) circle (0.05cm);
\fill[black] (1.2,0.3) circle (0.05cm);
\fill[black] (1.2,0.6) circle (0.05cm);
\fill[black] (1.2,0.9) circle (0.05cm);
\draw (1.2,0) -- (0.7,0) arc (-90:-120:0.15cm);
\draw (1.2,0.3) -- (0.7,0.3) arc (90:210:0.15cm);
\draw (1.2,0.6) -- (0.7,0.6) arc (-90:-120:0.15cm);
\draw (1.2,0.9) -- (0.7,0.9) arc (90:210:0.15cm);
\end{tikzpicture} 
\end{array}\,, \qquad 
\begin{array}{c}
\begin{tikzpicture}[thick]
\newcommand{\x}{0.6}
\fill[black] (0,0.0) circle (0.05cm);
\fill[black] (0,0.3) circle (0.05cm);
\fill[black] (0,0.6) circle (0.05cm);
\fill[black] (0,0.9) circle (0.05cm);
\draw (0,0) -- (0.5,0) arc (-90:30:0.15cm);
\draw (0,0.3) -- (0.5,0.3) arc (90:60:0.15cm);
\draw (0,0.6) -- (0.5,0.6) arc (-90:30:0.15cm);
\draw (0,0.9) -- (0.5,0.9) arc (90:60:0.15cm);
\fill[black] (1.2+\x,0.0) circle (0.05cm);
\fill[black] (1.2+\x,0.3) circle (0.05cm);
\fill[black] (1.2+\x,0.6) circle (0.05cm);
\fill[black] (1.2+\x,0.9) circle (0.05cm);
\draw (0.9,0) -- (0.7,0) arc (-90:-120:0.15cm);
\draw (0.9,0) -- (1.1,0) arc (-90:-60:0.15cm);
\draw (0.9,0.3) -- (0.7,0.3) arc (90:210:0.15cm);
\draw (0.9,0.3) -- (1.1,0.3) arc (90:-30:0.15cm);
\draw (0.9,0.6) -- (0.7,0.6) arc (-90:-120:0.15cm);
\draw (0.9,0.6) -- (1.1,0.6) arc (-90:-60:0.15cm);
\draw (0.9,0.9) -- (0.7,0.9) arc (90:210:0.15cm);
\draw (0.9,0.9) -- (1.1,0.9) arc (90:-30:0.15cm);
\draw (1.2+\x,0.3) -- (0.7+\x,0.3) arc (90:120:0.15cm);
\draw (1.2+\x,0.0) -- (0.7+\x,0.0) arc (-90:-210:0.15cm);
\draw (1.2+\x,0.9) -- (0.7+\x,0.9) arc (90:120:0.15cm);
\draw (1.2+\x,0.6) -- (0.7+\x,0.6) arc (-90:-210:0.15cm);
\end{tikzpicture} 
\end{array}\,, \qquad \ldots \qquad
\left(
\begin{array}{c}
\begin{tikzpicture}[thick]
\newcommand{\x}{0.6}
\fill[black] (0,0.0) circle (0.05cm);
\fill[black] (0,0.3) circle (0.05cm);
\fill[black] (0,0.6) circle (0.05cm);
\fill[black] (0,0.9) circle (0.05cm);
\draw (0,0) -- (0.5,0) arc (-90:30:0.15cm);
\draw (0,0.3) -- (0.5,0.3) arc (90:60:0.15cm);
\draw (0,0.6) -- (0.5,0.6) arc (-90:30:0.15cm);
\draw (0,0.9) -- (0.5,0.9) arc (90:60:0.15cm);
\fill[black] (1.2+\x,0.0) circle (0.05cm);
\fill[black] (1.2+\x,0.3) circle (0.05cm);
\fill[black] (1.2+\x,0.6) circle (0.05cm);
\fill[black] (1.2+\x,0.9) circle (0.05cm);
\draw (0.9,0) -- (0.7,0) arc (-90:-120:0.15cm);
\draw (0.9,0) -- (1.1,0) arc (-90:-60:0.15cm);
\draw (0.9,0.3) -- (0.7,0.3) arc (90:210:0.15cm);
\draw (0.9,0.3) -- (1.1,0.3) arc (90:-30:0.15cm);
\draw (0.9,0.6) -- (0.7,0.6) arc (-90:-120:0.15cm);
\draw (0.9,0.6) -- (1.1,0.6) arc (-90:-60:0.15cm);
\draw (0.9,0.9) -- (0.7,0.9) arc (90:210:0.15cm);
\draw (0.9,0.9) -- (1.1,0.9) arc (90:-30:0.15cm);
\draw (1.2+\x,0.3) -- (0.7+\x,0.3) arc (90:120:0.15cm);
\draw (1.2+\x,0.0) -- (0.7+\x,0.0) arc (-90:-210:0.15cm);
\draw (1.2+\x,0.9) -- (0.7+\x,0.9) arc (90:120:0.15cm);
\draw (1.2+\x,0.6) -- (0.7+\x,0.6) arc (-90:-210:0.15cm);
\end{tikzpicture} 
\end{array}
\right)^{2^{\ell-1}}\,.
\ee
In this family, every state is equivalent to the reduced density matrix of the previous state, which allows computing the entropies recursively. Specifically, one has
\begin{eqnarray}
|\Psi_0\rangle & = & s_0|00\rangle +c_0|11\rangle\,, \quad s_0 \ = \ (A^{4}+A^{-4})^2\,, \quad c_0 \ = \ (1-A^{-4})^2\,,\\
|\Psi_{\ell+1}\rangle & = & \frac{|s_\ell|^2}{|s_\ell|^2+|c_\ell|^2}|00\rangle +\frac{|c_\ell|^2}{|s_\ell|^2+|c_\ell|^2}|11\rangle\,,\quad \ell\ = \ 0,1,2\ldots\,, \\ S_\ell & = & - \frac{|s_\ell|^2}{|s_\ell|^2+|c_\ell|^2}\log\frac{|s_\ell|^2}{|s_\ell|^2+|c_\ell|^2} - \frac{|c_\ell|^2}{|s_\ell|^2+|c_\ell|^2}\log \frac{|s_\ell|^2}{|s_\ell|^2+|c_\ell|^2}\,.
\end{eqnarray}
The plots of the entropies for $\ell=0,1,2$, where by $\ell=0$ we understand the first diagram in~(\ref{chainstate}), as a function of the topological phase parameter $\theta=-i\log A$, which is related to coupling constant $k$ through~(\ref{qdef}) and~(\ref{Jonesrules}), are shown in figure~\ref{fig:chainent}. Values $\theta=\pm\pi/12$ and $\theta=\pi/4$ modulo $\pi/2$ are special TQFTs, in which the above states are equivalent to the maximally entangled one ($s_\ell=c_\ell$). For general values of $\theta$ the entropy drops with $\ell$. 

Note that states in family~(\ref{chainstate}) belong to the connectome class of the first diagram in~(\ref{2quadruples}), which describes a separable state. As compared to this state tangling increases entanglement, but the strongest entanglement is yet in the state with the least tangling. This counterintuitive feature is better explained with the use of the skein relation. By the latter, every subchain is a linear combination of two diagrams in the right hand side of~(\ref{skein}). However, expansion of a long chain produces only one connected element, while the number of disconnected elements grows with the length of the chain giving a higher weight to the disconnected elements. 

\begin{figure}
    $$\begin{array}{c}\includegraphics[width=0.2\linewidth]{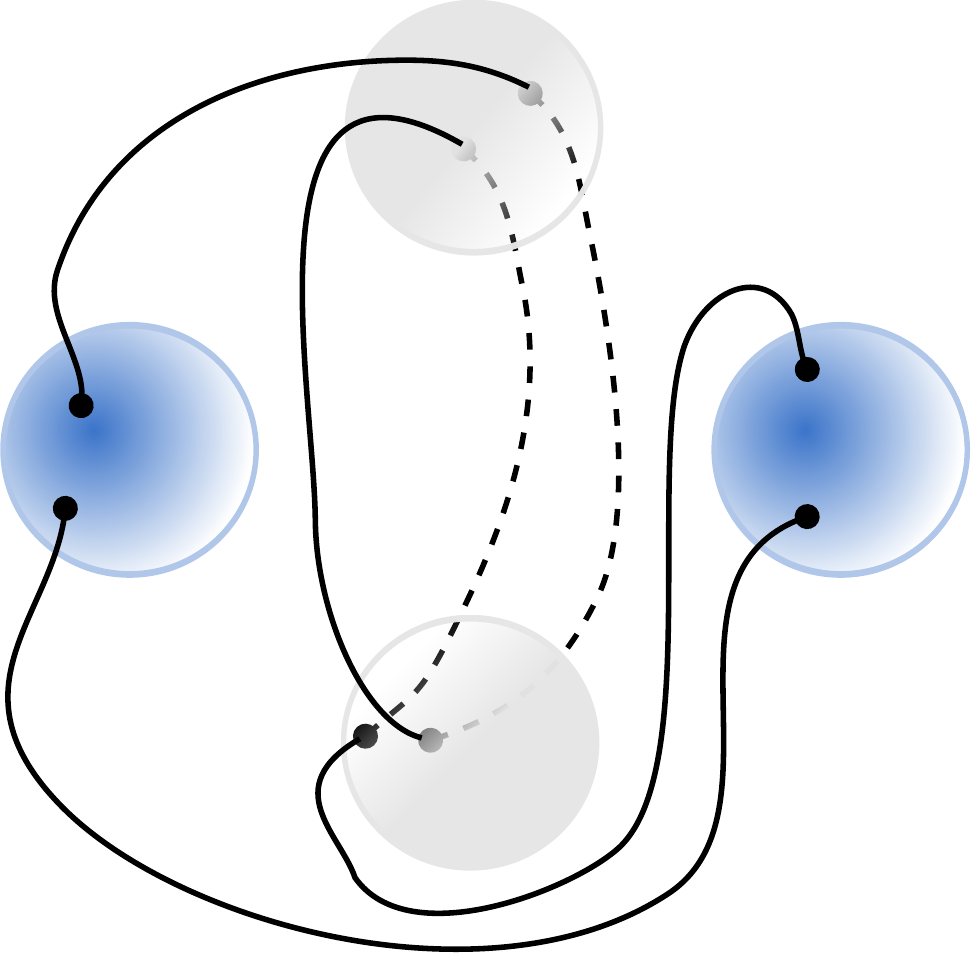}\end{array}
    {\qquad \Huge = \quad \sum\limits_R S^{R0}} \quad \begin{array}{c}\includegraphics[width=0.2\linewidth]{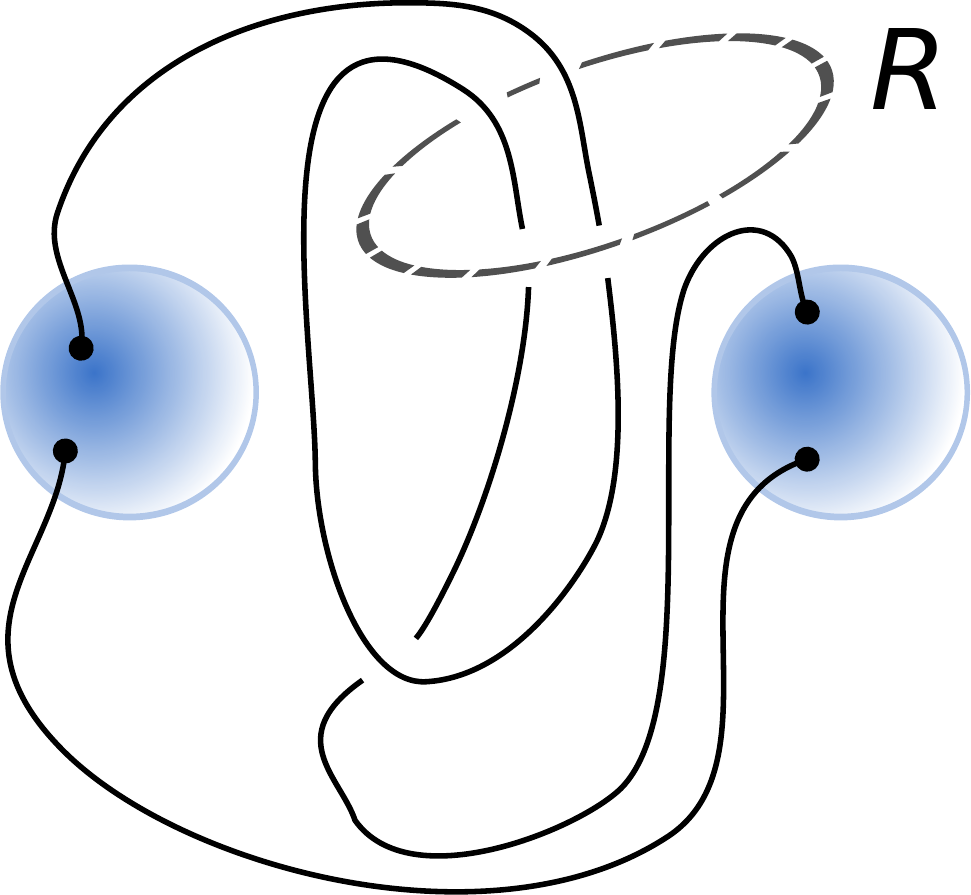}\end{array}$$
    \caption{Surgery allows to express non-trivial 3D topologies as linear combinations of simpler ones, in which 3D defects are replaced by 1D defects (Wilson lines). It is not possible to draw a picture of a 3D space with global 3D defects, for example $S^2\times S^1$, so the diagrams are heuristic depictions. Here the defect is shown as a pair of light gray spheres whose surfaces must be identified, so that lines can disappear at one sphere and reappear from the other one. Blue spheres are the boundaries. After the surgery, the defect is replaced by an additional Wilson loop in representation $R$. Coefficients $S^{R_1R_2}$ of the surgery operation are determined by the modular $S$ transformation in the basis labeled by irreducible representations~\cite{Witten:1988hf,Verlinde:1988sn}.}
    \label{fig:surgery}
\end{figure}

The effect of non-local tangling is similar to that of other types of defects, such as ``holes". By holes we mean non-trivial global 3D defects, which generalize the holes of Riemann surfaces to the case of 3-manifolds. Holes also tend to weaken entanglement if compared to simply connected spaces: heuristically they can be thought as of ruptures of space tying the quantum parties. In fact they can be reduced to a linear combination of tangles by the operation known as surgery. By a surgery one can close the 3D hole in the topology at the expense of insertion of additional Wilson loops, as illustrated by figure~\ref{fig:surgery}.

\section{Connectome states in the multipartite entanglement}
\label{sec:multipartite}

It would be interesting to extend the connectome classification to multipartite entanglement. For bipartite entanglement use of the planar connectomes (the connectomes with no line crossing) was successful: we showed how they describe the SLOCC entanglement classes. Moreover, the construction suggests that such planar connectomes correspond to maximal entanglement in each SLOCC class. In particular, introduction of tangling reduces the amount of entanglement. The first problem in the case of multipartite entanglement is that connectomes for a high number of parties are in general non-planar, so one has to refine the definition of the simplest representatives of each class. This can be done by requiring that a representative is ``simple'' if connecting endpoints of any line in the given topology results in a trivial loop.

For a triplet of qubits one can still use planarity as the criterion. However, in this case the classification diverges with the SLOCC one~\cite{Dur:2000zz}. For three qubits one can draw seven inequivalent planar diagrams up to local permutation of points and permutation of parties.
\be
\newcommand{\x}{0.75}
\begin{array}{c}
\scalebox{\x}{\begin{tikzpicture}[thick]
\fill[black] (0,0.0) circle (0.05cm);
\fill[black] (0,0.2) circle (0.05cm);
\fill[black] (0,0.4) circle (0.05cm);
\fill[black] (0,0.6) circle (0.05cm);
\draw (0,0) -- (0.2,0) arc (-90:90:0.1) -- (0,0.2);
\draw (0,0.4) -- (0.2,0.4) arc (-90:90:0.1) -- (0,0.6);
\fill[black] (1.8,0.0) circle (0.05cm);
\fill[black] (1.8,0.2) circle (0.05cm);
\fill[black] (1.8,0.4) circle (0.05cm);
\fill[black] (1.8,0.6) circle (0.05cm);
\draw (1.8,0.4) -- (1.6,0.4) arc (270:90:0.1) -- (1.8,0.6);
\draw (1.8,0.) -- (1.6,0.) arc (270:90:0.1) -- (1.8,0.2);
\fill[black] (0.6,1.2) circle (0.05cm);
\fill[black] (0.8,1.2) circle (0.05cm);
\fill[black] (1.,1.2) circle (0.05cm);
\fill[black] (1.2,1.2) circle (0.05cm);
\draw (0.6,1.2) -- (0.6,1.0) arc (-180:0:0.1) -- (0.8,1.2);
\draw (1.0,1.2) -- (1.0,1.0) arc (-180:0:0.1) -- (1.2,1.2);
\end{tikzpicture}} 
\end{array}
\quad
\begin{array}{c}
\scalebox{\x}{\begin{tikzpicture}[thick]
\fill[black] (0,0.0) circle (0.05cm);
\fill[black] (0,0.2) circle (0.05cm);
\fill[black] (0,0.4) circle (0.05cm);
\fill[black] (0,0.6) circle (0.05cm);
\draw (0,0.4) -- (0.2,0.4) arc (-90:90:0.1) -- (0,0.6);
\draw (0,0) -- (1.8,0.0);
\fill[black] (1.8,0.0) circle (0.05cm);
\fill[black] (1.8,0.2) circle (0.05cm);
\fill[black] (1.8,0.4) circle (0.05cm);
\fill[black] (1.8,0.6) circle (0.05cm);
\draw (1.8,0.2) -- (0,0.2);
\draw (1.8,0.4) -- (1.6,0.4) arc (270:90:0.1) -- (1.8,0.6);
\fill[black] (0.6,1.2) circle (0.05cm);
\fill[black] (0.8,1.2) circle (0.05cm);
\fill[black] (1.,1.2) circle (0.05cm);
\fill[black] (1.2,1.2) circle (0.05cm);
\draw (0.6,1.2) -- (0.6,1.0) arc (-180:0:0.1) -- (0.8,1.2);
\draw (1.0,1.2) -- (1.0,1.0) arc (-180:0:0.1) -- (1.2,1.2);
\end{tikzpicture}} 
\end{array}
\quad
\begin{array}{c}
\scalebox{\x}{\begin{tikzpicture}[thick]
\fill[black] (0,0.0) circle (0.05cm);
\fill[black] (0,0.2) circle (0.05cm);
\fill[black] (0,0.4) circle (0.05cm);
\fill[black] (0,0.6) circle (0.05cm);
\fill[black] (1.8,0.0) circle (0.05cm);
\fill[black] (1.8,0.2) circle (0.05cm);
\fill[black] (1.8,0.4) circle (0.05cm);
\fill[black] (1.8,0.6) circle (0.05cm);
\fill[black] (0.6,1.2) circle (0.05cm);
\fill[black] (0.8,1.2) circle (0.05cm);
\fill[black] (1.,1.2) circle (0.05cm);
\fill[black] (1.2,1.2) circle (0.05cm);
\draw (0.6,1.2) -- (0.6,1.0) arc (0:-90:0.4) -- (0.,0.6);
\draw (0.8,1.2) -- (0.8,1.0) arc (0:-90:0.6) -- (0.,0.4);
\draw (1.0,1.2) -- (1.0,1.0) arc (180:270:0.6) -- (1.8,0.4);
\draw (1.2,1.2) -- (1.2,1.0) arc (180:270:0.4) -- (1.8,0.6);
\draw (0,0) -- (0.2,0) arc (-90:90:0.1) -- (0,0.2);
\draw (1.8,0.) -- (1.6,0.) arc (270:90:0.1) -- (1.8,0.2);
\end{tikzpicture}}
\end{array}
\quad
\begin{array}{c}
\scalebox{\x}{\begin{tikzpicture}[thick]
\fill[black] (0,0.0) circle (0.05cm);
\fill[black] (0,0.2) circle (0.05cm);
\fill[black] (0,0.4) circle (0.05cm);
\fill[black] (0,0.6) circle (0.05cm);
\draw (0,0.4) -- (0.2,0.4) arc (90:-90:0.1) -- (0,0.2);
\draw (0,0) -- (1.8,0.0);
\fill[black] (1.8,0.0) circle (0.05cm);
\fill[black] (1.8,0.2) circle (0.05cm);
\fill[black] (1.8,0.4) circle (0.05cm);
\fill[black] (1.8,0.6) circle (0.05cm);
\draw (1.8,0.4) -- (1.6,0.4) arc (90:270:0.1) -- (1.8,0.2);
\fill[black] (0.6,1.2) circle (0.05cm);
\fill[black] (0.8,1.2) circle (0.05cm);
\fill[black] (1.,1.2) circle (0.05cm);
\fill[black] (1.2,1.2) circle (0.05cm);
\draw (0.6,1.2) -- (0.6,1.0) arc (0:-90:0.4) -- (0.,0.6);
\draw (0.8,1.2) -- (0.8,1.0) arc (-180:0:0.1) -- (1.0,1.2);
\draw (1.2,1.2) -- (1.2,1.0) arc (180:270:0.4) -- (1.8,0.6);
\end{tikzpicture}}
\end{array}
\quad
\begin{array}{c}
\scalebox{\x}{\begin{tikzpicture}[thick]
\fill[black] (0,0.0) circle (0.05cm);
\fill[black] (0,0.2) circle (0.05cm);
\fill[black] (0,0.4) circle (0.05cm);
\fill[black] (0,0.6) circle (0.05cm);
\fill[black] (1.8,0.0) circle (0.05cm);
\fill[black] (1.8,0.2) circle (0.05cm);
\fill[black] (1.8,0.4) circle (0.05cm);
\fill[black] (1.8,0.6) circle (0.05cm);
\draw (1.8,0.2) -- (0,0.2);
\draw (1.8,0.) -- (0,0.);
\draw (1.8,0.6) -- (0,0.6);
\draw (1.8,0.4) -- (0,0.4);
\fill[black] (0.6,1.2) circle (0.05cm);
\fill[black] (0.8,1.2) circle (0.05cm);
\fill[black] (1.,1.2) circle (0.05cm);
\fill[black] (1.2,1.2) circle (0.05cm);
\draw (0.6,1.2) -- (0.6,1.0) arc (-180:0:0.1) -- (0.8,1.2);
\draw (1.0,1.2) -- (1.0,1.0) arc (-180:0:0.1) -- (1.2,1.2);
\end{tikzpicture}}
\end{array}
\quad
\begin{array}{c}
\scalebox{\x}{\begin{tikzpicture}[thick]
\fill[black] (0,0.0) circle (0.05cm);
\fill[black] (0,0.2) circle (0.05cm);
\fill[black] (0,0.4) circle (0.05cm);
\fill[black] (0,0.6) circle (0.05cm);
\draw (0,0.4) -- (1.8,0.4);
\draw (0,0) -- (1.8,0.0);
\fill[black] (1.8,0.0) circle (0.05cm);
\fill[black] (1.8,0.2) circle (0.05cm);
\fill[black] (1.8,0.4) circle (0.05cm);
\fill[black] (1.8,0.6) circle (0.05cm);
\draw (0,0.2) -- (1.8,0.2);
\fill[black] (0.6,1.2) circle (0.05cm);
\fill[black] (0.8,1.2) circle (0.05cm);
\fill[black] (1.,1.2) circle (0.05cm);
\fill[black] (1.2,1.2) circle (0.05cm);
\draw (0.6,1.2) -- (0.6,1.0) arc (0:-90:0.4) -- (0.,0.6);
\draw (0.8,1.2) -- (0.8,1.0) arc (-180:0:0.1) -- (1.0,1.2);
\draw (1.2,1.2) -- (1.2,1.0) arc (180:270:0.4) -- (1.8,0.6);
\end{tikzpicture}}
\end{array}
\quad
\begin{array}{c}
\scalebox{\x}{\begin{tikzpicture}[thick]
\fill[black] (0,0.0) circle (0.05cm);
\fill[black] (0,0.2) circle (0.05cm);
\fill[black] (0,0.4) circle (0.05cm);
\fill[black] (0,0.6) circle (0.05cm);
\draw (0,0) -- (1.8,0.0);
\fill[black] (1.8,0.0) circle (0.05cm);
\fill[black] (1.8,0.2) circle (0.05cm);
\fill[black] (1.8,0.4) circle (0.05cm);
\fill[black] (1.8,0.6) circle (0.05cm);
\draw (0,0.2) -- (1.8,0.2);
\fill[black] (0.6,1.2) circle (0.05cm);
\fill[black] (0.8,1.2) circle (0.05cm);
\fill[black] (1.,1.2) circle (0.05cm);
\fill[black] (1.2,1.2) circle (0.05cm);
\draw (0.6,1.2) -- (0.6,1.0) arc (0:-90:0.4) -- (0.,0.6);
\draw (0.8,1.2) -- (0.8,1.0) arc (0:-90:0.6) -- (0.,0.4);
\draw (1.0,1.2) -- (1.0,1.0) arc (180:270:0.6) -- (1.8,0.4);
\draw (1.2,1.2) -- (1.2,1.0) arc (180:270:0.4) -- (1.8,0.6);
\end{tikzpicture}} 
\end{array}
\label{3partydiags}
\ee
As explained above, parties connected by only two fundamental Wilson lines are not entangled, so from the seven options only three are independent. Thus, the first four diagrams correspond to a fully separable state, the following pair to the biseparable generalization of the Bell type and only the last diagram to a genuinely tripartite entanglement. To determine the SLOCC class of the last state we can expand it in the basis~(\ref{4pointbasis}),
\begin{equation}
        \begin{array}{c}
\begin{tikzpicture}[thick]
\fill[black] (0,0.0) circle (0.05cm);
\fill[black] (0,0.2) circle (0.05cm);
\fill[black] (0,0.4) circle (0.05cm);
\fill[black] (0,0.6) circle (0.05cm);
\draw (0,0) -- (1.8,0.0);
\fill[black] (1.8,0.0) circle (0.05cm);
\fill[black] (1.8,0.2) circle (0.05cm);
\fill[black] (1.8,0.4) circle (0.05cm);
\fill[black] (1.8,0.6) circle (0.05cm);
\draw (0,0.2) -- (1.8,0.2);
\fill[black] (0.6,1.2) circle (0.05cm);
\fill[black] (0.8,1.2) circle (0.05cm);
\fill[black] (1.,1.2) circle (0.05cm);
\fill[black] (1.2,1.2) circle (0.05cm);
\draw (0.6,1.2) -- (0.6,1.0) arc (0:-90:0.4) -- (0.,0.6);
\draw (0.8,1.2) -- (0.8,1.0) arc (0:-90:0.6) -- (0.,0.4);
\draw (1.0,1.2) -- (1.0,1.0) arc (180:270:0.6) -- (1.8,0.4);
\draw (1.2,1.2) -- (1.2,1.0) arc (180:270:0.4) -- (1.8,0.6);
\end{tikzpicture} 
\end{array} \ =\  |000\rangle + \frac{1}{\sqrt{d^2 - 1}} |111\rangle \,.
    \end{equation}
This is a representative of the GHZ class in the SLOCC classification. One property of the GHZ class, which is usually cited, is the separability of the result of the measurement of any of the three qubits. This property can be seen in the diagram if one glues state $|0\rangle$ of~(\ref{4pointbasis}) to either of the ends of the diagram. The result of such an operation will be the second diagram of~(\ref{2quadruples}), which is a separable state.

Yet this analysis misses the W class of SLOCC. An obvious solution is to consider other connectomes, allowing for non-trivial tangling in the diagrams. This was attempted in~\cite{Melnikov:2022qyt} in particular, where a special tangled state was found to be at least numerically close to a W state for a specially tuned $k$, but no diagram that is generically of W type was found. Unlike GHZ state, W states are measure zero in the space of all three-qubit states, so a possibility remains that the connectome classes can only describe such states approximately, although with any desired precision.

More generally trivial connectomes rather describe different generalizations of GHZ states, either their versions for arbitrary number of parties, or embedding of the latter in partially separable situations. For four qubits the connectomes one still work with planar diagrams, and the same approach gives six inequivalent sets, of which two correspond to non-biseparable entanglement. The latter classes can be illustrated by the following diagrams
\begin{eqnarray}
\begin{array}{c}
     \begin{tikzpicture}[thick]
         \fill[black] (-0.15,0.15) circle (0.05cm);
         \fill[black] (-0.05,0.05) circle (0.05cm);
         \fill[black] (0.05,-0.05) circle (0.05cm);
         \fill[black] (0.15,-0.15) circle (0.05cm);
        \newcommand{\x}{1.0}
        \newcommand{\y}{0.0}
         \fill[black] (\x-0.15,\y-0.15) circle (0.05cm);
         \fill[black] (\x-0.05,\y-0.05) circle (0.05cm);
         \fill[black] (\x+0.05,\y+0.05) circle (0.05cm);
         \fill[black] (\x+0.15,\y+0.15) circle (0.05cm);
         \draw[thick] (0.15,-0.15) -- (\x-0.15,\y-0.15);
         \draw[thick] (0.05,-0.05) -- (\x-0.05,\y-0.05);
         \draw[thick] (-0.15,+0.15) -- (\y-0.15,\x-0.15);
         \draw[thick] (-0.05,+0.05) -- (\y-0.05,\x-0.05);
          \draw[thick] (0.15+\x,-0.15+\x) -- (\x+0.15,\y+0.15);
         \draw[thick] (0.05+\x,-0.05+\x) -- (\x+0.05,\y+0.05);
         \draw[thick] (-0.15+\x,+0.15+\x) -- (\y+0.15,\x+0.15);
         \draw[thick] (-0.05+\x,+0.05+\x) -- (\y+0.05,\x+0.05);
         \renewcommand{\x}{0.0}
        \renewcommand{\y}{1.0}
         \fill[black] (\x-0.15,\y-0.15) circle (0.05cm);
         \fill[black] (\x-0.05,\y-0.05) circle (0.05cm);
         \fill[black] (\x+0.05,\y+0.05) circle (0.05cm);
         \fill[black] (\x+0.15,\y+0.15) circle (0.05cm);
         \renewcommand{\x}{1.0}
        \renewcommand{\y}{1.0}
         \fill[black] (\x-0.15,\y+0.15) circle (0.05cm);
         \fill[black] (\x-0.05,\y+0.05) circle (0.05cm);
         \fill[black] (\x+0.05,\y-0.05) circle (0.05cm);
         \fill[black] (\x+0.15,\y-0.15) circle (0.05cm);
     \end{tikzpicture}
\end{array} & = & |0000\rangle + \frac{1}{d^2-1}|1111\rangle\,, \label{4qstate1}
\\
\begin{array}{c}
     \begin{tikzpicture}[thick]
         \fill[black] (-0.15,0.15) circle (0.05cm);
         \fill[black] (-0.05,0.05) circle (0.05cm);
         \fill[black] (0.05,-0.05) circle (0.05cm);
         \fill[black] (0.15,-0.15) circle (0.05cm);
        \newcommand{\x}{1.0}
        \newcommand{\y}{0.0}
         \fill[black] (\x-0.15,\y-0.15) circle (0.05cm);
         \fill[black] (\x-0.05,\y-0.05) circle (0.05cm);
         \fill[black] (\x+0.05,\y+0.05) circle (0.05cm);
         \fill[black] (\x+0.15,\y+0.15) circle (0.05cm);
         \draw[thick, rounded corners=2] (0.15,-0.15) -- (0.15,\y-0.05) -- (\x-0.05,\y-0.05);
         \draw[thick] (0.05,-0.05) -- (\x-0.05,\x+0.05);
         \draw[thick] (-0.15,+0.15) -- (\y-0.15,\x-0.15);
         \draw[thick] (-0.05,+0.05) -- (\y-0.05,\x-0.05);
          \draw[thick] (0.15+\x,-0.15+\x) -- (\x+0.15,\y+0.15);
         \draw[thick] (0.05+\x,-0.05+\x) -- (\x+0.05,\y+0.05);
         \draw[thick, rounded corners=2] (-0.15+\x,+0.15+\x) -- (-0.15+\x,0.05+\x) -- (0.05,\x+0.05);
         \draw[thick, rounded corners=2] (\y+0.15,\x+0.15) -- (\y+0.15,\x+0.25) -- (\x+0.25,\x+0.25) -- (\x+0.25,\y-0.15) -- (\x-0.15,\y-0.15);
         \renewcommand{\x}{0.0}
        \renewcommand{\y}{1.0}
         \fill[black] (\x-0.15,\y-0.15) circle (0.05cm);
         \fill[black] (\x-0.05,\y-0.05) circle (0.05cm);
         \fill[black] (\x+0.05,\y+0.05) circle (0.05cm);
         \fill[black] (\x+0.15,\y+0.15) circle (0.05cm);
         \renewcommand{\x}{1.0}
        \renewcommand{\y}{1.0}
         \fill[black] (\x-0.15,\y+0.15) circle (0.05cm);
         \fill[black] (\x-0.05,\y+0.05) circle (0.05cm);
         \fill[black] (\x+0.05,\y-0.05) circle (0.05cm);
         \fill[black] (\x+0.15,\y-0.15) circle (0.05cm);
     \end{tikzpicture}
\end{array}
& = & \frac{1}{d}|0000\rangle + \frac{1}{d}\left(|1100\rangle + |0011\rangle \right) - \frac{1}{d(d^2-1)}|1111\rangle\,. \label{4qstate2}
\end{eqnarray}
Here the first class is the four-qubit analog of the GHZ state. The second state differs from the GHZ one in that its parties share entanglement with all the remaining parties, while in the GHZ state entanglement is shared only with the two parties in the form of a chain. Breaking this chain at any segment, that is making a measurement of either of qubits, produces a product state. 

The above two classes of the four-partite entanglement can be compared with nine non-biseparable families of SLOCC in the classification found by~\cite{Verstraete:2002four}. The classification of~\cite{Verstraete:2002four} is not finite, since most of the families are defined using additional complex parameters that can vary continuously.  It turns out, that states~(\ref{4qstate1}) and~(\ref{4qstate2}) both belong to the same family dubbed $G_{abcd}$, for different choices of the parameters. Similarly to the GHZ states in the tripartite entanglement, states of the $G_{abcd}$ family exhibit strongest entanglement characteristics. This implies in particular, that such states are dense and can be used to approximate all other forms of entanglement, which makes them an important resource for quantum computation.


\section{Applications}
\label{sec:applications}

Connectome states and, more generally, the topological picture of quantum entanglement discussed in this paper are particularly suitable for discussing general properties and applications of entanglement. In this section we will discuss a few examples.

\subsection{Measures of entanglement}
\label{sec:measures}

The most common measure of entanglement is the von Neumann entropy. Let us briefly review a general result~\cite{Melnikov:2018zfn} for the entropy of an entanglement pair, which illustrates the convenience of the topological approach (to author's knowledge this idea was originally explored in~\cite{Dong:2008ft}, for an interesting realization on link complement states see~\cite{Balasubramanian:2016sro}).

Previously we have introduced a natural TQFT presentation for separable,
\be
\label{separable}
|\Psi_1\rangle \ = \ \begin{array}{c}\includegraphics[scale=0.25]{separable.pdf}\end{array},
\ee
and entangled,
\be
\label{EntState}
 |\Psi_2\rangle \ = \ \begin{array}{c}\includegraphics[scale=0.25]{entangled.pdf}\end{array},
\ee
states. The von Neumann entropies of these two states can be conveniently computed using the replica trick. First, one defines the reduced density matrices, say for subsystem $A$, by duplicating the diagrams representing the states and gluing the two copies along the boundary $\Sigma_B$. The result can be cast as the following diagrams, 

 \be
\label{rhoreduced}
 {\rho}_1(A) \ = \ \left[\begin{array}{c}\includegraphics[scale=0.2]{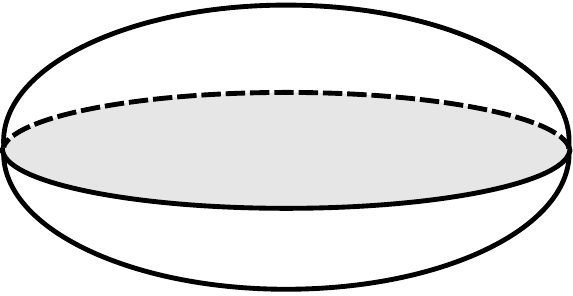}\end{array}\right]^{-1}  \begin{array}{c}\includegraphics[scale=0.25]{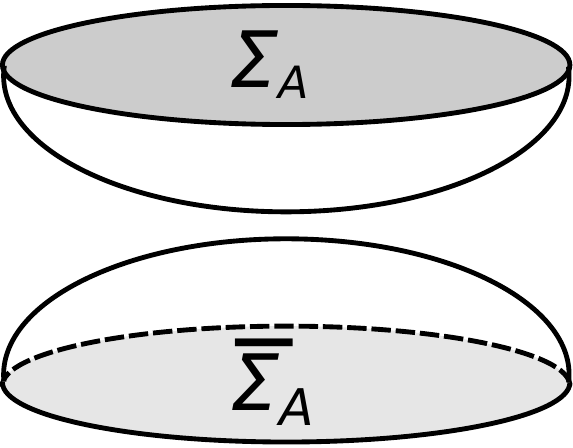}
 \end{array}, 
 \quad \text{or} \quad 
 {\rho}_2(A) \ = \ \left[\begin{array}{c}\includegraphics[scale=0.1]{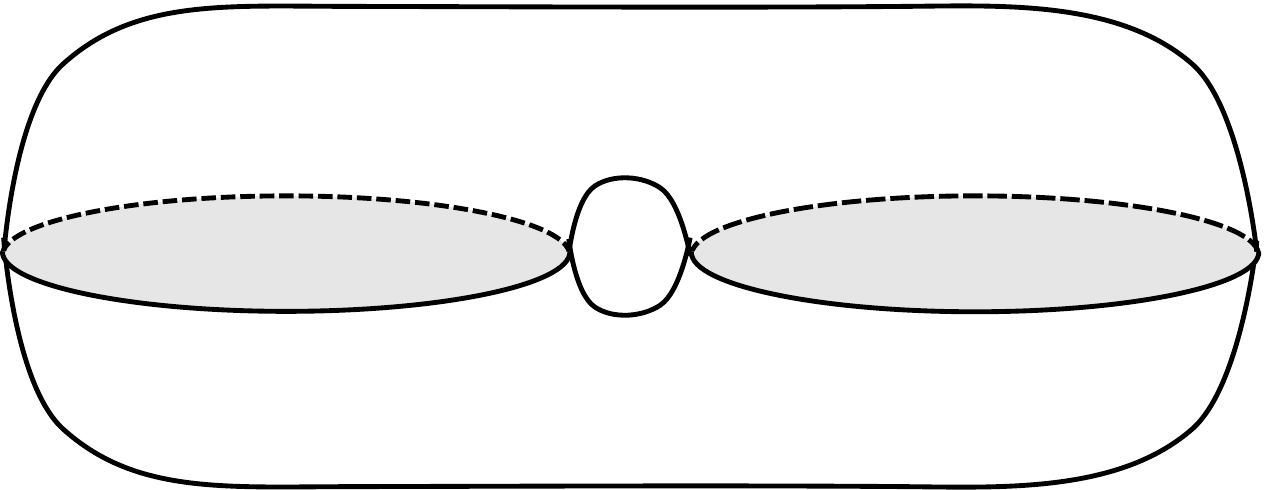}\end{array}\right]^{-1}  \begin{array}{c}\includegraphics[scale=0.25]{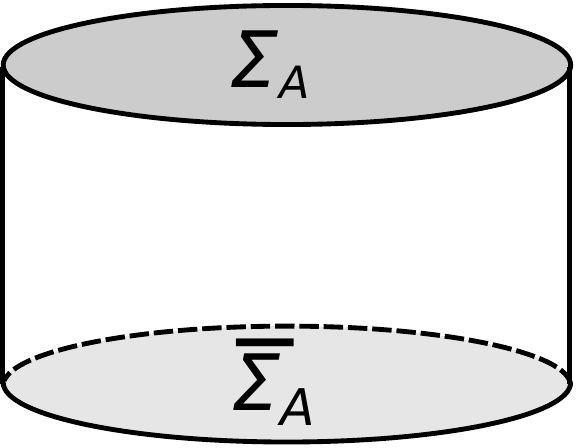}\end{array}.
 \ee
 Here the normalization factors in the square brackets have been added to ensure that the matrices have unit trace. In the replica trick one needs to compute $\tr \rho^n(A)$, analytically continue the result in $n$ and then compute
 \be
 S_{\rm E}(A) \ = \ - \lim\limits_{n\to 1} \frac{d}{dn}\,\tr \rho^n(A)\,.
 \ee
 Stacking replica of the diagrams in~(\ref{rhoreduced}) on top of themselves produce equivalent diagrams so, unless the interior of the spaces has some features, $n$ dependence can only occur in the normalization factors. In fact, in the first case $\tr \rho_1^n(A)=1$, and features of the interior are unimportant, so that the entropy vanishes, as expected for a separable state. In the second case one arrives at a formal result
\be
\label{SE}
 {S}_{{\rm E}}(A) \ = \ \log \left[\begin{array}{c}\includegraphics[scale=0.25]{donut.pdf}\end{array}\right]\,.
\ee

To get an actual number on needs to specify some further information about both the boundary $\Sigma_A$ and the topological features of the bulk of the above diagram. In the naive case of $\Sigma_A\simeq S^2$ and no features in the bulk, the above diagram equals unity and there is no entanglement between $A$ and $B$. This result was anticipated in the previous sections, since a trivial $S^2$ corresponds to a trivial Hilbert space. If $\Sigma_A$ is a torus $T^2$, or a higher genus Riemann surface, the Hilbert spaces are non-trivial and their dimensions non-trivially depend on $k$~\cite{Witten:1988hf,Dong:2008ft,Balasubramanian:2016sro}. In the absence of internal features, the diagram in~(\ref{SE}) counts the dimension of the Hilbert space, $\dim {\cal H}_{\Sigma_A}$.

Spheres with punctures also correspond to non-trivial Hilbert spaces. For $2n$ punctures generalization of states~(\ref{Bellstate}) with no extra features in the bulk have entropy
\be
S_{\rm E}(A) \ = \ \log\dim \Hc_{\Sigma_A} \ \xrightarrow[k>n-1] \ \log C_n\ \,.
\ee
These are maximally entangled states with reduced density matrices of maximal rank. Similar results apply to arbitrary planar connectome states with two 2-sphere boundaries, like~(\ref{2quadruples}) and~(\ref{6pointdiags}), since they have the property that their reduced matrices (and their powers) are proportional to the states themselves and the only problem is to correctly determine the normalization factor. For such states the entropy will be determined by the number of lines that connect $\Sigma_A$ to $\Sigma_B$. 

This property can be compared with the holographic formula for the entanglement entropy~\cite{Ryu:2006bv}, which states that the entropy is computed by the area of the minimum area surface in the bulk separating $\Sigma_A$ and $\Sigma_B$. Indeed, there is a close connection between the formulations of AdS/CFT and TQFT, and the latter can be seen as a fundamental version of the former, where the bulk space only has topological rather than geometric structure. The notion of the area is then replaced by the discrete counting of Wilson lines (flux units) piercing the space, with approximately $\log 2$ entropy per unit flux.

If the states have additional bulk features, like tangling in~(\ref{chainedstate0}), the entropy counting is more involved. Equation~(\ref{SE}) is not valid in general, but the steps of the replica procedure are well defined and allow computing the entropy at least in principle. We still expect the area law to control the entropy in this case, in particular by providing an upper bound.

The topological approach gives a very convenient way of understanding the entanglement entropy. The property of the entropy of being defined by the part of the space (section) that has the minimum number of piercing Wilson lines suggest a way to construct generalizations of this measure to the case of multiple parties.  For example, for an indicator of genuine tripartite entanglement between parties $A$, $B$ and $C$ one can consider the following. Let us schematically denote a tripartite state as
\be
|\Psi\rangle \ = \! 
\begin{array}{c}
\begin{tikzpicture}[thick]
\newcommand{\x}{1.4}
\newcommand{\y}{0.8}
\fill[black] (0,0.0) circle (0.05cm);
\fill[black] (0,0.2) circle (0.05cm);
\draw[rounded corners=2] (0.0,0.2) -- (0.4,0.2) -- (0.4,0.5);
\draw[rounded corners=2] (1.0,0.2) -- (0.6,0.2) -- (0.6,0.5);
\fill[black] (0.6,0.5) circle (0.05cm);
\fill[black] (0.4,0.5) circle (0.05cm);
\fill[black] (1,0.2) circle (0.05cm);
\fill[black] (1,0) circle (0.05cm);
\draw (0,0) -- (1.0,0.0);
\draw (-0.3,0.1) node {\small $A$};
\draw (1.3,0.1) node {\small $C$};
\draw (0.15,0.5) node {\small $B$};
\end{tikzpicture} 
\end{array}\!\,.
\ee
It does not necessarily stand for a genuinely tripartite entanglement and can be separable. Consider the following object acting in $\Hc_{A}\times\Hc_{A}$,
\be
\hat L \ = \!\!
\begin{array}{c}
\begin{tikzpicture}[thick]
\newcommand{\x}{1.4}
\newcommand{\y}{0.8}
\fill[black] (0,0.0) circle (0.05cm);
\fill[black] (0,0.2) circle (0.05cm);
\fill[black] (0,0.8) circle (0.05cm);
\fill[black] (0,1) circle (0.05cm);
\draw (0,0.2) -- (0.4,0.2) arc (-90:0:0.1) -- (0.5,0.7) arc (0:90:0.1) -- (0,0.8);
\draw (0.8,0.2) -- (1.2,0.2) arc (-90:0:0.1) -- (1.3,0.7) arc (0:90:0.1) -- (0.8,0.8) arc (90:180:0.1) -- (0.7,0.3) arc (180:270:0.1);
\fill[black] (0.5,0.5) circle (0.05cm);
\fill[black] (0.7,0.5) circle (0.05cm);
\fill[black] (1,0.8) circle (0.05cm);
\fill[black] (1,1) circle (0.05cm);
\fill[black] (1,0.2) circle (0.05cm);
\fill[black] (1,0) circle (0.05cm);
\draw (0.8+\y,0.2) -- (1.2+\y,0.2) arc (-90:0:0.1) -- (1.3+\y,0.7) arc (0:90:0.1) -- (0.8+\y,0.8) arc (90:180:0.1) -- (0.7+\y,0.3) arc (180:270:0.1);
\fill[black] (0.5+\y,0.5) circle (0.05cm);
\fill[black] (0.7+\y,0.5) circle (0.05cm);
\fill[black] (1+\y,0.8) circle (0.05cm);
\fill[black] (1+\y,1) circle (0.05cm);
\fill[black] (1+\y,0.2) circle (0.05cm);
\fill[black] (1+\y,0) circle (0.05cm);
\draw (0,0) -- (1.4+\x,0.0);
\fill[black] (1.3+\y,0.5) circle (0.05cm);
\fill[black] (1.5+\y,0.5) circle (0.05cm);
\fill[black] (1.4+\x,0.0) circle (0.05cm);
\fill[black] (1.4+\x,0.2) circle (0.05cm);
\fill[black] (1.4+\x,0.8) circle (0.05cm);
\fill[black] (1.4+\x,1) circle (0.05cm);
\draw (1.4+\x,1) -- (0,1);
\draw (1.4+\x,0.2) -- (1.+\x,0.2) arc (270:180:0.1) -- (0.9+\x,0.7) arc (180:90:0.1) -- (1.4+\x,0.8);
\draw (-0.3,0.1) node {\small $A$};
\draw (-0.3,0.9) node {\small $A$};
\draw (3.1,0.1) node {\small $A$};
\draw (3.1,0.9) node {\small $A$};
\draw (0.3,0.5) node {\small $B$};
\draw (1.1,0.5) node {\small $A$};
\draw (1.9,0.5) node {\small $C$};
\draw (1.,-0.25) node {\small $C$};
\draw (1.,1.25) node {\small $C$};
\draw (1.8,-0.25) node {\small $B$};
\draw (1.8,1.25) node {\small $B$};
\end{tikzpicture} 
\end{array}
\!,
\ee
This ladder is constructed in such a way that if either party $A$, $B$ or $C$ is disconnected, the ladder breaks into disconnected pieces. This object is intended to substitute the reduced one-party density matrix. The latter has the form
\be
\hat\rho \ = \!\!
\begin{array}{c}
\begin{tikzpicture}[thick]
\newcommand{\x}{1.4}
\newcommand{\y}{0.8}
\fill[black] (0,0.0) circle (0.05cm);
\fill[black] (0,0.2) circle (0.05cm);
\draw (0,0.2) -- (0.4,0.2) arc (-90:0:0.1) -- (0.5,0.9) arc (180:90:0.1) -- (1.4,1.0) arc (90:0:0.1) -- (1.5,0.3) arc (180:270:0.1) -- (2.0,0.2);
\draw (0.8,0.2) -- (1.2,0.2) arc (-90:0:0.1) -- (1.3,0.7) arc (0:90:0.1) -- (0.8,0.8) arc (90:180:0.1) -- (0.7,0.3) arc (180:270:0.1);
\fill[black] (1,0.8) circle (0.05cm);
\fill[black] (1,1) circle (0.05cm);
\fill[black] (1,0.2) circle (0.05cm);
\fill[black] (1,0) circle (0.05cm);
\fill[black] (2.0,0.0) circle (0.05cm);
\fill[black] (2.0,0.2) circle (0.05cm);
\draw (0,0) -- (2.0,0.0);
\draw (-0.3,0.1) node {\small $A$};
\draw (2.3,0.1) node {\small $A$};
\draw (1.,-0.25) node {\small $C$};
\draw (1.,1.25) node {\small $B$};
\end{tikzpicture} 
\end{array}
\!,
\ee
up to normalization, and only breaks when both parties $B$ and $C$ are separable. The analog of von Neumann entropy for appropriately normalized $L$ defines a measure of tripartite entanglement, 
\be
\label{tau3}
\tau_3(A) \ = - \Tr_{\Hc_A\times\Hc_A} L\log L\,, \qquad L  = \ \frac{\hat L}{\Tr_{\Hc_A\times\Hc_A}\hat{L}}\,, \qquad
\ee
One can introduce the same ladders for $B$ and $C$, and the computation of the entropy will be non-zero, for either $\tau_3(A)$, $\tau_3(B)$ or $\tau_3(C)$, only if there is three-way entanglement between $A$, $B$ and $C$. 

Let us compute $\tau_3(A)$ for the principle representatives of the SLOCC classes of tripartite entanglement. One easily finds the following up to obvious permutations,
\begin{eqnarray}
    |\Psi\rangle & = & |000\rangle\,, \qquad \tau_3(A) \ = \ 0\,,\\
    |\Psi\rangle & = & \frac{1}{\sqrt{2}}\left(|000\rangle+|011\rangle\right) \,, \qquad \tau_3(A) \ = \ 0\,, \qquad \tau_3(B) \ = \ 0 \,,  \\
    |\Psi\rangle & = & \frac{1}{\sqrt{2}}\left(|000\rangle+|111\rangle\right) \,, \qquad \tau_3(A) \ = \ \log2\,, \\
    |\Psi\rangle & = & \frac{1}{\sqrt{3}}\left(|001\rangle+|010\rangle+|100\rangle\right) \,, \quad \tau_3(A) \ = \ \frac{1}{4} \left(i \left(\sqrt{5}-1\right) \pi +4 \log4 -2 \sqrt{5}\, {\rm arcsinh}\,2\right)\!.
\end{eqnarray}
As expected, $\tau_3$ vanishes on biseparable states, while GHZ and W states produce non-vanishing values. Note that the values of $\tau_3$ are in general complex, because matrix $L$ is not Hermitian by construction. One can improve the result by introducing a more complicated but Hermitian matrix 
\be
\label{HermL}
\hat L \ = \!\!
\begin{array}{c}
\begin{tikzpicture}[thick]
\newcommand{\x}{1.4}
\newcommand{\y}{0.8}
\fill[black] (0,0.0) circle (0.05cm);
\fill[black] (0,0.2) circle (0.05cm);
\fill[black] (0,0.8) circle (0.05cm);
\fill[black] (0,1) circle (0.05cm);
\draw (0,0.2) -- (0.4,0.2) arc (-90:0:0.1) -- (0.5,0.7) arc (0:90:0.1) -- (0,0.8);
\draw (0.8,0.2) -- (1.2,0.2) arc (-90:0:0.1) -- (1.3,0.7) arc (0:90:0.1) -- (0.8,0.8) arc (90:180:0.1) -- (0.7,0.3) arc (180:270:0.1);
\fill[black] (0.5,0.5) circle (0.05cm);
\fill[black] (0.7,0.5) circle (0.05cm);
\fill[black] (1,0.8) circle (0.05cm);
\fill[black] (1,1) circle (0.05cm);
\fill[black] (1,0.2) circle (0.05cm);
\fill[black] (1,0) circle (0.05cm);
\draw (0.8+\y,0.2) -- (1.2+\y,0.2) arc (-90:0:0.1) -- (1.3+\y,0.7) arc (0:90:0.1) -- (0.8+\y,0.8) arc (90:180:0.1) -- (0.7+\y,0.3) arc (180:270:0.1);
\fill[black] (0.5+\y,0.5) circle (0.05cm);
\fill[black] (0.7+\y,0.5) circle (0.05cm);
\fill[black] (1+\y,0.8) circle (0.05cm);
\fill[black] (1+\y,1) circle (0.05cm);
\fill[black] (1+\y,0.2) circle (0.05cm);
\fill[black] (1+\y,0) circle (0.05cm);
\draw (0,0) -- (1.4+\x+\y,0.0);
\fill[black] (1.3+\y,0.5) circle (0.05cm);
\fill[black] (1.5+\y,0.5) circle (0.05cm);
\draw (1.0+\x,0.2) -- (1.4+\x,0.2) arc (-90:0:0.1) -- (1.5+\x,0.7) arc (0:90:0.1) -- (1.0+\x,0.8) arc (90:180:0.1) -- (0.9+\x,0.3) arc (180:270:0.1);
\fill[black] (1.7+\x,0.5) circle (0.05cm);
\fill[black] (1.5+\x,0.5) circle (0.05cm);
\fill[black] (1.4+\x+\y,0.0) circle (0.05cm);
\fill[black] (1.4+\x+\y,0.2) circle (0.05cm);
\fill[black] (1.4+\x+\y,0.8) circle (0.05cm);
\fill[black] (1.4+\x+\y,1) circle (0.05cm);
\draw (1.4+\x+\y,1) -- (0,1);
\draw (1.4+\x+\y,0.2) -- (1.+\x+\y,0.2) arc (270:180:0.1) -- (0.9+\x+\y,0.7) arc (180:90:0.1) -- (1.4+\x+\y,0.8);
\fill[black] (1.2+\x,0.0) circle (0.05cm);
\fill[black] (1.2+\x,0.2) circle (0.05cm);
\fill[black] (1.2+\x,0.8) circle (0.05cm);
\fill[black] (1.2+\x,1) circle (0.05cm);
\draw (-0.3,0.1) node {\small $A$};
\draw (-0.3,0.9) node {\small $A$};
\draw (3.1+\y,0.1) node {\small $A$};
\draw (3.1+\y,0.9) node {\small $A$};
\draw (0.3,0.5) node {\small $B$};
\draw (1.1,0.5) node {\small $A$};
\draw (1.9,0.5) node {\small $A$};
\draw (1.,-0.25) node {\small $C$};
\draw (1.,1.25) node {\small $C$};
\draw (1.8,-0.25) node {\small $B$};
\draw (1.8,1.25) node {\small $B$};
\draw (1.2+\x,-0.25) node {\small $C$};
\draw (1.2+\x,1.25) node {\small $C$};
\draw (1.9+\y,0.5) node {\small $B$};
\end{tikzpicture} 
\end{array}
\!,
\ee
The updated definition only changes the value on the W state, now given by
\be
\text{W:}\qquad \tau_3(A) \ = \ \frac{1}{9} \left(13 \log 3-\sqrt{13}\,{\rm arccoth}\left(\frac{7}{\sqrt{13}}\right)\right) \simeq 1.36 \ > \ \log 2\,.
\ee

\begin{figure}
    \centering
    \includegraphics[width=0.45\linewidth]{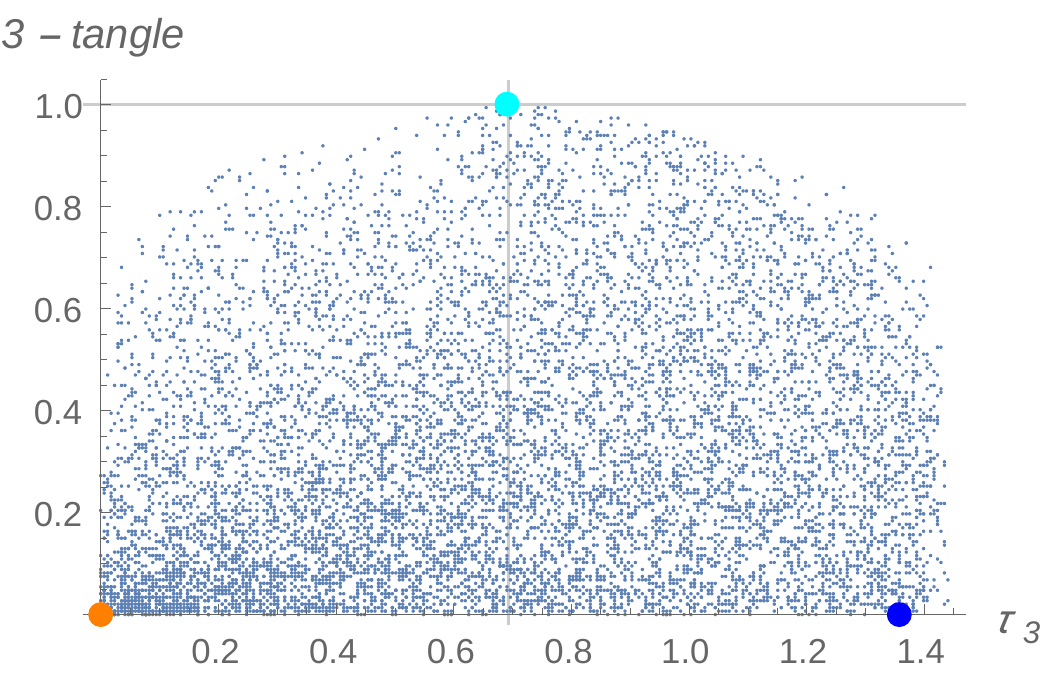}
    \caption{Correlation between the tripartite entanglement measure (\ref{tau3}), defined with respect to the Hermitian ladder~(\ref{HermL}), and the 3-tangle~\cite{Coffman:1999jd} calculated for $10^4$ three-qubit states with random real coefficients. The dots are showing the locations of biseparable states (orange), GHZ state (cyan) and the W state (blue).}
    \label{fig:laddertangle}
\end{figure}

In figure~\ref{fig:laddertangle} we show the correlation of the ladder $\tau_3$~(\ref{HermL}) with the measure of triparite entanglement known as 3-tangle~\cite{Coffman:1999jd}. The plot shows the following properties: $\tau_3$ is bounded from above by $\sim 1.44$; the W state almost maximizes $\tau_3$ and the GHZ, which maximizes the 3-tangle has only a half of the maximum value of $\tau_3$; states with low $\tau_3$ tend to have low 3-tangle; although the 3-tangle appears bounded as $\tau_3$ approaches zero, its value on some states may still be considerable.

The discussed measure of tripartite entanglement can be generalized to higher number of parties. Moreover, it is straightforward to extend the topological interpretation to the Renyi entropies and other related measures of entanglement.

\subsection{Basic properties of entanglement} 
\label{sec:properties}

In quantum computation entanglement is viewed as a resource shared between parts of a quantum system. The topological picture provides an intuitive interpretation of this nature of entanglement. The resource are the strings connecting points distributed between parts. One obvious property of entanglement in this picture is monogamy. For example, if a pair of qudits (with the same $d$) are maximally entangled, they must share all the strings with each other and cannot be entangled with any other qudit. Let us compare two situations described by the following diagrams,
\be
\newcommand{\x}{1.0}
\begin{array}{c}
\scalebox{\x}{\begin{tikzpicture}[thick]
\fill[black] (0,0.0) circle (0.05cm);
\fill[black] (0,0.2) circle (0.05cm);
\fill[black] (0,0.4) circle (0.05cm);
\fill[black] (0,0.6) circle (0.05cm);
\fill[black] (1.8,0.0) circle (0.05cm);
\fill[black] (1.8,0.2) circle (0.05cm);
\fill[black] (1.8,0.4) circle (0.05cm);
\fill[black] (1.8,0.6) circle (0.05cm);
\draw (1.8,0.2) -- (0,0.2);
\draw (1.8,0.) -- (0,0.);
\draw (1.8,0.6) -- (0,0.6);
\draw (1.8,0.4) -- (0,0.4);
\fill[black] (0.6,1.2) circle (0.05cm);
\fill[black] (0.8,1.2) circle (0.05cm);
\fill[black] (1.,1.2) circle (0.05cm);
\fill[black] (1.2,1.2) circle (0.05cm);
\draw (0.6,1.2) -- (0.6,1.0) arc (-180:0:0.1) -- (0.8,1.2);
\draw (1.0,1.2) -- (1.0,1.0) arc (-180:0:0.1) -- (1.2,1.2);
\draw (-0.3,0.3) node {$A$};
\draw (2.1,0.3) node {$B$};
\draw (1.5,1.1) node {$C$};
\end{tikzpicture}}
\end{array} \qquad \text{and} \qquad 
\begin{array}{c}
\scalebox{\x}{\begin{tikzpicture}[thick]
\draw (1.8,0.2) -- (0,0.2);
\draw (1.8,0.) -- (0,0.);
\draw (1.8,0.6) -- (0,0.6);
\draw (1.8,0.4) -- (0,0.4);
\draw[draw=white,double=black,very thick] (0.6,1.2) -- (0.6,0.4) arc (-180:-30:0.1);
\draw[draw=white,double=black] (1.0,1.2) -- (1.0,0.4) arc (-180:-30:0.1);
\draw (0.8,0.45) -- (0.8,0.55);
\draw (0.8,0.65) -- (0.8,1.2); 
\draw (1.2,0.45) -- (1.2,0.55);
\draw (1.2,0.65) -- (1.2,1.2); 
\fill[black] (0,0.0) circle (0.05cm);
\fill[black] (0,0.2) circle (0.05cm);
\fill[black] (0,0.4) circle (0.05cm);
\fill[black] (0,0.6) circle (0.05cm);
\fill[black] (1.8,0.0) circle (0.05cm);
\fill[black] (1.8,0.2) circle (0.05cm);
\fill[black] (1.8,0.4) circle (0.05cm);
\fill[black] (1.8,0.6) circle (0.05cm);
\fill[black] (0.6,1.2) circle (0.05cm);
\fill[black] (0.8,1.2) circle (0.05cm);
\fill[black] (1.,1.2) circle (0.05cm);
\fill[black] (1.2,1.2) circle (0.05cm);
\draw (-0.3,0.3) node {$A$};
\draw (2.1,0.3) node {$B$};
\draw (1.5,1.1) node {$C$};
\end{tikzpicture}}
\end{array}
\ee
In both cases $A$ and $B$ share the maximal possible number of connections, but in the second case part $C$ is topologically entangled with the pair $AB$. As a consequence, in the second case, $C$ has some amount of quantum entanglement with $A$ and $B$. But in this case the entanglement between $A$ and $B$ is not maximal. One can show, using skein relations~(\ref{skein}), for example, that the second state is a linear combination of a state with maximally entangled $AB$ and other states with less entanglement in the same pair.

As another example of the properties of quantum entanglement in the topological realization we will discuss the subadditivity of the von Neumann entropy. The regular subadditivity is the inequality for the entropy of two subsystems
\be
\label{subadditivity}
S(\rho_{AB}) \ \leq \ S(\rho_A) + S(\rho_B)\,,
\ee
where $\rho_{\cal S}$ are reduced density matrices of subsystems ${\cal S}=A,B,AB$, respectively. To compare this with the topological picture, let us use the fact that the dimension of the Hilbert space of a sphere with $n$ punctures scales approximately as $4^{n/2}$ if one assumes $k>n$, that is the von Neumann entropy of maximally entangled pairs is approximately linear in the number of lines that connect them. In other words, the number of lines connecting two systems is a measure of entropy of entanglement shared by them. 

Let us denote as $C\equiv\overline{AB}$ the complement of $AB$ ($C=\emptyset$ if $AB$ describes a pure state, otherwise the state is mixed). If $N_A$ and $N_B$ are the numbers of lines that emanate from $A$ and $B$, respectively, $\ell_{AB}$ is the number of lines connecting $A$ and $B$, and $N_{\overline{AB}}$ is the number of lines connecting both $A$ and $B$ to $C$, then the following inequality holds,
\be
\label{SAconnectome}
N_A + N_B \ = \ N_{\overline{AB}} + 2\ell_{AB} \ \geq \ N_{\overline{AB}}\,,
\ee
which is the ``connectome'' version of~(\ref{subadditivity}). To be precise, we do not include in $N_A$ and $N_B$ lines that have both endpoints belonging to the same subsystem. For planar connectome states it is straightforward to promote this relation to the statement about the entropies. As we discussed in the previous section, for such states the entanglement entropy for any subsystem ${\cal S}$ is given by the logarithm of the dimension of the Hilbert space $\Hc_{\rm min}({\cal S})$ of a surface that separates this subsystem from its complement and contains the minimum number of punctures. In our case such surface is simply a 2-sphere in ${\mathbb{R}}^3$ that encircles ${\cal S}$. Inequality~(\ref{SAconnectome}) then implies a statement about the dimensions,
\be
\dim\Hc_{\rm min}(A)\cdot\dim\Hc_{\rm min}(B)\  \geq \ \dim\Hc_{\rm min}(AB) \,,
\ee
where $\Hc_{\rm min}$ is defined with respect to the mentioned ``minimal" surface. In other words, $\Hc_{\rm min}(AB)$ in general contains less degrees of freedom than $\Hc_{AB}$, which is defined as a product $\Hc_A\otimes\Hc_B$. Then, for connectome states, inequality~(\ref{subadditivity}) follows from the above statement about the dimensions. Of course, the inequality must hold for arbitrary TQFT states, but we will not give a topological proof of this fact.

Similar discussion applies to the case of the strong subadditivity property, which is defined for three subsystems and spells
\be
\label{SSA}
S(\rho_{ABC}) + S(\rho_B) \leq S(\rho_{AB}) + S(\rho_{BC})\,.
\ee
The connectome version of this inequality reads
\be
\label{SSAconnectome}
N_{\overline{ABC}} \ = \  N_{\overline{AB}} + N_{\overline{BC}} - 2\ell_{AC} - N_B \leq  N_{\overline{AB}} + N_{\overline{BC}} - N_B\,,
\ee
where $N_{\overline{ABC}}$ is the total number of lines that connect $A$, $B$ or $C$ with the complement of their union, $N_{\overline{AB}}$ and $N_{\overline{BC}}$ count the number of lines connecting the respective unions with their complements, $N_B$ is the number of lines emanating from $B$, and $\ell_{AC}$ is the number of lines connecting $A$ and $C$. Again, for planar connectome states~(\ref{SSAconnectome}) implies~(\ref{SSA}) through the statement about the dimensions of the Hilbert spaces.

It is straightforward to generalize inequalities like~(\ref{SAconnectome}) and~(\ref{SSAconnectome}) to a larger number of parties. For example, for a system with parts $A$, $B$, $C$ and $D$ one possibility would be
\be
N_{\overline{ABCD}} \ \leq \ N_{\overline{ABC}}  + N_{\overline{BCD}} - N_B - N_C 
\ee
The corresponding relation for the entropies must be satisfied on the connectome states.

We note again the similarity of the connectome states with quantum states in holographic models, which comes through the connection of entanglement entropy with minimal surfaces. The inequalities used above are based on a discrete version of the area counting, which also appears in the tensor network models of holography~\cite{Pastawski:2015qua}. In the context of holographic theories, the inequalities for entanglement entropy and other measures were extensively studied, starting from~\cite{Headrick:2007km}. In~\cite{Bao:2015bfa} a comprehensive study of the inequalities was performed, which is very reminiscent of the topological approach. Among other results it was shown in that study that holographic states satisfy a set of other commonly known inequalities, such as the monogamy of the mutual information, Zhang-Yeung inequality and Ingleton inequality. Below we list these inequalities and their connectome versions.

The monogamy of mutual information~\cite{Hayden:2011ag} can be cast in the following way,
\be
\label{MImonogamy}
S(AB) + S(BC) + S(AC) \ \geq \ S(ABC) + S(A) + S(B) + S(C)\,.
\ee
Constructing the connectome analog one finds that the connectomes actually saturate this inequality
\be
N_{\overline{ABC}} \ = \  N_{\overline{AB}} + N_{\overline{BC}} + N_{\overline{AC}} - N_A - N_B - N_C\,.
\ee
A simple way to see this is to write~(\ref{MImonogamy}) in terms of mutual information $I(A:B)=S(A)+S(B)-S(AB)$,
\be
I(A:BC) \ \geq\ I(A:B) + I(A:C)\,. 
\ee
In the connectome version, the mutual information $I(A:B)\sim2\ell_{AB}$, just the double of the number of lines connecting $A$ and $B$. Then the number of lines connecting $A$ with the union of $B$ and $C$ is simply $\ell_{AB}+\ell_{AC}$.

The Zhang-Yeung inequality~\cite{Zhang:1998cha} is
\be
2I(C:D) \ \leq \ I(A:B) + I(A:CD) + 3I(C:D|A) + I(C:D|B)\,,
\ee
where $I(A:B|C)=S(AC)+S(BC)-S(ABC)-S(C)$. The latter quantity also characterizes the correlations between $A$ and $B$, but in the presence of an additional subsystem $C$. Its connectome version is just equivalent to the ordinary mutual information $I(A:B|C)=I(A:B)$. Hence the above inequality is satisfied trivially by the connectome states.

The Ingleton inequality~\cite{Ingleton:1971rep} reads
\be
\label{Ingleton}
I(A:B|C)+I(A:B|D) + I(C:D) \ \geq I(A:B)\,.
\ee
It is also trivially satisfied by the connectomes due to positivity of mutual information and the above interpretation, $I(A:B)\sim 2\ell_{AB}$ and $I(A:B|C)\sim 2\ell_{AB}$. In fact, connectome states must satisfy a stronger version of~(\ref{Ingleton}), 
\be
I(A:B|C)+I(A:B|D) + 2I(C:D) \ \geq 2I(A:B)\,.
\ee

Finally, we can comment on other inequalities, derived in~\cite{Bao:2015bfa} for the first time. It turns out that the ``cyclic entropy inequalities'' for $n\geq 2k+l$ subsystems,
\be
\sum\limits_{i=1}^n S(A_i\cdots A_{i+l-1}|A_{i+l}\cdots A_{i+k+l-1}) \ \geq \ S(A_1\cdots A_n)\,,
\ee
are satisfied by the connectome states. Here $S(A|B)=S(AB)-S(B)$ is the conditional entropy, and the indices are defined modulo $n$. For $n=2$ and $l=2$ and $k=0$ this is just the regular subadditivity and for $n=3$, $k=l=1$ this is the monogamy of mutual information. As stated in~\cite{Bao:2015bfa}, the most interesting case is $l=1$ and $n=2k+1$, because all other cases follow from this one and strong subadditivity. It easy to check that the connectome states saturate the inequality, that is
\be
\sum\limits_{i=1}^{2k+1} S(A_i|A_{i+1}\cdots A_{i+k}) \ = \ S(A_1\cdots A_{2k+1})\,.
\ee
A simple argument to demonstrate this is to note that a measure of the conditional entropy $S(A|B)$ is the difference between the number of lines connecting $A$ to the complement of $AB$ and the number of lines connecting $A$ to $B$. In the sum over all the subsystems $A_i$ all of them contribute $\ell_{A_iA_j}$ twice: first  when $A_j$ is in a subsystem with $A_i$ and second, when it is in a complement. Hence, the only contribution to the left hand side is the number of lines connecting $A_1\cdots A_n$ to its complement, which is precisely the right hand side.

Overall, the comparison with the study of holographic states~\cite{Bao:2015bfa} shows that connectomes are a subclass of holographic states, which likely correspond to simply-connected topologies.

\begin{figure}
    \centering
    \begin{minipage}{0.45\linewidth}
        \begin{tabular}{cccccc}
           \Trule\Brule $b_1$ & $b_2$ & \scalebox{0.8}{$\begin{array}{c}
                \begin{tikzpicture}
                \newcommand{\y}{0.5}
                    \fill[lightgray] (0,0+\y) ellipse (0.9cm and 0.2cm);
                    \draw[thick] (0,0+\y) ellipse (0.9cm and 0.2cm);
                    \fill[lightgray] (-0.9,\y) rectangle (0.9,\y+1.3);
                    \draw[thick] (0.9,\y) -- (0.9,\y+1.3);
                    \renewcommand{\y}{1.8}
                    \fill[gray] (0,0+\y) ellipse (0.9cm and 0.2cm);
                    \draw[dashed,teal] (-0.6,0+\y) -- (-0.6,\y-0.2);
                    \draw[dashed,teal] (-0.2,0+\y) -- (-0.2,\y-0.2);
                    \draw[dashed] (0.2,0+\y) -- (0.2,\y-0.2);
                    \draw[dashed] (0.6,0+\y) -- (0.6,\y-0.2);
                    \fill[black] (-0.6,0+\y) circle (0.05);
                    \fill[black] (-0.2,0+\y) circle (0.05);
                    \fill[black] (0.2,0+\y) circle (0.05);
                    \fill[black] (0.6,0+\y) circle (0.05);
                \end{tikzpicture} 
           \end{array}$} & $b_1$ & $b_2$ & \scalebox{0.8}{$\begin{array}{c}
                \begin{tikzpicture}
                \newcommand{\y}{0.5}
                    \fill[lightgray] (0,0+\y) ellipse (0.9cm and 0.2cm);
                    \draw[thick] (0,0+\y) ellipse (0.9cm and 0.2cm);
                    \fill[lightgray] (-0.9,\y) rectangle (0.9,\y+1.3);
                    \draw[thick] (0.9,\y) -- (0.9,\y+1.3);
                    \renewcommand{\y}{1.8}
                    \fill[gray] (0,0+\y) ellipse (0.9cm and 0.2cm);
                    \draw[dashed,teal] (-0.6,0+\y) -- (-0.6,\y-0.2);
                    \draw[dashed,teal] (-0.2,0+\y) -- (-0.2,\y-0.2);
                    \draw[dashed] (0.2,0+\y) -- (0.2,\y-0.2);
                    \draw[dashed] (0.6,0+\y) -- (0.6,\y-0.2);
                    \fill[black] (-0.6,0+\y) circle (0.05);
                    \fill[black] (-0.2,0+\y) circle (0.05);
                    \fill[black] (0.2,0+\y) circle (0.05);
                    \fill[black] (0.6,0+\y) circle (0.05);
                \end{tikzpicture} 
           \end{array}$} \\ [20pt]
           \Trule\Brule 0  &  0 &  \scalebox{0.8}{$\begin{array}{c} 
                \begin{tikzpicture}
                \newcommand{\y}{0.5}
                    \fill[gray] (0,0+\y) ellipse (0.9cm and 0.2cm);
                    \draw[very thick,teal] (-0.6,\y) -- (-0.6,\y+1.2);
                    \draw[very thick,teal] (-0.2,\y) -- (-0.2,\y+1.2);
                    \draw[very thick] (0.2,\y) -- (0.2,\y+1.2);
                    \draw[very thick] (0.6,\y) -- (0.6,\y+1.2);
                    \fill[black] (-0.6,0+\y) circle (0.05);
                    \fill[black] (-0.2,0+\y) circle (0.05);
                    \fill[black] (0.2,0+\y) circle (0.05);
                    \fill[black] (0.6,0+\y) circle (0.05);
                    \renewcommand{\y}{1.8}
                    \fill[gray] (0,0+\y) ellipse (0.9cm and 0.2cm);
                    \draw[dashed,teal] (-0.6,0+\y) -- (-0.6,\y-0.2);
                    \draw[dashed,teal] (-0.2,0+\y) -- (-0.2,\y-0.2);
                    \draw[dashed] (0.2,0+\y) -- (0.2,\y-0.2);
                    \draw[dashed] (0.6,0+\y) -- (0.6,\y-0.2);
                    \fill[black] (-0.6,0+\y) circle (0.05);
                    \fill[black] (-0.2,0+\y) circle (0.05);
                    \fill[black] (0.2,0+\y) circle (0.05);
                    \fill[black] (0.6,0+\y) circle (0.05);
                \end{tikzpicture} 
           \end{array}$} & 0  & 1  & \scalebox{0.8}{$\begin{array}{c} 
                \begin{tikzpicture}
                \newcommand{\y}{0.5}
                    \fill[gray] (0,0+\y) ellipse (0.9cm and 0.2cm);
                    \draw[very thick,teal] (-0.6,\y) -- (-0.6,\y+1.2);
                    \draw[very thick,teal] (-0.2,\y) -- (-0.2,\y+1.2);
                    \draw[very thick,rounded corners=1.5] (0.6,\y) -- (0.6,\y+0.55) -- (0.2,\y+0.7)  -- (0.2,\y+1.2);
                    \draw[very thick,white] (0.5,0.2875+\y+0.3) -- (0.3,0.3625+\y+0.3);
                    \draw[very thick,rounded corners=1.5] (0.2,\y) -- (0.2,\y+0.55) -- (0.6,\y+0.7) -- (0.6,\y+1.2);
                    \fill[black] (-0.6,0+\y) circle (0.05);
                    \fill[black] (-0.2,0+\y) circle (0.05);
                    \fill[black] (0.2,0+\y) circle (0.05);
                    \fill[black] (0.6,0+\y) circle (0.05);
                    \renewcommand{\y}{1.8}
                    \fill[gray] (0,0+\y) ellipse (0.9cm and 0.2cm);
                    \draw[dashed,teal] (-0.6,0+\y) -- (-0.6,\y-0.2);
                    \draw[dashed,teal] (-0.2,0+\y) -- (-0.2,\y-0.2);
                    \draw[dashed] (0.2,0+\y) -- (0.2,\y-0.2);
                    \draw[dashed] (0.6,0+\y) -- (0.6,\y-0.2);
                    \fill[black] (-0.6,0+\y) circle (0.05);
                    \fill[black] (-0.2,0+\y) circle (0.05);
                    \fill[black] (0.2,0+\y) circle (0.05);
                    \fill[black] (0.6,0+\y) circle (0.05);
                \end{tikzpicture} 
           \end{array}$}\\ [20pt]
           1  &  0 & \scalebox{0.8}{$\begin{array}{c} 
                \begin{tikzpicture}
                \newcommand{\y}{0.5}
                    \fill[gray] (0,0+\y) ellipse (0.9cm and 0.2cm);
                   \draw[very thick,rounded corners=1.5,teal] (-0.6,\y) -- (-0.6,\y+0.55) -- (-0.2,\y+0.7) -- (-0.2,\y+1.2);
                    \draw[very thick,white] (-0.5,0.2875+\y+0.3) -- (-0.3,0.3625+\y+0.3);
                    \draw[very thick,rounded corners=1.5,teal] (-0.2,\y) -- (-0.2,\y+0.55) -- (-0.6,\y+0.7)  -- (-0.6,\y+1.2);
                    \draw[very thick] (0.2,\y) -- (0.2,\y+1.2);
                    \draw[very thick] (0.6,\y) -- (0.6,\y+1.2);
                    \fill[black] (-0.6,0+\y) circle (0.05);
                    \fill[black] (-0.2,0+\y) circle (0.05);
                    \fill[black] (0.2,0+\y) circle (0.05);
                    \fill[black] (0.6,0+\y) circle (0.05);
                    \renewcommand{\y}{1.8}
                    \fill[gray] (0,0+\y) ellipse (0.9cm and 0.2cm);
                    \draw[dashed,teal] (-0.6,0+\y) -- (-0.6,\y-0.2);
                    \draw[dashed,teal] (-0.2,0+\y) -- (-0.2,\y-0.2);
                    \draw[dashed] (0.2,0+\y) -- (0.2,\y-0.2);
                    \draw[dashed] (0.6,0+\y) -- (0.6,\y-0.2);
                    \fill[black] (-0.6,0+\y) circle (0.05);
                    \fill[black] (-0.2,0+\y) circle (0.05);
                    \fill[black] (0.2,0+\y) circle (0.05);
                    \fill[black] (0.6,0+\y) circle (0.05);
                \end{tikzpicture} 
           \end{array}$} & 1  & 1  & \scalebox{0.8}{$\begin{array}{c} 
                \begin{tikzpicture}
                \newcommand{\y}{0.5}
                    \fill[gray] (0,0+\y) ellipse (0.9cm and 0.2cm);
                    \draw[very thick,rounded corners=1.5,teal] (-0.6,\y) -- (-0.6,\y+0.55) -- (-0.2,\y+0.7) -- (-0.2,\y+1.2);
                    \draw[very thick,white] (-0.5,0.2875+\y+0.3) -- (-0.3,0.3625+\y+0.3);
                    \draw[very thick,rounded corners=1.5,teal] (-0.2,\y) -- (-0.2,\y+0.55) -- (-0.6,\y+0.7)  -- (-0.6,\y+1.2);
                    \draw[very thick,rounded corners=1.5] (0.6,\y) -- (0.6,\y+0.55) -- (0.2,\y+0.7)  -- (0.2,\y+1.2);
                    \draw[very thick,white] (0.5,0.2875+\y+0.3) -- (0.3,0.3625+\y+0.3);
                    \draw[very thick,rounded corners=1.5] (0.2,\y) -- (0.2,\y+0.55) -- (0.6,\y+0.7) -- (0.6,\y+1.2);
                    \fill[black] (-0.6,0+\y) circle (0.05);
                    \fill[black] (-0.2,0+\y) circle (0.05);
                    \fill[black] (0.2,0+\y) circle (0.05);
                    \fill[black] (0.6,0+\y) circle (0.05);
                    \renewcommand{\y}{1.8}
                    \fill[gray] (0,0+\y) ellipse (0.9cm and 0.2cm);
                    \draw[dashed,teal] (-0.6,0+\y) -- (-0.6,\y-0.2);
                    \draw[dashed,teal] (-0.2,0+\y) -- (-0.2,\y-0.2);
                    \draw[dashed] (0.2,0+\y) -- (0.2,\y-0.2);
                    \draw[dashed] (0.6,0+\y) -- (0.6,\y-0.2);
                    \fill[black] (-0.6,0+\y) circle (0.05);
                    \fill[black] (-0.2,0+\y) circle (0.05);
                    \fill[black] (0.2,0+\y) circle (0.05);
                    \fill[black] (0.6,0+\y) circle (0.05);
                \end{tikzpicture} 
           \end{array}$}
        \end{tabular}
    \end{minipage}\quad 
    \begin{minipage}{0.5\linewidth}
\begin{tikzpicture}       
\fill[pink] (-1.5,-1) rectangle (5.5,0.25);
\fill[cyan] (-1.5,0.25) rectangle (1.5,2.05);
\fill[lime] (1.5,0.25) rectangle (5.5,2.05);
\fill[lime] (-1.5,2.05) rectangle (5.5,4.0);
\newcommand{\x}{0}
\draw[thick,teal] (-0.6+\x,0) -- (-0.6+\x,-0.65) arc (180:270:0.15) -- (0.45+\x+4,-0.8) arc (-90:0:0.15) -- (0.6+\x+4,0);
\draw[thick,teal] (-0.2+\x,0) -- (-0.2+\x,-0.45) arc (180:270:0.15) -- (0.05+\x+4,-0.6) arc (-90:0:0.15) -- (0.2+\x+4,0);
\draw[thick] (0.2+\x,0) -- (0.2+\x,-0.25) arc (180:270:0.15) -- (-0.35+\x+4,-0.4) arc (-90:0:0.15) -- (-0.2+\x+4,0);
\draw[thick] (0.6+\x,0) -- (0.6+\x,-0.05) arc (180:270:0.15) -- (-0.75+\x+4,-0.2) arc (-90:0:0.15) -- (-0.6+\x+4,0);
\fill[gray] (0+\x,0) ellipse (0.9cm and 0.2cm);
\draw[dashed,teal] (-0.6+\x,0) -- (-0.6+\x,-0.2);
\draw[dashed,teal] (-0.2+\x,0) -- (-0.2+\x,-0.2);
\draw[dashed] (0.2+\x,0) -- (0.2+\x,-0.2);
\draw[dashed] (0.6+\x,0) -- (0.6+\x,-0.05) arc (180:270:0.15);
\fill[black] (-0.6+\x,0) circle (0.05);
\fill[black] (-0.2+\x,0) circle (0.05);
\fill[black] (0.2+\x,0) circle (0.05);
\fill[black] (0.6+\x,0) circle (0.05);
\renewcommand{\x}{4}
\fill[gray] (0+\x,0) ellipse (0.9cm and 0.2cm);
\draw[dashed] (-0.6+\x,0) -- (-0.6+\x,-0.05) arc (0:-90:0.15);
\draw[dashed] (-0.2+\x,0) -- (-0.2+\x,-0.2);
\draw[dashed,teal] (0.2+\x,0) -- (0.2+\x,-0.2);
\draw[dashed,teal] (0.6+\x,0) -- (0.6+\x,-0.2);
\fill[black] (-0.6+\x,0) circle (0.05);
\fill[black] (-0.2+\x,0) circle (0.05);
\fill[black] (0.2+\x,0) circle (0.05);
\fill[black] (0.6+\x,0) circle (0.05);
\newcommand{\y}{0.5}
\fill[lightgray] (0,0+\y) ellipse (0.9cm and 0.2cm);
\draw[thick] (0,0+\y) ellipse (0.9cm and 0.2cm);
\fill[lightgray] (-0.9,\y) rectangle (0.9,\y+1.3);
\draw[thick] (0.9,\y) -- (0.9,\y+1.3);
\fill[gray] (0+\x,0+\y) ellipse (0.9cm and 0.2cm);
\draw[thick,rounded corners=1.5,teal] (0.2+\x,0+\y) -- (0.2+\x,0.25+\y) -- (-0.2+\x,0.4+\y) -- (-0.2+\x,0.55+\y) -- (-0.6+\x,0.7+\y) -- (-0.6+\x,1.2+\y);
\draw[very thick,lime] (0.1+\x,0.2875+\y) -- (-0.1+\x,0.3625+\y);
\draw[very thick,lime] (-0.3+\x,0.2875+\y+0.3) -- (-0.5+\x,0.3625+\y+0.3);
\draw[thick,rounded corners=1.5,teal] (0.6+\x,0+\y) -- (0.6+\x,0.55+\y) -- (0.2+\x,0.7+\y) -- (0.2+\x,0.85+\y) -- (-0.2+\x,1.+\y) -- (-0.2+\x,1.2+\y);
\draw[very thick,lime] (0.1+\x,0.2875+\y+0.6) -- (-0.1+\x,0.3625+\y+0.6);
\draw[very thick,lime] (0.5+\x,0.2875+\y+0.3) -- (0.3+\x,0.3625+\y+0.3);
\draw[thick,rounded corners=1.5] (-0.2+\x,0+\y) -- (-0.2+\x,0.25+\y) -- (0.2+\x,0.4+\y) -- (0.2+\x,0.55+\y) -- (0.6+\x,0.7+\y) -- (0.6+\x,1.2+\y);
\draw[thick,rounded corners=1.5] (-0.6+\x,0+\y) -- (-0.6+\x,0.55+\y) -- (-0.2+\x,0.7+\y) -- (-0.2+\x,0.85+\y) -- (0.2+\x,1.0+\y) -- (0.2+\x,1.2+\y);
\fill[black] (-0.6+\x,0+\y) circle (0.05);
\fill[black] (-0.2+\x,0+\y) circle (0.05);
\fill[black] (0.2+\x,0+\y) circle (0.05);
\fill[black] (0.6+\x,0+\y) circle (0.05);
\renewcommand{\y}{1.8}
\fill[gray] (0,0+\y) ellipse (0.9cm and 0.2cm);
\draw[dashed,teal] (-0.6,0+\y) -- (-0.6,\y-0.2);
\draw[dashed,teal] (-0.2,0+\y) -- (-0.2,\y-0.2);
\draw[dashed] (0.2,0+\y) -- (0.2,\y-0.2);
\draw[dashed] (0.6,0+\y) -- (0.6,\y-0.2);
\fill[black] (-0.6,0+\y) circle (0.05);
\fill[black] (-0.2,0+\y) circle (0.05);
\fill[black] (0.2,0+\y) circle (0.05);
\fill[black] (0.6,0+\y) circle (0.05);
\fill[gray] (0+\x,0+\y) ellipse (0.9cm and 0.2cm);
\draw[dashed,teal] (-0.6+\x,0+\y) -- (-0.6+\x,\y-0.2);
\draw[dashed,teal] (-0.2+\x,0+\y) -- (-0.2+\x,\y-0.2);
\draw[dashed] (0.2+\x,0+\y) -- (0.2+\x,\y-0.2);
\draw[dashed] (0.6+\x,0+\y) -- (0.6+\x,\y-0.2);
\fill[black] (-0.6+\x,0+\y) circle (0.05);
\fill[black] (-0.2+\x,0+\y) circle (0.05);
\fill[black] (0.2+\x,0+\y) circle (0.05);
\fill[black] (0.6+\x,0+\y) circle (0.05);
\renewcommand{\y}{2.3}
\renewcommand{\x}{4}
\fill[gray] (0,0+\y) ellipse (0.9cm and 0.2cm);
\fill[gray] (0+\x,0+\y) ellipse (0.9cm and 0.2cm);
\draw[thick,rounded corners=1.5] (0.2+\x,\y) -- (0.2+\x,\y+0.55) -- (0.6+\x,\y+0.7) -- (0.6+\x,\y+1.2);
\draw[very thick,lime] (0.3+\x,0.2875+\y+0.3) -- (0.5+\x,0.3625+\y+0.3);
\draw[thick,rounded corners=1.5,teal] (-0.2,0+\y) -- (-0.2,\y+0.55) -- (-0.6,\y+0.7) -- (-0.6,1.2+\y);
\draw[very thick,lime] (-0.3,0.2875+\y+0.3) -- (-0.5,0.3625+\y+0.3);
\draw[thick,rounded corners=1.5] (0.2,0+\y) -- (0.2,0.35+\y) arc (180:90:0.15) -- (\x/2-0.5,0.5+\y) -- (\x/2-0.3,0.7+\y)-- (\x/2-0.1,0.7+\y) -- (\x/2+0.1,0.9+\y) -- (\x-0.75,0.9+\y) arc (-90:0:0.15) -- (-0.6+\x,1.2+\y);
\draw[thick,rounded corners=1.5] (0.6,0+\y) -- (0.6,0.15+\y) arc (180:90:0.15) -- (\x/2-0.1,0.3+\y) -- (\x/2+0.1,0.5+\y) -- (\x/2+0.3,0.5+\y) -- (\x/2+0.5,0.7+\y) -- (\x-0.35,0.7+\y) arc (-90:0:0.15)  -- (\x+0.2,1.0+\y) -- (\x+0.2,1.2+\y);
\draw[very thick,lime] (-0.1+\x,0.2875+\y+0.6) -- (0.1+\x,0.3625+\y+0.6);
 \draw[thick,rounded corners=1.5] (0.6+\x,\y) -- (0.6+\x,\y+0.55) -- (0.2+\x,\y+0.7)  -- (0.2+\x,\y+0.85) -- (-0.2+\x,\y +1.) -- (-0.2+\x,\y+1.2);
\draw[thick,teal,rounded corners=1.5] (-0.6+\x,0+\y) -- (-0.6+\x,0.15+\y) arc (0:90:0.15) -- (\x/2+0.1,0.3+\y) -- (\x/2+0.05,0.35+\y);
\draw[thick,teal,rounded corners=1.5] (\x/2-0.05,0.45+\y) -- (\x/2-0.1,0.5+\y) -- (\x/2-0.3,0.5+\y) -- (\x/2-0.35,0.55+\y);
\draw[thick,rounded corners=1.5,teal] (\x/2-0.45,0.65+\y) -- (\x/2-0.5,0.7+\y) -- (0.35,0.7+\y) arc (-90:-180:0.15) -- (-0.2,1.0+\y) -- (-0.2,1.2+\y);
\draw[very thick,lime] (0.1,0.2875+\y+0.6) -- (-0.1,0.3625+\y+0.6);
 \draw[thick,teal,rounded corners=1.5] (-0.6,0+\y) -- (-0.6,\y+0.55) -- (-0.2,\y+0.7)  -- (-0.2,\y+0.85) -- (0.2,\y +1.) -- (0.2,1.2+\y);
\draw[thick,teal] (-0.2+\x,0+\y) -- (-0.2+\x,0.35+\y) arc (0:90:0.15) -- (\x/2+0.5,0.5+\y) -- (\x/2+0.45,0.55+\y);
\draw[thick,teal,rounded corners=1.5] (\x/2+0.35,0.65+\y) -- (\x/2+0.3,0.7+\y) -- (\x/2+0.1,0.7+\y) -- (\x/2+0.05,0.75+\y);
\draw[thick,teal,rounded corners=1.5] (\x/2-0.05,0.85+\y) -- (\x/2-0.1,0.9+\y) -- (0.75,0.9+\y) arc (-90:-180:0.15) -- (0.6,1.2+\y);
\fill[black] (-0.6,0+\y) circle (0.05);
\fill[black] (-0.2,0+\y) circle (0.05);
\fill[black] (0.2,0+\y) circle (0.05);
\fill[black] (0.6,0+\y) circle (0.05);
\fill[black] (-0.6+\x,0+\y) circle (0.05);
\fill[black] (-0.2+\x,0+\y) circle (0.05);
\fill[black] (0.2+\x,0+\y) circle (0.05);
\fill[black] (0.6+\x,0+\y) circle (0.05);
\renewcommand{\y}{3.6}
\fill[gray] (0,0+\y) ellipse (0.9cm and 0.2cm);
\draw[dashed,teal] (-0.6,0+\y) -- (-0.6,-0.15+\y);
\draw[dashed,teal] (-0.2,0+\y) -- (-0.2,-0.2+\y);
\draw[dashed,teal] (0.2,0+\y) -- (0.2,-0.2+\y);
\draw[dashed,teal] (0.6,0+\y) -- (0.6,-0.15+\y);
\fill[black] (-0.6,0+\y) circle (0.05);
\fill[black] (-0.2,0+\y) circle (0.05);
\fill[black] (0.2,0+\y) circle (0.05);
\fill[black] (0.6,0+\y) circle (0.05);
\fill[gray] (0+\x,0+\y) ellipse (0.9cm and 0.2cm);
\fill[black] (-0.6+\x,0+\y) circle (0.05);
\fill[black] (-0.2+\x,0+\y) circle (0.05);
\fill[black] (0.2+\x,0+\y) circle (0.05);
\fill[black] (0.6+\x,0+\y) circle (0.05);
\draw[dashed] (-0.6+\x,0+\y) -- (-0.6+\x,-0.15+\y);
\draw[dashed] (-0.2+\x,0+\y) -- (-0.2+\x,-0.2+\y);
\draw[dashed] (0.2+\x,0+\y) -- (0.2+\x,-0.2+\y);
\draw[dashed] (0.6+\x,0+\y) -- (0.6+\x,-0.2+\y);
    \end{tikzpicture}
    \end{minipage}
    \caption{The topological version of the dense coding protocol. Charlie (pink background) prepares a pair of entangled qubits and passes them to Alice (light blue) and Bob (lime). Alice codes a pair of classical bits by applying one of the four operations shown on the left in the place indicated by the gray cylinder. After that Bob receives Alice's qubit. He applies one local transformation on his original qubit and a non-local disentangling transformation on the pair of qubits~\cite{Kauffman:2004bra}, separating black and blue lines. In the final state the two unentangled qubits encode the pair of Alice's classical bits.}
    \label{fig:densecoding}
\end{figure}
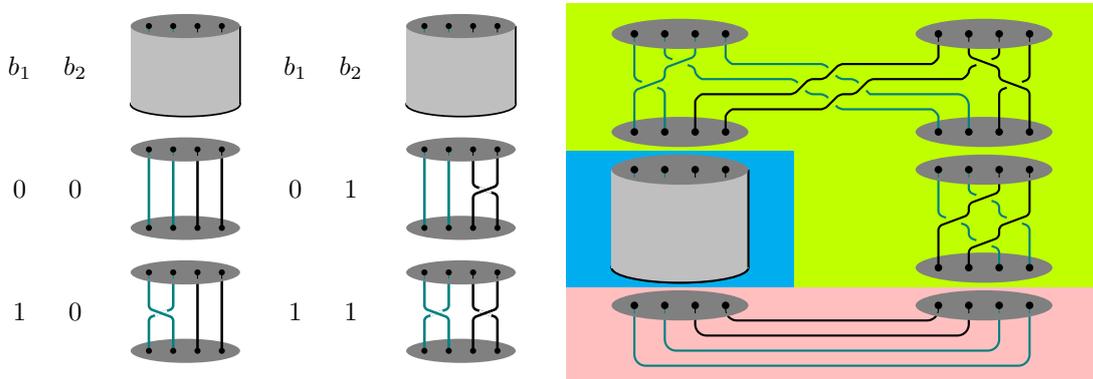

\subsection{Basic quantum algorithms}
\label{sec:algorithms}

\subsubsection{Dense coding} 

Let us consider a few examples of quantum algorithms as seen by the topological approach. We will work out two examples of communication algorithms based on pre-sharing entanglement between the parties. The first example is superdense (or simply dense) coding~\cite{Bennett:1992zzb}.

In the simplest version of dense coding Alice and Bob share a pair of entangled qubits. Alice can use her qubit to code a pair of classical bits and send the result to Bob via a quantum communication channel. By performing operations and measurements on his and Alice's qubits Bob can recover the classical bits of Alice without classically communicating with her. This algorithm is an example of a quantum cryptography protocol, in which the security of information is protected by the distribution of entanglement between the parties.

We will explain the dense coding algorithm directly through its topological version. The necessary ingredients include a maximally entangled pair and a measurement basis. From section~\ref{sec:connectome} we know that a maximally entangled pair can be represented by diagrams like~(\ref{Bellstate}). For the measurement basis it is more convenient to choose the non-orthonormal basis~(\ref{4pointbasis0}). To code the classical bits Alice can use the braiding gates.

The dense coding protocol is shown in figure~\ref{fig:densecoding}. On the left of the figure a table shows the gates that Alice needs to apply to code classical qubit pairs $00$, $01$, $10$ and $11$, so that Bob would measure states $|00\rangle$, $|01\rangle$, $|10\rangle$ and $|11\rangle$ respectively, here coded by basis~(\ref{4pointbasis0}), with unit probability. These gates must be substituted for the gray cylinder in the circuit on the right half of the figure.

The essence of the protocol is in the existence of two topologically separable subcircuits (blue and black in the figure). At different stages these subcircuits are shared between the parties (qubits) or separated. The fact that a planar connectome state is used simplifies the protocol: one should not worry about additional tangling of the subcircuits.

\begin{figure}
    \centering
    \begin{minipage}{0.55\linewidth}
\begin{tikzpicture}
\fill[pink] (1.5,-1.0) rectangle (7.2,0.25);
\fill[lime] (-1.2,-1.0) rectangle (1.5,0.25);
\fill[lime] (-1.2,0.25) rectangle (4.5,3.4);
\fill[cyan] (4.5,0.25) rectangle (7.2,3.4);
\fill[black] (-0.9,0) arc (-180:0:0.9);
\draw[gray,ultra thick] (-0.85,0) arc (-180:-100:0.9);
\draw[gray,ultra thick] (-0.8,0) arc (-180:-100:0.9);
\fill[gray] (0,0) ellipse (0.9cm and 0.2cm);
\fill[black] (-0.6,0) circle (0.05);
\fill[black] (-0.2,0) circle (0.05);
\fill[black] (0.2,0) circle (0.05);
\fill[black] (0.6,0) circle (0.05);
\draw[dashed] (-0.6,0) -- (-0.6,-0.2);
\draw[dashed] (-0.2,0) -- (-0.2,-0.2);
\draw[dashed] (0.2,0) -- (0.2,-0.2);
\draw[dashed] (0.6,0) -- (0.6,-0.2);
\newcommand{\y}{0.5}
\newcommand{\x}{3}
\draw[thick] (-0.6+\x,0) -- (-0.6+\x,-0.65) arc (180:270:0.15) -- (0.45+\x+3,-0.8) arc (-90:0:0.15) -- (0.6+\x+3,0);
\draw[thick] (-0.2+\x,0) -- (-0.2+\x,-0.45) arc (180:270:0.15) -- (0.05+\x+3,-0.6) arc (-90:0:0.15) -- (0.2+\x+3,0);
\draw[thick] (0.2+\x,0) -- (0.2+\x,-0.25) arc (180:270:0.15) -- (-0.35+\x+3,-0.4) arc (-90:0:0.15) -- (-0.2+\x+3,0);
\draw[thick] (0.6+\x,0) -- (0.6+\x,-0.05) arc (180:270:0.15) -- (-0.75+\x+3,-0.2) arc (-90:0:0.15) -- (-0.6+\x+3,0);
\fill[gray] (0+\x,0) ellipse (0.9cm and 0.2cm);
\fill[black] (-0.6+\x,0) circle (0.05);
\fill[black] (-0.2+\x,0) circle (0.05);
\fill[black] (0.2+\x,0) circle (0.05);
\fill[black] (0.6+\x,0) circle (0.05);
\draw[dashed] (-0.6+\x,0) -- (-0.6+\x,-0.2);
\draw[dashed] (-0.2+\x,0) -- (-0.2+\x,-0.2);
\draw[dashed] (0.2+\x,0) -- (0.2+\x,-0.2);
\draw[dashed] (0.6+\x,0) -- (0.6+\x,-0.05) arc (180:270:0.15);
\fill[gray] (0,0+\y) ellipse (0.9cm and 0.2cm);
\fill[gray] (0+\x,0+\y) ellipse (0.9cm and 0.2cm);
\draw[thick,rounded corners=1.5] (0.2+\x,\y) -- (0.2+\x,\y+0.55) -- (0.6+\x,\y+0.7) -- (0.6+\x,\y+1.2);
\draw[very thick,lime] (0.3+\x,0.2875+\y+0.3) -- (0.5+\x,0.3625+\y+0.3);
\draw[thick,rounded corners=1.5] (-0.2,0+\y) -- (-0.2,\y+0.55) -- (-0.6,\y+0.7) -- (-0.6,1.2+\y);
\draw[very thick,lime] (-0.3,0.2875+\y+0.3) -- (-0.5,0.3625+\y+0.3);
\draw[thick,rounded corners=1.5] (0.2,0+\y) -- (0.2,0.35+\y) arc (180:90:0.15) -- (\x/2-0.5,0.5+\y) -- (\x/2-0.3,0.7+\y)-- (\x/2-0.1,0.7+\y) -- (\x/2+0.1,0.9+\y) -- (\x-0.75,0.9+\y) arc (-90:0:0.15) -- (-0.6+\x,1.2+\y);
\draw[thick,rounded corners=1.5] (0.6,0+\y) -- (0.6,0.15+\y) arc (180:90:0.15) -- (\x/2-0.1,0.3+\y) -- (\x/2+0.1,0.5+\y) -- (\x/2+0.3,0.5+\y) -- (\x/2+0.5,0.7+\y) -- (\x-0.35,0.7+\y) arc (-90:0:0.15)  -- (\x+0.2,1.0+\y) -- (\x+0.2,1.2+\y);
\draw[very thick,lime] (-0.1+\x,0.2875+\y+0.6) -- (0.1+\x,0.3625+\y+0.6);
 \draw[thick,rounded corners=1.5] (0.6+\x,\y) -- (0.6+\x,\y+0.55) -- (0.2+\x,\y+0.7)  -- (0.2+\x,\y+0.85) -- (-0.2+\x,\y +1.) -- (-0.2+\x,\y+1.2);
\draw[thick,rounded corners=1.5] (-0.6+\x,0+\y) -- (-0.6+\x,0.15+\y) arc (0:90:0.15) -- (\x/2+0.1,0.3+\y) -- (\x/2+0.05,0.35+\y);
\draw[thick,rounded corners=1.5] (\x/2-0.05,0.45+\y) -- (\x/2-0.1,0.5+\y) -- (\x/2-0.3,0.5+\y) -- (\x/2-0.35,0.55+\y);
\draw[thick,rounded corners=1.5] (\x/2-0.45,0.65+\y) -- (\x/2-0.5,0.7+\y) -- (0.35,0.7+\y) arc (-90:-180:0.15) -- (-0.2,1.0+\y) -- (-0.2,1.2+\y); 
\draw[very thick,lime] (0.1,0.2875+\y+0.6) -- (-0.1,0.3625+\y+0.6);
 \draw[thick,rounded corners=1.5] (-0.6,0+\y) -- (-0.6,\y+0.55) -- (-0.2,\y+0.7)  -- (-0.2,\y+0.85) -- (0.2,\y +1.) -- (0.2,1.2+\y);
\draw[thick] (-0.2+\x,0+\y) -- (-0.2+\x,0.35+\y) arc (0:90:0.15) -- (\x/2+0.5,0.5+\y) -- (\x/2+0.45,0.55+\y);
\draw[thick,rounded corners=1.5] (\x/2+0.35,0.65+\y) -- (\x/2+0.3,0.7+\y) -- (\x/2+0.1,0.7+\y) -- (\x/2+0.05,0.75+\y);
\draw[thick,rounded corners=1.5] (\x/2-0.05,0.85+\y) -- (\x/2-0.1,0.9+\y) -- (0.75,0.9+\y) arc (-90:-180:0.15) -- (0.6,1.2+\y);
\fill[black] (-0.6,0+\y) circle (0.05);
\fill[black] (-0.2,0+\y) circle (0.05);
\fill[black] (0.2,0+\y) circle (0.05);
\fill[black] (0.6,0+\y) circle (0.05);
\fill[black] (-0.6+\x,0+\y) circle (0.05);
\fill[black] (-0.2+\x,0+\y) circle (0.05);
\fill[black] (0.2+\x,0+\y) circle (0.05);
\fill[black] (0.6+\x,0+\y) circle (0.05);
\renewcommand{\y}{1.8}
\fill[gray] (0,0+\y) ellipse (0.9cm and 0.2cm);
\fill[black] (-0.6,0+\y) circle (0.05);
\fill[black] (-0.2,0+\y) circle (0.05);
\fill[black] (0.2,0+\y) circle (0.05);
\fill[black] (0.6,0+\y) circle (0.05);
\fill[gray] (0+\x,0+\y) ellipse (0.9cm and 0.2cm);
\fill[black] (-0.6+\x,0+\y) circle (0.05);
\fill[black] (-0.2+\x,0+\y) circle (0.05);
\fill[black] (0.2+\x,0+\y) circle (0.05);
\fill[black] (0.6+\x,0+\y) circle (0.05);
\draw[dashed] (-0.6+\x,0+\y) -- (-0.6+\x,-0.15+\y);
\draw[dashed] (-0.2+\x,0+\y) -- (-0.2+\x,-0.2+\y);
\draw[dashed] (0.2+\x,0+\y) -- (0.2+\x,-0.2+\y);
\draw[dashed] (0.6+\x,0+\y) -- (0.6+\x,-0.2+\y);
\draw[dashed] (-0.6,0+\y) -- (-0.6,-0.15+\y);
\draw[dashed] (-0.2,0+\y) -- (-0.2,-0.2+\y);
\draw[dashed] (0.2,0+\y) -- (0.2,-0.2+\y);
\draw[dashed] (0.6,0+\y) -- (0.6,-0.15+\y);
\renewcommand{\y}{2.3}
\fill[brown] (0,0+\y) ellipse (0.9cm and 0.2cm);
\draw[black] (0,0+\y) ellipse (0.9cm and 0.2cm);
\draw[black,ultra thick] (0.9,0+\y) arc (0:80:0.85);
\fill[brown] (-0.9,0+\y) arc (180:0:0.9);
\fill[olive] (0+\x,0+\y) ellipse (0.9cm and 0.2cm);
\draw[black] (0+\x,0+\y) ellipse (0.9cm and 0.2cm);
\draw[black,ultra thick] (0.9+\x,0+\y) arc (0:80:0.85);
\fill[olive] (-0.9+\x,0+\y) arc (180:0:0.9);
\renewcommand{\x}{6}
\fill[gray] (0+\x,0) ellipse (0.9cm and 0.2cm);
\fill[black] (-0.6+\x,0) circle (0.05);
\fill[black] (-0.2+\x,0) circle (0.05);
\fill[black] (0.2+\x,0) circle (0.05);
\fill[black] (0.6+\x,0) circle (0.05);
\draw[dashed] (-0.6+\x,0) -- (-0.6+\x,-0.05) arc (0:-90:0.15);
\draw[dashed] (-0.2+\x,0) -- (-0.2+\x,-0.2);
\draw[dashed] (0.2+\x,0) -- (0.2+\x,-0.2);
\draw[dashed] (0.6+\x,0) -- (0.6+\x,-0.2);
\renewcommand{\y}{0.5}
                    \fill[lightgray] (0+\x,0+\y) ellipse (0.9cm and 0.2cm);
                    \draw[thick] (0+\x,0+\y) ellipse (0.9cm and 0.2cm);
                    \fill[lightgray] (-0.9+\x,\y) rectangle (0.9+\x,\y+1.3);
                    \draw[thick] (0.9+\x,\y) -- (0.9+\x,\y+1.3);
                    \renewcommand{\y}{1.8}
                    \fill[gray] (0+\x,0+\y) ellipse (0.9cm and 0.2cm);
                    \draw[dashed] (-0.6+\x,0+\y) -- (-0.6+\x,\y-0.2);
                    \draw[dashed] (-0.2+\x,0+\y) -- (-0.2+\x,\y-0.2);
                    \draw[dashed] (0.2+\x,0+\y) -- (0.2+\x,\y-0.2);
                    \draw[dashed] (0.6+\x,0+\y) -- (0.6+\x,\y-0.2);
                    \fill[black] (-0.6+\x,0+\y) circle (0.05);
                    \fill[black] (-0.2+\x,0+\y) circle (0.05);
                    \fill[black] (0.2+\x,0+\y) circle (0.05);
                    \fill[black] (0.6+\x,0+\y) circle (0.05);
\end{tikzpicture}
    \end{minipage}
    \begin{minipage}{0.4\linewidth}
        \begin{tabular}{ccc}
             \scalebox{0.6}{$\begin{array}{c}
                  \begin{tikzpicture}
                      \fill[brown] (0,0) ellipse (0.9cm and 0.2cm);
                    \draw[black] (0,0) ellipse (0.9cm and 0.2cm);
                    \draw[black,ultra thick] (0.9,0) arc (0:80:0.85);
                    \fill[brown] (-0.9,0) arc (180:0:0.9);
                  \end{tikzpicture}
             \end{array}$} &  \scalebox{0.6}{$\begin{array}{c}
                  \begin{tikzpicture}
                      \fill[olive] (0,0) ellipse (0.9cm and 0.2cm);
                    \draw[black] (0,0) ellipse (0.9cm and 0.2cm);
                    \draw[black,ultra thick] (0.9,0) arc (0:80:0.85);
                    \fill[olive] (-0.9,0) arc (180:0:0.9);
                  \end{tikzpicture}
             \end{array}$} & \scalebox{0.6}{$\begin{array}{c}
                \begin{tikzpicture}
                \newcommand{\y}{0.5}
                    \fill[lightgray] (0,0+\y) ellipse (0.9cm and 0.2cm);
                    \draw[thick] (0,0+\y) ellipse (0.9cm and 0.2cm);
                    \fill[lightgray] (-0.9,\y) rectangle (0.9,\y+1.3);
                    \draw[thick] (0.9,\y) -- (0.9,\y+1.3);
                    \renewcommand{\y}{1.8}
                    \fill[gray] (0,0+\y) ellipse (0.9cm and 0.2cm);
                    \draw[dashed] (-0.6,0+\y) -- (-0.6,\y-0.2);
                    \draw[dashed] (-0.2,0+\y) -- (-0.2,\y-0.2);
                    \draw[dashed] (0.2,0+\y) -- (0.2,\y-0.2);
                    \draw[dashed] (0.6,0+\y) -- (0.6,\y-0.2);
                    \fill[black] (-0.6,0+\y) circle (0.05);
                    \fill[black] (-0.2,0+\y) circle (0.05);
                    \fill[black] (0.2,0+\y) circle (0.05);
                    \fill[black] (0.6,0+\y) circle (0.05);
                \end{tikzpicture} 
           \end{array}$}\\ [15pt]
            \scalebox{0.8}{$\begin{array}{c}
                  \begin{tikzpicture}
                      \fill[gray] (0,0) ellipse (0.9cm and 0.2cm);
\fill[black] (-0.6,0) circle (0.05);
\fill[black] (-0.2,0) circle (0.05);
\fill[black] (0.2,0) circle (0.05);
\fill[black] (0.6,0) circle (0.05);
\draw[thick] (-0.6,0) -- (-0.6,0.5) arc (180:0:0.2) -- (-0.2,0);
\draw[thick] (0.2,0) -- (0.2,0.5) arc (180:0:0.2) -- (0.6,0);
                  \end{tikzpicture}
             \end{array}$} &  \scalebox{0.8}{$\begin{array}{c}
                  \begin{tikzpicture}
                      \fill[gray] (0,0) ellipse (0.9cm and 0.2cm);
\fill[black] (-0.6,0) circle (0.05);
\fill[black] (-0.2,0) circle (0.05);
\fill[black] (0.2,0) circle (0.05);
\fill[black] (0.6,0) circle (0.05);
\draw[thick] (-0.6,0) -- (-0.6,0.5) arc (180:0:0.2) -- (-0.2,0);
\draw[thick] (0.2,0) -- (0.2,0.5) arc (180:0:0.2) -- (0.6,0);
                  \end{tikzpicture}
             \end{array}$} & \scalebox{0.7}{$\begin{array}{c} 
                \begin{tikzpicture}
                \newcommand{\y}{0.5}
                    \fill[gray] (0,0+\y) ellipse (0.9cm and 0.2cm);
                    \draw[very thick] (-0.6,\y) -- (-0.6,\y+1.2);
                    \draw[very thick] (-0.2,\y) -- (-0.2,\y+1.2);
                    \draw[very thick] (0.2,\y) -- (0.2,\y+1.2);
                    \draw[very thick] (0.6,\y) -- (0.6,\y+1.2);
                    \fill[black] (-0.6,0+\y) circle (0.05);
                    \fill[black] (-0.2,0+\y) circle (0.05);
                    \fill[black] (0.2,0+\y) circle (0.05);
                    \fill[black] (0.6,0+\y) circle (0.05);
                    \renewcommand{\y}{1.8}
                    \fill[gray] (0,0+\y) ellipse (0.9cm and 0.2cm);
                    \draw[dashed] (-0.6,0+\y) -- (-0.6,\y-0.2);
                    \draw[dashed] (-0.2,0+\y) -- (-0.2,\y-0.2);
                    \draw[dashed] (0.2,0+\y) -- (0.2,\y-0.2);
                    \draw[dashed] (0.6,0+\y) -- (0.6,\y-0.2);
                    \fill[black] (-0.6,0+\y) circle (0.05);
                    \fill[black] (-0.2,0+\y) circle (0.05);
                    \fill[black] (0.2,0+\y) circle (0.05);
                    \fill[black] (0.6,0+\y) circle (0.05);
                \end{tikzpicture} 
           \end{array}$} \\ [15pt]
            \scalebox{0.8}{$\begin{array}{c}
                  \begin{tikzpicture}
                      \fill[gray] (0,0) ellipse (0.9cm and 0.2cm);
\fill[black] (-0.6,0) circle (0.05);
\fill[black] (-0.2,0) circle (0.05);
\fill[black] (0.2,0) circle (0.05);
\fill[black] (0.6,0) circle (0.05);
\draw[thick] (-0.6,0) -- (-0.6,0.5) arc (180:0:0.2) -- (-0.2,0);
\draw[thick] (0.2,0) -- (0.2,0.5) arc (180:0:0.2) -- (0.6,0);
                  \end{tikzpicture}
             \end{array}$} & \scalebox{0.8}{$\begin{array}{c}
                  \begin{tikzpicture}
                      \fill[gray] (0,0) ellipse (0.9cm and 0.2cm);
\fill[black] (-0.6,0) circle (0.05);
\fill[black] (-0.2,0) circle (0.05);
\fill[black] (0.2,0) circle (0.05);
\fill[black] (0.6,0) circle (0.05);
\draw[thick] (-0.2,0) -- (-0.2,0.35) arc (180:90:0.15) -- (0.05,0.5) arc (90:0:0.15) -- (0.2,0);
\draw[thick] (-0.6,0) -- (-0.6,0.55) arc (180:90:0.15) -- (0.45,0.7) arc (90:0:0.15) -- (0.6,0);
                  \end{tikzpicture}
             \end{array}$}  & \scalebox{0.7}{$\begin{array}{c} 
                \begin{tikzpicture}
                \newcommand{\y}{0.5}
                    \fill[gray] (0,0+\y) ellipse (0.9cm and 0.2cm);
                    \draw[very thick,rounded corners=1.5] (0.2,\y) -- (0.2,\y+1.2);
                    \draw[very thick,rounded corners=1.5] (0.6,\y) -- (0.6,\y+1.2);
                    \draw[very thick,rounded corners=1.5] (-0.2,\y) -- (-0.2,\y+0.55) -- (-0.6,\y+0.7)  -- (-0.6,\y+1.2);
                    \draw[very thick,white] (-0.3,0.2875+\y+0.3) -- (-0.5,0.3625+\y+0.3);
                    \draw[very thick,rounded corners=1.5] (-0.6,\y) -- (-0.6,\y+0.55) -- (-0.2,\y+0.7) -- (-0.2,\y+1.2);
                    \fill[black] (-0.6,0+\y) circle (0.05);
                    \fill[black] (-0.2,0+\y) circle (0.05);
                    \fill[black] (0.2,0+\y) circle (0.05);
                    \fill[black] (0.6,0+\y) circle (0.05);
                    \renewcommand{\y}{1.8}
                    \fill[gray] (0,0+\y) ellipse (0.9cm and 0.2cm);
                    \draw[dashed] (-0.6,0+\y) -- (-0.6,\y-0.2);
                    \draw[dashed] (-0.2,0+\y) -- (-0.2,\y-0.2);
                    \draw[dashed] (0.2,0+\y) -- (0.2,\y-0.2);
                    \draw[dashed] (0.6,0+\y) -- (0.6,\y-0.2);
                    \fill[black] (-0.6,0+\y) circle (0.05);
                    \fill[black] (-0.2,0+\y) circle (0.05);
                    \fill[black] (0.2,0+\y) circle (0.05);
                    \fill[black] (0.6,0+\y) circle (0.05);
                \end{tikzpicture} 
           \end{array}$} \\ [15pt]
             \scalebox{0.8}{$\begin{array}{c}
                  \begin{tikzpicture}
                      \fill[gray] (0,0) ellipse (0.9cm and 0.2cm);
\fill[black] (-0.6,0) circle (0.05);
\fill[black] (-0.2,0) circle (0.05);
\fill[black] (0.2,0) circle (0.05);
\fill[black] (0.6,0) circle (0.05);
\draw[thick] (-0.2,0) -- (-0.2,0.35) arc (180:90:0.15) -- (0.05,0.5) arc (90:0:0.15) -- (0.2,0);
\draw[thick] (-0.6,0) -- (-0.6,0.55) arc (180:90:0.15) -- (0.45,0.7) arc (90:0:0.15) -- (0.6,0);
                  \end{tikzpicture}
             \end{array}$} & \scalebox{0.8}{$\begin{array}{c}
                  \begin{tikzpicture}
                      \fill[gray] (0,0) ellipse (0.9cm and 0.2cm);
\fill[black] (-0.6,0) circle (0.05);
\fill[black] (-0.2,0) circle (0.05);
\fill[black] (0.2,0) circle (0.05);
\fill[black] (0.6,0) circle (0.05);
\draw[thick] (-0.6,0) -- (-0.6,0.5) arc (180:0:0.2) -- (-0.2,0);
\draw[thick] (0.2,0) -- (0.2,0.5) arc (180:0:0.2) -- (0.6,0);
                  \end{tikzpicture}
             \end{array}$} & \scalebox{0.7}{$\begin{array}{c} 
                \begin{tikzpicture}
                \newcommand{\y}{0.5}
                    \fill[gray] (0,0+\y) ellipse (0.9cm and 0.2cm);
                    \draw[very thick,rounded corners=1.5] (0.2,\y) -- (0.2,\y+0.55) -- (0.6,\y+0.7) -- (0.6,\y+1.2);
                    \draw[very thick,white] (0.3,0.2875+\y+0.3) -- (0.5,0.3625+\y+0.3);
                    \draw[very thick,rounded corners=1.5] (0.6,\y) -- (0.6,\y+0.55) -- (0.2,\y+0.7)  -- (0.2,\y+1.2);
                    \draw[very thick,rounded corners=1.5] (-0.2,\y) -- (-0.2,\y+1.2);
                    \draw[very thick,rounded corners=1.5] (-0.6,\y) -- (-0.6,\y+1.2);
                    \fill[black] (-0.6,0+\y) circle (0.05);
                    \fill[black] (-0.2,0+\y) circle (0.05);
                    \fill[black] (0.2,0+\y) circle (0.05);
                    \fill[black] (0.6,0+\y) circle (0.05);
                    \renewcommand{\y}{1.8}
                    \fill[gray] (0,0+\y) ellipse (0.9cm and 0.2cm);
                    \draw[dashed] (-0.6,0+\y) -- (-0.6,\y-0.2);
                    \draw[dashed] (-0.2,0+\y) -- (-0.2,\y-0.2);
                    \draw[dashed] (0.2,0+\y) -- (0.2,\y-0.2);
                    \draw[dashed] (0.6,0+\y) -- (0.6,\y-0.2);
                    \fill[black] (-0.6,0+\y) circle (0.05);
                    \fill[black] (-0.2,0+\y) circle (0.05);
                    \fill[black] (0.2,0+\y) circle (0.05);
                    \fill[black] (0.6,0+\y) circle (0.05);
                \end{tikzpicture} 
           \end{array}$} \\ [15pt]
            \scalebox{0.8}{$\begin{array}{c}
                  \begin{tikzpicture}
                      \fill[gray] (0,0) ellipse (0.9cm and 0.2cm);
\fill[black] (-0.6,0) circle (0.05);
\fill[black] (-0.2,0) circle (0.05);
\fill[black] (0.2,0) circle (0.05);
\fill[black] (0.6,0) circle (0.05);
\draw[thick] (-0.2,0) -- (-0.2,0.35) arc (180:90:0.15) -- (0.05,0.5) arc (90:0:0.15) -- (0.2,0);
\draw[thick] (-0.6,0) -- (-0.6,0.55) arc (180:90:0.15) -- (0.45,0.7) arc (90:0:0.15) -- (0.6,0);
                  \end{tikzpicture}
             \end{array}$} & \scalebox{0.8}{$\begin{array}{c}
                  \begin{tikzpicture}
                      \fill[gray] (0,0) ellipse (0.9cm and 0.2cm);
\fill[black] (-0.6,0) circle (0.05);
\fill[black] (-0.2,0) circle (0.05);
\fill[black] (0.2,0) circle (0.05);
\fill[black] (0.6,0) circle (0.05);
\draw[thick] (-0.2,0) -- (-0.2,0.35) arc (180:90:0.15) -- (0.05,0.5) arc (90:0:0.15) -- (0.2,0);
\draw[thick] (-0.6,0) -- (-0.6,0.55) arc (180:90:0.15) -- (0.45,0.7) arc (90:0:0.15) -- (0.6,0);
                  \end{tikzpicture}
             \end{array}$} & \scalebox{0.7}{$\begin{array}{c} 
                \begin{tikzpicture}
                \newcommand{\y}{0.5}
                    \fill[gray] (0,0+\y) ellipse (0.9cm and 0.2cm);
                    \draw[very thick,rounded corners=1.5] (0.2,\y) -- (0.2,\y+0.55) -- (0.6,\y+0.7) -- (0.6,\y+1.2);
                    \draw[very thick,white] (0.3,0.2875+\y+0.3) -- (0.5,0.3625+\y+0.3);
                    \draw[very thick,rounded corners=1.5] (0.6,\y) -- (0.6,\y+0.55) -- (0.2,\y+0.7)  -- (0.2,\y+1.2);
                    \draw[very thick,rounded corners=1.5] (-0.2,\y) -- (-0.2,\y+0.55) -- (-0.6,\y+0.7)  -- (-0.6,\y+1.2);
                    \draw[very thick,white] (-0.3,0.2875+\y+0.3) -- (-0.5,0.3625+\y+0.3);
                    \draw[very thick,rounded corners=1.5] (-0.6,\y) -- (-0.6,\y+0.55) -- (-0.2,\y+0.7) -- (-0.2,\y+1.2);
                    \fill[black] (-0.6,0+\y) circle (0.05);
                    \fill[black] (-0.2,0+\y) circle (0.05);
                    \fill[black] (0.2,0+\y) circle (0.05);
                    \fill[black] (0.6,0+\y) circle (0.05);
                    \renewcommand{\y}{1.8}
                    \fill[gray] (0,0+\y) ellipse (0.9cm and 0.2cm);
                    \draw[dashed] (-0.6,0+\y) -- (-0.6,\y-0.2);
                    \draw[dashed] (-0.2,0+\y) -- (-0.2,\y-0.2);
                    \draw[dashed] (0.2,0+\y) -- (0.2,\y-0.2);
                    \draw[dashed] (0.6,0+\y) -- (0.6,\y-0.2);
                    \fill[black] (-0.6,0+\y) circle (0.05);
                    \fill[black] (-0.2,0+\y) circle (0.05);
                    \fill[black] (0.2,0+\y) circle (0.05);
                    \fill[black] (0.6,0+\y) circle (0.05);
                \end{tikzpicture} 
           \end{array}$}
        \end{tabular}
    \end{minipage}
    \caption{The topological version of the quantum teleportation protocol~\cite{Melnikov:2022vij}. Alice (lime background) possesses a qubit in an unspecified state (black cup). She also receives a qubit from an entangled pair (pink). Brown and olive caps denote projections of the results of Alice's operations on a two-qubit basis in the table on the right. To retrieve Alice's black qubit Bob (blue) needs to apply an operation denoted by the gray cylinder, according to the outcome of Alice's measurement, as instructed by the table.}
    \label{fig:teleport}
\end{figure}
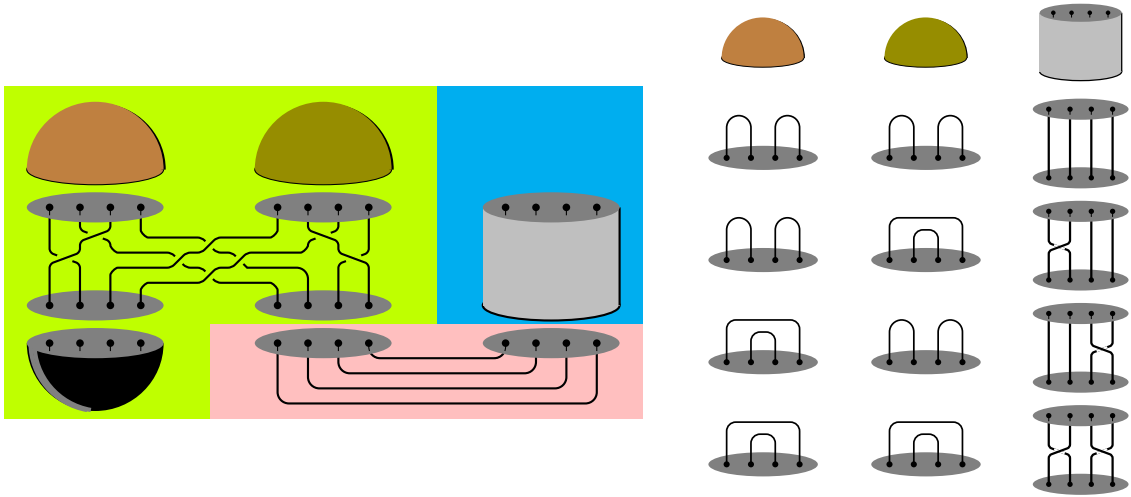

\subsubsection{Quantum teleportation} 

Quantum teleportation is another basic quantum protocol in which Alice has an unknown quantum state, which she would like to pass to Bob~\cite{Bennett:1992tv}. As in the case of dense coding, Alice and Bob pre-share a pair of entangled qubits. In order to transfer the unknown qubit, Alice first entangles it with her half of the shared pair and then measures both qubits. The result of the measurement is communicated classically to Bob, who can recover the unknown qubit on his half of the shared pair after applying transformations defined by the received classical information.

The topological version of the teleportation protocol is shown in figure~\ref{fig:teleport}. Again we use~(\ref{4pointbasis0}) as the measurement basis. In her manipulations Alice can use the same two-qubit entangling gate as Bob used in the dense coding protocol, cf. figure~\ref{fig:densecoding}. Figure~\ref{fig:teleport} shows which transformations Bob needs to apply on his qubit for each one of four possible outputs of Alice's measurement. In fact, they are the same transformations as the ones Alice would use in the dense coding up to the one qubit transformation used by Bob in the same protocol.

\subsection{Model of unitary evaporation} 
\label{sec:blackhole}

Quantum teleportation algorithm, especially in its topological version, highlights some properties of entanglement of interacting particles. Interactions modify the entanglement which results in a transfer of some particle properties, which we generally refer to as information (about states or particles), from one particle state to a distant particle state. The information, however, is not accessible unless there is a classical channel that provides details about the interaction. Otherwise the information is encrypted.

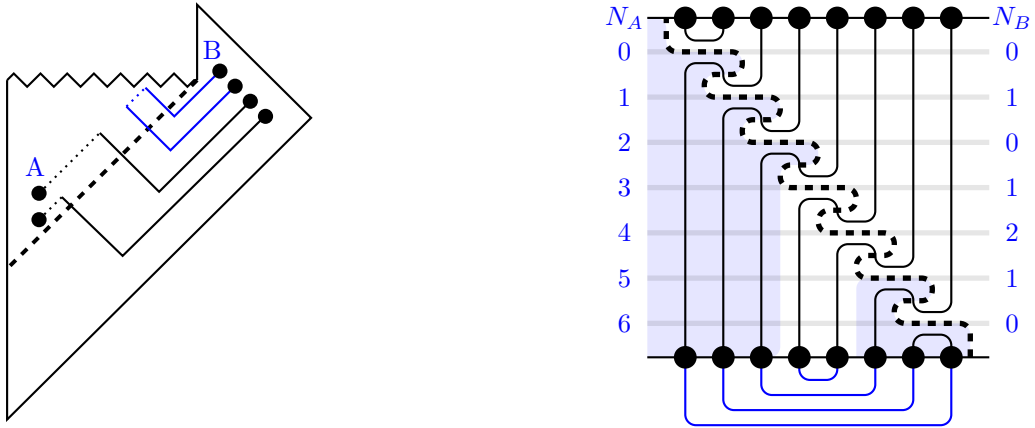
\begin{figure}[htb]
\begin{minipage}{0.45\linewidth}
    \centering
    \begin{tikzpicture}
    \draw[thick] (0,4.5) -- (0,0) -- (4,4) -- (2.5,5.5) -- (2.5,4.5);
    \draw[thick,decorate,decoration=zigzag] (0,4.5) -- (2.5,4.5);
    \draw[thick,dashed,line width=1.5] (2.5,4.5) -- (0,2);
    \fill[black] (0.42,3) circle (0.1);
    \fill[black] (3.2,4.22) circle (0.1); 
    \draw[thick,dotted] (0.42,3) -- (1.22,3.8);
    \draw[thick] (1.22,3.8) -- (2.0,3.02) -- (3.2,4.22);
    \fill[black] (0.42,2.65) circle (0.1);
    \draw[thick,dotted] (0.42,2.65) -- (0.72,2.95);
    \draw[thick] (0.72,2.95) -- (1.52,2.17) -- (3.4,4.03);
    \fill[black] (3.4,4.02) circle (0.1);
    \draw[thick,blue] (2.8,4.62) -- (2.2,4.02) -- (1.82,4.4);
    \draw[thick,blue] (3.0,4.42) -- (2.15,3.57) -- (1.57,4.15); 
    \draw[thick,blue,dotted] (1.57,4.15) -- (1.82,4.4);
    \fill[black] (3.0,4.42) circle (0.1);
    \fill[black] (2.8,4.62) circle (0.1);
    \draw[blue] (2.7,4.9) node {B};
    \draw[blue] (0.37,3.35) node {A};
    \end{tikzpicture}
    \end{minipage}
    \hfill{
    \begin{minipage}{0.45\linewidth}
        \centering
        \begin{tikzpicture}
        \fill[blue,rounded corners=4,opacity=0.1] (4.25,0) -- (4.25,0.45) -- (3.25,0.45) -- (3.25,0.75) -- (3.75,0.75) -- (3.75,1.05) -- (2.75,1.05) -- (2.75,0.0);
        \fill[blue,rounded corners=4,opacity=0.1] (0,0) -- (1.75,0) -- (1.75,2.55) -- (2.25,2.55) -- (2.25,2.85) -- (1.25,2.85) -- (1.25,3.15) -- (1.75,3.15) -- (1.75,3.45) -- (0.75,3.45) -- (0.75,3.75) -- (1.25,3.75) -- (1.25,4.05) -- (0.25,4.05) -- (0.25,4.5) -- (0,4.5);
        \foreach \a in {0.45,1.05,...,4.65}
            \draw[line width=2,opacity=0.1] (0,\a) -- (4.5,\a);
        \draw[thick] (0,0) -- (4.5,0);
        \draw[thick] (0,4.5) -- (4.5,4.5);
        \draw[blue,thick,rounded corners=4] (2,0) -- (2,-0.3) -- (2.5,-0.3) -- (2.5,0);
        \draw[blue,thick,rounded corners=4] (1.5,0) -- (1.5,-0.5) -- (3.0,-0.5) -- (3.0,0);
        \draw[blue,thick,rounded corners=4] (1.,0) -- (1.,-0.7) -- (3.5,-0.7) -- (3.5,0);
        \draw[blue,thick,rounded corners=4] (0.5,0) -- (0.5,-0.9) -- (4.0,-0.9) -- (4.0,0);
        \foreach \a in {0.5,1.0,...,4.0}
            \fill[black] (\a,0) circle (0.15);
        \foreach \a in {0.5,1.0,...,4.0}
            \fill[black] (\a,4.5) circle (0.15);
        \draw[thick,rounded corners=4] (3.5,0) -- (3.5,0.3) -- (4.0,0.3) -- (4.0,0);
        \draw[thick,rounded corners=4] (0.5,4.5) -- (0.5,4.2) -- (1.0,4.2) -- (1.0,4.5);
        \newcommand{\x}{3.0}
        \newcommand{\y}{0.9}
        \draw[thick,rounded corners=4] (\x,0) -- (\x,\y) -- (\x+0.5,\y) -- (\x+0.5,\y-0.3) -- (\x+1,\y-0.3) -- (\x+1,4.5);
        \renewcommand{\x}{2.5}
        \renewcommand{\y}{1.5}
        \draw[thick,rounded corners=4] (\x,0) -- (\x,\y) -- (\x+0.5,\y) -- (\x+0.5,\y-0.3) -- (\x+1,\y-0.3) -- (\x+1,4.5);
        \renewcommand{\x}{2.0}
        \renewcommand{\y}{2.1}
        \draw[thick,rounded corners=4] (\x,0) -- (\x,\y) -- (\x+0.5,\y) -- (\x+0.5,\y-0.3) -- (\x+1,\y-0.3) -- (\x+1,4.5);
        \renewcommand{\x}{1.5}
        \renewcommand{\y}{2.7}
        \draw[thick,rounded corners=4] (\x,0) -- (\x,\y) -- (\x+0.5,\y) -- (\x+0.5,\y-0.3) -- (\x+1,\y-0.3) -- (\x+1,4.5);
        \renewcommand{\x}{1.0}
        \renewcommand{\y}{3.3}
        \draw[thick,rounded corners=4] (\x,0) -- (\x,\y) -- (\x+0.5,\y) -- (\x+0.5,\y-0.3) -- (\x+1,\y-0.3) -- (\x+1,4.5);
        \renewcommand{\x}{0.5}
        \renewcommand{\y}{3.9}
        \draw[thick,rounded corners=4] (\x,0) -- (\x,\y) -- (\x+0.5,\y) -- (\x+0.5,\y-0.3) -- (\x+1,\y-0.3) -- (\x+1,4.5);
        \draw[dashed,rounded corners=4,line width=2] (4.25,0) -- (4.25,0.45) -- (3.25,0.45) -- (3.25,0.75) -- (3.75,0.75) -- (3.75,1.05) -- (2.75,1.05) -- (2.75,1.35) -- (3.25,1.35) -- (3.25,1.65) -- (2.25,1.65) -- (2.25,1.95) -- (2.75,1.95) -- (2.75,2.25) -- (1.75,2.25) -- (1.75,2.55) -- (2.25,2.55) -- (2.25,2.85) -- (1.25,2.85) -- (1.25,3.15) -- (1.75,3.15) -- (1.75,3.45) -- (0.75,3.45) -- (0.75,3.75) -- (1.25,3.75) -- (1.25,4.05) -- (0.25,4.05) -- (0.25,4.5);
        \draw[blue] (-0.3,4.5) node {$N_A$};
        \draw[blue] (4.8,4.5) node {$N_B$};
        \draw[blue] (-0.3,0.45) node {$6$};
        \draw[blue] (4.8,0.45) node {$0$};
        \draw[blue] (-0.3,1.05) node {$5$};
        \draw[blue] (4.8,1.05) node {$1$};
        \draw[blue] (-0.3,1.65) node {$4$};
        \draw[blue] (4.8,1.65) node {$2$};
        \draw[blue] (-0.3,2.25) node {$3$};
        \draw[blue] (4.8,2.25) node {$1$};
        \draw[blue] (-0.3,2.85) node {$2$};
        \draw[blue] (4.8,2.85) node {$0$};
        \draw[blue] (-0.3,3.45) node {$1$};
        \draw[blue] (4.8,3.45) node {$1$};
        \draw[blue] (-0.3,4.05) node {$0$};
        \draw[blue] (4.8,4.05) node {$0$};
        \end{tikzpicture}
    \end{minipage}
    }
    \caption{(Left) A causal diagram of an evaporating black hole. Solid straight lines show the trajectories of Hawking pairs. Interactions of the Hawking quanta in the interior result in the entanglement of the Hawking radiation with the internal modes of the black hole (dotted lines) but they can also result in the teleportation of the interior states to the exterior (blue lines). (Right) The evaporation of the black hole in the topological model. The blue lines show the initial scrambled state of the black hole. The dashed line shows the location of the horizon, which separates the interior (shaded) and exterior, and contracts as long as the evaporation progresses. The shape of this line is adapted to reflect the fact that the on-shell particles cross it in one direction, while the off-shell ones in the opposite. Solid black lines show evolution histories of the black hole degrees of freedom, which include pair creation-annihilation processes prescribed by the S matrix~(\ref{evolop}). The void in the interior corresponds to the formation of the island, as explained in the text. Numbers on the sides count the number of degrees of freedom contained in the interior ($N_A$) and entangled with the Hawking radiation ($N_B$) at discrete steps of evolution.}
    \label{fig:causald}
\end{figure}

This mechanism of the transfer of properties is a possible way to explain, how the information contained in a causally disconnected region, such as the interior of a black hole, can be retrieved in the presence of entanglement with that region~\cite{Melnikov:2022vij}. In the case of a black hole Hawking pairs is the source of the shared entanglement between the exterior and interior~\cite{Hawking:1976ra}. Hawking pairs are created by the gravitational field of the black hole, with one particle of the entangled pair falling inside the horizon and the other escaping the black hole and being collected by a distant observer~\cite{Hawking:1975vcx}. Trajectories of such pairs are shown as solid straight lines in the causal diagram in the left panel of figure~\ref{fig:causald}. 

Particles inside the black hole must interact. Creation of the Hawking pairs themselves is one of the results of these interactions. The other type of interactions expected inside the black hole is the one that guarantees maximal scrambling of the information in the interior~\cite{Hayden:2007cs}. These kind of interactions can engage the quantum teleportation protocol so that the information of the interacting quanta are transferred to a state of their entangled cousins outside the black hole. The latter situation is illustrated by the blue lines in figure~\ref{fig:causald} (left).

The essential features of such processes can be illustrated by the following simple topological model, considered in~\cite{Melnikov:2022vij}. Let us assume that system $A$ is causally disconnected from system $B$. We assume that $A$ and $B$ are separated by a horizon-like interface that on-shell particles can only cross in one direction, from $B$ to $A$. In the meantime, virtual particles can cross the interface in either directions. Let us use solid lines for the trajectories of the on-shell particles. Virtual particles, which are intermediate states in the interactions of the on-shell particles will have no lines, but will rather appear as voids. A toy two-particle S matrix of a relevant interaction can be chosen to be the R matrix,
\be
\begin{array}{c}
\scalebox{0.7}{\begin{tikzpicture}
\draw[line width=2.0,rounded corners=4] (0,0) -- (0,0.55) -- (0.75,0.85) -- (0.75,1.4);
\draw[line width=2,rounded corners=4] (0,1.4) -- (0,0.85) -- (0.25,0.75);
\draw[line width=2,rounded corners=4] (0.5,0.65) -- (0.75,0.55) -- (0.75,0);
\end{tikzpicture}}
\end{array}
\ = \ A
\begin{array}{c}
\scalebox{0.7}{\begin{tikzpicture}
\draw[line width=2,rounded corners=4] (0,0) -- (0.,1.4);
\draw[line width=2,rounded corners=4] (0.75,1.4) -- (0.75,0);
\end{tikzpicture}}
\end{array} \ + \ A^{-1}
\begin{array}{c}
\scalebox{0.7}{\begin{tikzpicture}
\draw[line width=2,rounded corners=4] (0,0) -- (0,0.55) -- (0.75,0.55) -- (0.75,0);
\draw[line width=2,rounded corners=4] (0,1.4) -- (0,0.85) -- (0.75,0.85) -- (0.75,1.4);
\end{tikzpicture}}
\end{array}\,,
\label{evolop}
\ee
where the first diagram on the right is a trivial scattering and the second is the particle production via annihilation with a virtual particle in the intermediate state. This is of course the skein relation~(\ref{skein}). 

Let us assume that particles in the interior interact via~(\ref{evolop}) close to the horizon in such a way that the virtual particle crosses the interface and a pair is created in the exterior. This is the evaporation of the black hole~\cite{Hawking:1974rv,Hawking:1975vcx}. A series of such evaporation events is shown in the right diagram of figure~\ref{fig:causald}, where the horizon is shown as a dashed line. 

One of the created on-shell particles returns to the interior, crossing the interface in the allowed direction. As a result entanglement is formed between the exterior and the interior in the sense discussed in this paper. By the same kind of interaction, the in-falling particle gets ``scrambled'' in the interior at the initial steps of the evaporation. In the topological setup the scrambling is modeled by entangling modes in the deep interior and the modes close to the horizon. In this way the out-going particle gets entangled with the particles deep in the interior. 

While this process continues, and more Hawking pairs are created, the entanglement of the escaped particles with the interior grows. Horizontal lines show discrete time steps in the diagram of figure~\ref{fig:causald} (right). At approximately half of the evaporation one has the external particles connected with all the remaining particles in the interior. At this moment the number of internal degrees of freedom of the black hole is equal to the number of entangled Hawking quanta (the black hole entropy and the entanglement entropy are the same). After that moment, the in-falling quanta cannot entangle with the particles in the interior without destroying their entanglement with the exterior, as a consequence of monogamy. From that moment the number of pairs connecting the interior and the exterior begins to fall, and no such pairs remain when the black hole evaporates completely. 

In this model the evaporation happens unitarily and no information paradox appears~\cite{Hawking:1976ra,Mathur:2009hf}. In particular, one can measure the entanglement by counting the lines connecting particles in the interior and exterior, which as we argued before, gives a measure of entanglement entropy. The entropy of the black hole is defined by the number of lines in the interior at a given moment of evaporation. The entropy of the Hawking radiation is counted by the number of lines connecting the exterior and interior. Clearly the latter cannot be bigger than the former. In terms of the example of figure~\ref{fig:causald} (right), this fact is reflected in the inequality $N_A\geq N_B$. While $N_A$ decreases with time, $N_B$ may increase initially, but is eventually bounded by $N_A$, providing a unitary example of the Page curve~\cite{Page:1993wv}.

The notion of the ``island''~\cite{Almheiri:2019hni}, which appears in the recent proposals of the solution of the information paradox~\cite{Penington:2019npb,Almheiri:2019psf,Penington:2019kki} can be illustrated by the topological model. In figure~\ref{fig:causald} (right) the moment of the first engagement of quantum teleportation occurs after the third Hawking pair is created. A pair of entangled particles in the center is teleported to the exterior and the corresponding void is indicated by the unshaded part of the interior. The complement of the unshaded part at the given moment of time is the island.

\section{Conclusions}
\label{sec:conclusions}

In this paper we have shown how topology can encode correlations in quantum systems. One of the messages that was conveyed is that topology can provide an intuitive understanding of quantum entanglement and its properties and can assist in developing new theoretical tools and applications. Let us summarize some possible further questions that can be addressed in the topological approach.

In the study of the classification of entanglement we have found that planar representatives of connectome classes are only sufficient for the description of the SLOCC classes of bipartite entanglement. For multipartite entanglement the class of topologies must be extended. It remains to be verified, whether the connectome classes contain all the SLOCC entanglement classes or they can only approximate the latter. A basic version of this question is whether there is a topological representation of W entanglement of three qubits.

We have argued that planar connectome states are those realizing maximal entanglement in each class of the bipartite entanglement. We have also demonstrated how tangling reduces entanglement. It would be interesting to perform a more systematic study of the effects of tangling or non-trivial 3D topology on entanglement. Such a study can shed additional light on the problem of classification.

Topological realization gives an intuitive interpretation not only to quantum states, but more generally to correlations and measures of quantum entanglement. We used this advantage to introduce indicators of multipartite entanglement constructing analogs of reduced density matrices for the multipartite case. Although a more detailed study is necessary to show that such indicators are useful measures of entanglement, the topological approach seems to be a very simple tool for engineering specifically tailored measures.

We have seen that certain properties of entanglement are especially transparent in the topological interpretation. Planar connectomes are particularly simple quantum states, which illustrate the properties of entanglement and its measures. We have noticed that planar connectome states have similarities with the holographic quantum states, which includes the ``minimal area'' formula for the entanglement entropy and different inequalities that the entanglement entropy satisfies on both classes of states. Yet, the connectome states seem to be a more restricted class, saturating some of the inequalities of the holographic states. Therefore, it would be interesting to understand how the connectomes can be generalized to match the properties of the holographic states. Presumably this can be done by allowing non-trivial topologies, but one should also remember that in most studies holographic states are described by classical geometries, so there is a tantalizing perspective that generic topologies can encode quantum geometries, that is states of quantum gravity. In this respect the present work touches upon some active areas of research, including tensor network models of quantum gravity and low dimensional gravities, which are themselves topological theories.

A somewhat related problem is the generalization of the entropy formula for planar connectome states. We have argued that on this class counting Wilson lines gives a faithful measure of entanglement entropy. This formula should somehow be corrected on generic states -- a relevant problem for the holographic states as well.

A model of information retrieval from a causally disconnected region viewed here as a toy model of black hole evaporation was also motivated by recent discussion and progress in quantum gravity. The topological model makes the information transfer particularly manifest. The main question is whether such a mechanism can be consistently embedded in a theory with dynamical gravity. We can note that the specific interactions considered in the model were only chosen to make explicit the action of the teleportation protocol, but the protocol itself should work in a similar way for arbitrary particle interactions.

Finally, we have shown that basic quantum algorithms have a very intuitive realization in the topological setup. The dense coding algorithm is based on the realization of shared entanglement as a physical resource (Wilson lines) that can be manipulated and redistributed between parties. The quantum teleportation is based on the topological equivalence: topological sets can be deformed violating locality. It would of course be interesting to know if such topological tricks can be played to construct yet unknown quantum algorithms.

\paragraph{Acknowledgments.} This work was supported by RSF grant No.~18-71-10073.

\bibliographystyle{JHEP}
\bibliography{refs}

\end{document}